\documentclass[twoside, twocolumn, 9pt]{article}
\usepackage{extsizes}
\usepackage[super,sort&compress,comma]{natbib} 
\usepackage[version=3]{mhchem}
\usepackage[left=1.5cm, right=1.5cm, top=1.785cm, bottom=2.0cm]{geometry}
\usepackage{balance}
\usepackage{mathptmx}
\usepackage{graphicx} 
\usepackage{sectsty}
\usepackage{lastpage}
\usepackage[format=plain,justification=justified,singlelinecheck=false,font={stretch=1.125,small,sf},labelfont=bf,labelsep=space]{caption}
\usepackage{float}
\usepackage{fancyhdr}
\usepackage{fnpos}
\usepackage[english]{babel}
\addto{\captionsenglish}{%
  
}
\usepackage{array}
\usepackage{droidsans}
\usepackage{charter}
\usepackage[T1]{fontenc}
\usepackage[usenames,dvipsnames]{xcolor}
\usepackage{setspace}
\usepackage[compact]{titlesec}
\usepackage{hyperref}
\usepackage{wrapfig}
\usepackage{amssymb}
\usepackage{import}
\usepackage{amsmath}
\usepackage{soul}

\usepackage{physics}
\usepackage{epstopdf}%This line makes .eps figures into .pdf - please comment out if not required.

\newcommand*{\citen}[1]{%
  \begingroup
    \romannumeral-`\x % remove space at the beginning of \setcitestyle
    \setcitestyle{numbers}%
    \citep{#1}%
  \endgroup   
}

\definecolor{cream}{RGB}{222,217,201}

\begin{document}

\pagestyle{fancy}
\thispagestyle{plain}
\fancypagestyle{plain}{
%%%HEADER%%%
\renewcommand{\headrulewidth}{0pt}
}
%%%END OF HEADER%%%

%%%PAGE SETUP - Please do not change any commands within this section%%%
\makeFNbottom
\makeatletter
\renewcommand\LARGE{\@setfontsize\LARGE{15pt}{17}}
\renewcommand\Large{\@setfontsize\Large{12pt}{14}}
\renewcommand\large{\@setfontsize\large{10pt}{12}}
\renewcommand\footnotesize{\@setfontsize\footnotesize{7pt}{10}}
\makeatother

\renewcommand{\thefootnote}{\fnsymbol{footnote}}
\renewcommand\footnoterule{\vspace*{1pt}% 
\color{cream}\hrule width 3.5in height 0.4pt \color{black}\vspace*{5pt}} 
\setcounter{secnumdepth}{5}

\makeatletter 
\renewcommand\@biblabel[1]{#1}            
\renewcommand\@makefntext[1]% 
{\noindent\makebox[0pt][r]{\@thefnmark\,}#1}
\makeatother 
\renewcommand{\figurename}{\small{Fig.}~}
\sectionfont{\sffamily\Large}
\subsectionfont{\normalsize}
\subsubsectionfont{\bf}
\setstretch{1.125} %In particular, please do not alter this line.
\setlength{\skip\footins}{0.8cm}
\setlength{\footnotesep}{0.25cm}
\setlength{\jot}{10pt}
\titlespacing*{\section}{0pt}{4pt}{4pt}
\titlespacing*{\subsection}{0pt}{15pt}{1pt}
%%%END OF PAGE SETUP%%%

%%%FOOTER%%%
\fancyfoot{}
\fancyfoot[LO,RE]{\vspace{-7.1pt}\includegraphics[height=9pt]{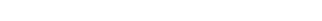}}
\fancyfoot[CO]{\vspace{-7.1pt}\hspace{13.2cm}\includegraphics{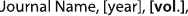}}
\fancyfoot[CE]{\vspace{-7.2pt}\hspace{-14.2cm}\includegraphics{head_foot/RF}}
\fancyfoot[RO]{\footnotesize{\sffamily{1--\pageref{LastPage} ~\textbar  \hspace{2pt}\thepage}}}
\fancyfoot[LE]{\footnotesize{\sffamily{\thepage~\textbar\hspace{3.45cm} 1--\pageref{LastPage}}}}
\fancyhead{}
\renewcommand{\headrulewidth}{0pt} 
\renewcommand{\footrulewidth}{0pt}
\setlength{\arrayrulewidth}{1pt}
\setlength{\columnsep}{6.5mm}
\setlength\bibsep{1pt}
%%%END OF FOOTER%%%

%%%FIGURE SETUP - please do not change any commands within this section%%%
\makeatletter 
\newlength{\figrulesep} 
\setlength{\figrulesep}{0.5\textfloatsep} 

\newcommand{\topfigrule}{\vspace*{-1pt}% 
\noindent{\color{cream}\rule[-\figrulesep]{\columnwidth}{1.5pt}} }

\newcommand{\botfigrule}{\vspace*{-2pt}% 
\noindent{\color{cream}\rule[\figrulesep]{\columnwidth}{1.5pt}} }

\newcommand{\dblfigrule}{\vspace*{-1pt}% 
\noindent{\color{cream}\rule[-\figrulesep]{\textwidth}{1.5pt}} }

\makeatother
%%%END OF FIGURE SETUP%%%

%%%TITLE, AUTHORS AND ABSTRACT%%%
\twocolumn[
  \begin{@twocolumnfalse}
{\includegraphics[height=50pt]{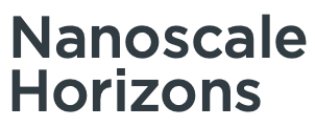}\hfill\raisebox{0pt}[0pt][0pt]{\includegraphics[height=55pt]{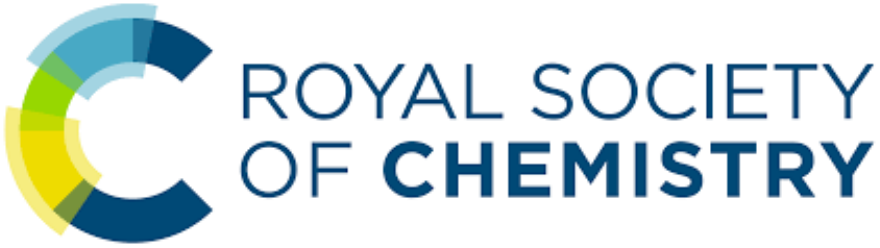}}\\[1ex]
\includegraphics[width=19.cm]{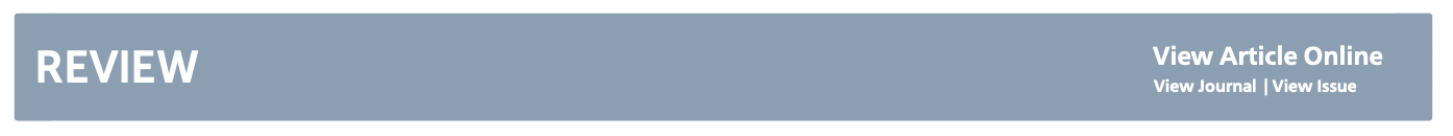}}
\par
\vspace{1em}
\sffamily
\begin{tabular}{m{4.5cm} p{13.5cm} }

\includegraphics{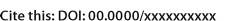} & 
\noindent\LARGE{\textbf{Spintronic devices and applications using noncollinear \textcolor{black}{chiral} antiferromagnets}} \\%Article title goes here instead of the text "This is the title"
\vspace{0.3cm} & \vspace{0.3cm} \\

 & \noindent\large{Ankit Shukla,\textit{$^{a}$} Siyuan Qian,\textit{$^{b}$} and Shaloo Rakheja\textit{$^{\ast}$}} \\%Author names go here instead of "Full name", etc.

\includegraphics{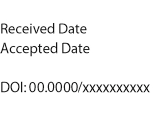} & \noindent\normalsize{
Antiferromagnetic materials have several unique properties, such as 
a vanishingly small net magnetization, which generates weak dipolar fields and makes them robust against perturbation from external magnetic fields and rapid magnetization dynamics, as dictated by the geometric mean of their exchange and anisotropy energies. However, experimental and theoretical techniques to detect and manipulate the antiferromagnetic order in a fully electrical manner must be developed to enable advanced
spintronic devices with antiferromagnets as their active spin-dependent elements. Among the various antiferromagnetic materials, conducting antiferromagnets offer high electrical and thermal conductivities and strong electron-spin-phonon interactions. Noncollinear metallic antiferromagnets with negative chirality, including Mn$_3$Sn, Mn$_3$Ge, and Mn$_3$GaN, offer rich physics of spin momentum locking, topologically protected surface states, large spin Hall conductivity, and magnetic spin Hall effect that arises from their topology. 
In this review article, we introduce the crystal structure and the physical phenomena, including anomalous Hall and Nernst effects, spin Hall effect, and magneto-optic Kerr effect, observed in negative chirality antiferromagnets. Experimental advances related to spin-orbit torque-induced dynamics and the impact of the torque on the microscopic spin structure of Mn$_3$Sn are also discussed. 
\textcolor{black}{
Recent experimental demonstrations of a finite room-temperature tunneling magnetoresistance in tunnel junctions with chiral antiferromagnets opens the prospect of developing spintronic devices with fully electrical readout.
Applications of chiral antiferromagnets, including non-volatile memory, high-frequency signal generator/detector, neuro-synaptic emulator, probabilistic bit, thermoelectric device, and Josephson junction, are highlighted.}
% We then present a potential antiferromagnetic spintronic device that can serve as a non-volatile memory, high-frequency signal generator/detector, neuron emulator, probabilistic bit, and even a Josephson junction, depending on the design parameters and the input stimuli, \emph{i.e.}, amplitude and pulse width of the injected spin current and the externally applied magnetic field. In this device, spin-orbit torques can be used to manipulate the order parameter, while the device state can be read via tunneling magnetoresistance which has been recently experimentally demonstrated in all-Mn$_3$Sn tunnel junctions at room temperature. 
We also present analytic models that relate the performance characteristics of the device with its design parameters, thus enabling a rapid technology-device assessment. Effects of Joule heating and thermal  noise on the device characteristics are briefly discussed. We close the paper by summarizing the status of research and present our outlook in this rapidly evolving research field.

%The abstract should be a single paragraph which summarises the content of the article. Any references in the abstract should be written out in full \textit{e.g.}\ [Surname \textit{et al., Journal Title}, 2000, \textbf{35}, 3523].

} \\%The abstrast goes here instead of the text "The abstract should be..."

\end{tabular}

 \end{@twocolumnfalse} \vspace{0.6cm}

  ]
%%%END OF TITLE, AUTHORS AND ABSTRACT%%%

%%%FONT SETUP - please do not change any commands within this section
\renewcommand*\rmdefault{bch}\normalfont\upshape
\rmfamily
\section*{}
\vspace{-1cm}

%%%FOOTNOTES%%%

\footnotetext{\textit{$^{\ast}$Electrical and Computer Engineering, The Grainger College of Engineering, University of Illinois Urbana-Champaign, Urbana, USA. Tel: 217-244-3616; E-mail:rakheja@illinois.edu}}
%\footnotetext{\textit{$^{b}$~Address, Address, Town, Country. }}

%Please use \dag to cite the ESI in the main text of the article.
%If you article does not have ESI please remove the the \dag symbol from the title and the footnotetext below.
%\footnotetext{\dag~Electronic Supplementary Information (ESI) available: [details of any supplementary information available should be included here]. See DOI: 00.0000/00000000.}
%additional addresses can be cited as above using the lower-case letters, c, d, e... If all authors are from the same address, no letter is required

% \footnotetext{\ddag~Additional footnotes to the title and authors can be included \textit{e.g.}\ `Present address:' or `These authors contributed equally to this work' as above using the symbols: \ddag, \textsection, and \P. Please place the appropriate symbol next to the author's name and include a \texttt{\textbackslash footnotetext} entry in the the correct place in the list.}

%%%END OF FOOTNOTES%%%

%%%MAIN TEXT%%%%
% The main text of the article\cite{Mena2000} should appear here.

% \subsection{This is the subsection heading style}
% Section headings can be typeset with and without numbers.\cite{Abernethy2003}

% \subsubsection{This is the subsubsection style.~~} These headings should end in a full point.  

% \paragraph{This is the next level heading.~~} For this level please use \texttt{\textbackslash paragraph}. These headings should also end in a full point.

\section{Introduction}
%\textcolor{black}{TODO: SR}

The field of spintronics utilizes information stored in the magnetization vector of magnetically ordered medium. Typically effects, such as spin-transfer torque (STT) and spin-orbit torque (SOT) in conjunction with a magnetic field, are used to switch the magnetic state.~\cite{slonczewski1996current, berger1996emission, gambardella2011current, brataas2012current, sinova2015spin, kent2015new, ramaswamy2018recent, song2021spin, shao2021roadmap, han2021spin, pinarbasi2022perspectives} 
The non-volatility and scalability of spintronics devices can allow us to implement information processing devices with reduced resources (\emph{i.e.}, energy, area, time). Spintronics devices can complement the functionality of silicon CMOS circuits for beyond von Neumann computing applications including probabilistic and stochastic computing, bio-inspired computing, and Joule heating free transfer of information.~\cite{fong2015spin, fong2016spin, camsari2017stochastic, camsari2019p, borders2019integer, joshi2020mtj, dieny2020opportunities, grollier2020neuromorphic, hirohata2020review, guo2021spintronics, houshang2022phase, misra2023probabilistic, shukla2023true, litvinenko2023spinwave, chowdhury2023full, chen2023spintronic} 

Typically, the active element in spintronics devices is a ferromagnet (FM) in which spins on the neighboring ions are aligned in a parallel manner, resulting in a net macroscopic moment. A two-terminal (2T) device composed of thin FM layers separated by a non-magnetic insulating spacer, referred to as the magnetic tunnel junction (MTJ), is considered as the building block in spintronics computing. The net resistance of the MTJ stack depends on the relative magnetic orientation of the two FM leads: a low resistance state is obtained for parallel orientation of the FMs, while a high resistance state is obtained for anti-parallel orientation of the FMs. In a 2T MTJ, STT is used to change the orientation of one of the FMs, referred to as the \emph{free} layer, while the other FM has a pinned orientation, whose magnetization is unperturbed in the presence of STT.    
A three-terminal (3T) MTJ in which the SOT is used to control the orientation of the free FM layer offers a higher degree of flexibility and energy efficiency for spintronics computing applications. The on-off ratio, \emph{i.e.}, the difference between the high and low resistance states, of the MTJ is read out via the tunneling magnetoresistance (TMR), which can reach values as high as 631\% at room temperature in epitaxial devices employing MgO as the tunneling barrier~\cite{scheike2023631}.   

More recently, antiferromagnetic-based spintronics devices have gained significant attention from the research community. Unlike FMs, antiferromagnets (AFMs) do not have a macroscopic magnetic moment which creates both challenges and opportunities for their future computing paradigms.~\cite{gomonay2014spintronics, jungwirth2016antiferromagnetic, gomonay2018antiferromagnetic, baltz2018antiferromagnetic, bai2020functional, siddiqui2020metallic, xiong2022antiferromagnetic} 
For example, the lack of dipolar fields in AFMs can enable their ultra-dense integration in circuits without crosstalk issues and related security vulnerabilities. However, as AFMs lack a macroscopic moment, it is difficult to read their magnetic state in device geometries. In this regard, octupole moment in chirally ordered noncollinear AFMs (\emph{e.g.}, Mn$_3$Sn, Mn$_3$Ge, Mn$_3$GaN, \emph{etc.}) can be used as an order parameter in an all-AFM MTJs that exhibit finite TMR and have negligible stray fields and the potential for high-speed operation.
The octupole moment is a rank-3 multipole moment of Mn atoms averaged over the entire magnetic unit cell.~\cite{suzuki2017cluster} 
The octupole moment has the same irreducible representation as the dipole moment in FMs.
Thus, chiral noncollinear AFMs combine the benefits of FMs and other collinear AFMs, setting the stage for advanced information processing devices using the octupole moment as the computational state variable.~\cite{higo2022thin, wang2022noncollinear} 
\textcolor{black}{
Interested readers can refer to Ref.~\citen{yang2021chiral} for a detailed review on chiral spintronics.
Although not the focus of this work, interested readers may also refer to Refs.~\citen{vsmejkal2022emerging, mazin2022altermagnetism, mazin2023altermagnetism, cheong2024altermagnetism, chakraborty2024strain, mcclarty2024landau, niu2024electrically, smolyanyuk2024tool, han2024observation} for detailed overview of altermagnetism, which is a newly discovered phase in magnetic materials.
Altermagnets are interesting candidates for spintronics research since they share at least two properties in common with chiral AFMs.
Firstly, altermagnets exhibit antiferromagnetic properties due to the opposite arrangement of neighboring atomic spins, similar to that in collinear AFMs. Secondly, they exhibit ferromagnetic properties owing to certain broken symmetries (\emph{e.g.}, time-reversal symmetry), which arise due to the alternating variation in the local atomic structure.}

In this paper, we review recent experimental and theoretical progress in noncollinear AFMs with negative chirality, in which the octupole moment gives rise to non-degenerate spin-polarized bands at the Fermi level and a finite TMR at room temperature. 
The octupole moment in these AFMs can be manipulated via SOT from a proximal non-magnetic material such as a heavy metal (\emph{e.g.}, W, Pt, \emph{etc.}), and high-speed order parameter dynamics including chiral oscillations and switching can be initiated depending on the energy landscape of the AFM and the input SOT and magnetic field. 
\textcolor{black}{As compared to recent review articles on AFM spintronics that either provide an in-depth picture of a few phenomena~\cite{vsmejkal2022anomalous, han2023coherent, shao2024antiferromagnetic} or present brief overviews of various magnetic material systems, phenomena, and applications,~\cite{xiong2022antiferromagnetic, higo2022thin, khalili2024prospects, chen2024emerging} our work primarily focuses on the recent developments pertaining to detection and manipulation of the magnetic order in the noncollinear chiral AFMs.~\cite{wang2022noncollinear} A brief overview of the various possible device applications, driven by chiral AFMs, is also presented in this work.}

{We present a discussion of the crystal structure and the observed and predicted physical phenomena in chiral AFMs in} Sec.~\ref{sec:noncollinear}. 
{Details regarding order parameter dynamics in six-fold and two-fold symmetric chiral AFMs, particularly in Mn$_3$Sn, are discussed in} Sec.~\ref{sec:devices}. 
{These dynamics along with fully electrical readout of the order parameter, the TMR, form the basis of realizing new types of spintronic devices with chiral AFM serving as the active element. 
Such devices include high-frequency signal generators and detectors, bio-mimetic neurons, and non-volatile memory. The role of Joule heating and thermal noise in influencing the SOT-driven response of the system is also briefly presented in} Sec.~\ref{sec:devices}. 
{As discussed in} Sec.~\ref{sec:application}, \textcolor{black}{{chiral AFMs
can be used for non-volatile memory, artificial synapses, spiking neurons, and as probabilistic bits, which serve as the building blocks of hardware designed to solve machine learning and hard optimization problems. Additionally, chiral AFMs exhibit anomalous Nernst effect (ANE), which leads to a voltage signal that is transverse to the thermal gradient and the magnetization of the material. ANE in chiral AFMs can be exploited for thermoelectric generation, on-chip cooling, and as temperature gradient sensors. A brief discussion on quantum computing utilizing chiral AFM-based superconducting Josephson junction is also included.}}
{The paper concludes with a summary of key findings and outlook in the field of condensed matter physics and future chiral AFM spintronic devices.}

\section{Noncollinear Chiral Antiferromagnets}\label{sec:noncollinear}

\subsection{Crystal Structure}\label{sec_crystal}
\begin{figure}[h!]
    \centering
    \includegraphics[width = \columnwidth, clip = true, trim = 2mm 2mm 10mm 5mm]{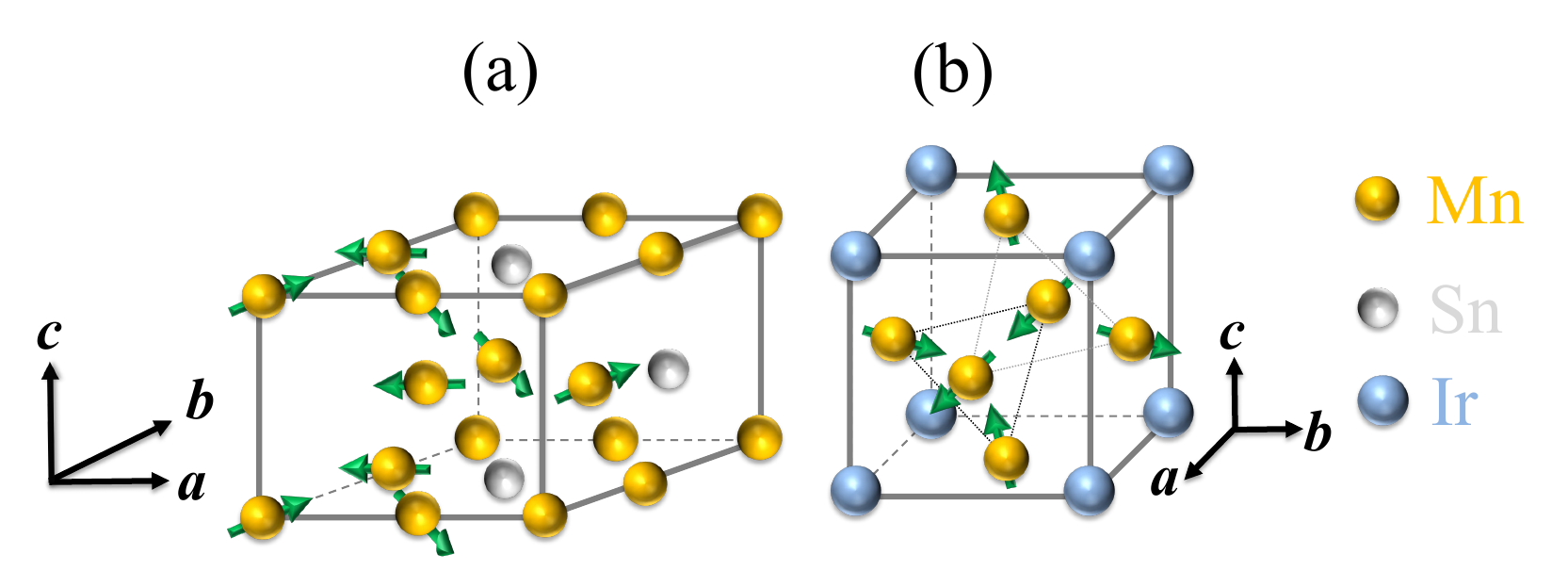}
    \caption{Crystal and spin structure in (a) D0$_{19}$-Mn$_3$Sn and (b) L1$_2$-ordered Mn$_3$Ir, respectively.}
    \label{fig:crystal}
\end{figure}

%\textcolor{black}{TODO: SR with figures from AS}
Chiral AFMs are materials with
lattice spins oriented in a noncollinear and chiral
configuration. In AFMs like Mn$_3$Sn, Mn$_3$Ga and Mn$_3$Ge, the atomic moments are arranged in the basal plane of the hexagonal crystal, while in AFMs like Mn$_3$Ir and Mn$_3$Pt, the atomic spins are in the (111) plane of the face-centered cubic crystal. 
Figure~\ref{fig:crystal} shows the crystal and the triangular spin structures of D0$_{19}$-Mn$_3$Sn and L1$_2$-ordered Mn$_3$Ir AFMs. 
% The Mn$_3$Sn (Mn$_3$Ir) AFM has negative (positive) chirality: hopping between Mn atoms in an anticlockwise (clockwise) sense, the atomic moment rotates by 120$^\circ$ in a clockwise (clockwise) sense. 
\textcolor{black}{The Mn$_3$Sn (Mn$_3$Ir) AFM has negative (positive) chirality: hopping between Mn atoms in a clockwise sense, the atomic moment rotates by 120$^\circ$ in an anticlockwise (a clockwise) sense.} 

The noncollinear moments introduce Dzyaloshinskii-Moriya interaction (DMI), which stabilizes the triangular spin structure with a given chirality. As DMI is a super-exchange interaction, it is present in crystal structures with broken inversion symmetry. DMI introduces a term given as $ -\bf{D}_{12}\cdot (\bf{S}_1\times \bf{S}_2)$ ($\bf{D}_{12}$ is the DMI coefficient and $\bf{S_1}$ and $\bf{S_2}$ are the atomic spins) in the Hamiltonian and thus favors a certain orientation of $\bf{S}_1$ with respect to $\bf{S_2}$. This means that DMI introduces handedness or chirality to the spin structure of the material. 
Although Mn$_3$Sn and Mn$_3$Ir
are globally centrosymmetric crystals, the noncollinearly arranged moments on Mn atoms break inversion symmetry locally, which introduces DMI in the crystal. 
Indeed, neutron diffraction measurements carried out on single-crystal Mn$_3$Sn in 1982 revealed weak ferromagnetism~\cite{tomiyoshi1982magnetic} which was later explained in terms of the DMI mechanism, which leads to canting of moments in the material~\cite{nagamiya1982triangular}.
The chirality of the spin structure breaks the mirror and time-reversal symmetries and lifts degeneracies in the materials' energy-momentum dispersion, which creates band crossings.  

The triangular spin structure in chiral AFMs possess momentum-space Berry curvature, \textcolor{black}{which can be viewed as an effective magnetic field in the reciprocal space~\cite{gustafsson2013time, kubler2014non} and is obtained as
%\begin{equation}
$   {\Omega}(\bf{k}) = \nabla_{\bf{k}}\cross \mathcal{A}(\bf{k})$, where $\mathcal{A}(\bf{k}) = i \sum_{n\in \mathrm{occ}} \langle u_{n\bf{k}}\lvert \nabla_{\bf{k}}\rvert u_{n\bf{k}}\rangle$ is the Berry connection. Here, $u_{n\bf{k}}$ is the eigenfunction with wave-vector $\bf{k}$ and band index $n$, and the sum extends over all occupied states. Crystal-periodic eigenfunctions from the density functional theory calculations must be used to evaluate the Berry connection.
The Berry curvature is then employed to obtain the anomalous Hall conductivity per
\begin{equation}
    \sigma_{lm}^p = \frac{e^2}{\hslash} \int_{\mathrm{BZ}} \frac{d{\bf{k}}}{(2\pi)^3}\Omega_p({\bf{k}})f({\bf{k}}),
\end{equation}
where $f({\bf{k}})$ is the Fermi distribution function, and the directions $l,\,m,\,p$ denote $x,\,y,\,z$ in a cyclic manner. $\sigma_{lm}^p$ can be considered a pseudo-vector pointing along $p$ that is perpendicular to both $l$ and $m$ and corresponds to anomalous Hall conductivity.~\cite{nayak2016large} That is, when current flows along $l$, it generates a Hall voltage along $m$. The symmetry of the AFM structure dictates which components of the anomalous Hall conductivity tensor are non-zero. 
}

Berry curvature is
concentrated around hot spots in the reciprocal space, 
while the integral of the Berry curvature over the Brillouin zone modifies the magneto-transport properties of the AFMs.~\cite{kubler2014non} The sign of the Berry curvature flips with the chirality of the noncollinear AFM. Thus, the sign of magneto-transport signals that are odd with respect to the Berry curvature also flips in chiral AFMs. 
Theoretical calculations have shown multiple Weyl points of type II in the bulk bandstructure and Fermi arcs on the surface connecting Weyl points of opposite chirality in Mn$_3$Sn and Mn$_3$Ge.~\cite{yang2017topological, kubler2018weyl} 
\textcolor{black}{Weyl points are topological objects that arise when the spin degeneracy of a Dirac point is lifted, 
and they act as the monopole sources/sinks of Berry curvature (\emph{i.e.}, a singularity of Berry curvature).~\cite{shekhar2015extremely} Experimental evidence of buried Weyl points in Mn$_3$Sn was presented in Ref.~\citen{kuroda2017evidence}, where angle-resolved photoemission spectroscopy (ARPES) was used to determine that Mn$_3$Sn is a strongly correlated magnetic material, while the observations of large AHE, positive longitudinal magnetoconductance, and negative transverse magnetoconductance were utilized to provide evidence for the magnetic Weyl points in the material.}
Thus, the bandstructure features in chiral AFMs are described as topological in nature and have launched the research field of topological spintronics. 
Table~\ref{tab:materials} lists the material properties of various chiral AFMs. 
\begin{table}[h]
\small
  \caption{List of chiral antiferromagnetic materials $\mathrm{Mn_{3}X}$ and their associated parameters. Here $T_N$ is the N\'eel temperature, $M_{s}$ is the saturation magnetization, $K_u$ is the single-ion uniaxial anisotropy constant, $J_E$ is the exchange constant, , is in $\mathrm{MJ/m^{3}}$. Sign of DMI constant, $D_M$, which decides the chirality is also mentioned.}
  \label{tab:materials}
  \begin{tabular*}{0.48\textwidth}{@{\extracolsep{\fill}}cccccc}
    \hline
    X  & $T_N~(\mathrm{K})$ & $M_{s}~(\mathrm{T})$  & $K_u~(\mathrm{kJ/m^3})$ & $J_E~(\mathrm{MJ/m^3})$ & $D_M$ \\
    \hline
    $\mathrm{Ir}$~\cite{zhang2017strong, yamane2019dynamics} & $960 \pm 10$ & $1.63$ & $3000$ & $240$ & -- \\
    $\mathrm{Pt}$~\cite{kren1968magnetic, zhang2017strong} & $473 \pm 10$ & $1.37$ & $10$ & $280$ & -- \\
    $\mathrm{Rh}$~\cite{kren1968magnetic, feng2015large, zhang2017strong} & $853 \pm 10$ & $2.00$ & $10$ & $230$ & -- \\
    {$\mathrm{Ga}$}~\cite{kurt2011mn3, zhang2017strong, nyari2019weak, seyd2020mn3ge} & $470$ & $0.54$ & $100$ & $110$ & + \\
    {$\mathrm{Sn}$}~\cite{tsai2020electrical, liu2017anomalous, zhang2017strong, nyari2019weak} & $420-430$ & $0.50$ & $110$ & $59$ & + \\
    {$\mathrm{Ge}$}~\cite{yamada1988magnetic, zhang2017strong, nyari2019weak, seyd2020mn3ge} & $365$ & $0.28$ & $1320$ & $77$ & + \\
    {$\mathrm{GaN}$}~\cite{gurung2020spin, chen2017electronic} & $345$ & $0.69$ & $10$ & $280$ & -- \\
    {$\mathrm{NiN}$}~\cite{gurung2020spin, chen2017electronic} & $262$ & $1.54$ & $10$ & $177$ & -- \\
    \hline
  \end{tabular*}
\end{table}

\subsection{Physical Phenomena}\label{sec_physphen}
\textcolor{black}{
Recent magneto-transport measurements have shown finite 
anomalous Hall effect (AHE),~\cite{kubler2014non, chen2014anomalous, nakatsuji2015large, nayak2016large}} the spin Hall effect (SHE),~\cite{zhang2016giant} the anomalous Nernst effect (ANE),~\cite{narita2017anomalous} and the topological Hall effect (THE) in chiral AFMs. 
Despite its vanishingly small net magnetic moment $0.002~\mu_B$ ($\mu_B$ is the Bohr magneton) per Mn atom, a large anomalous Hall conductivity (AHC) of $20~\mathrm{(\Omega.cm)^{-1}}$ at room temperature and $100~\mathrm{(\Omega.cm)^{-1}}$ at low temperatures has been experimentally measured in Mn$_3$Sn.~\cite{nakatsuji2015large} Although it is more challenging to observe AHE in cubic chiral AFMs due to weaker spontaneous ferromagnetism and stronger magnetic anisotropy, uniaxial stress and magnetic field were used in Ref.~\citen{zuniga2023observation} to align the domains in Mn$_3$Pt, leading to a substantial AHC ($\sim 20-30~\mathrm{(\Omega.cm)^{-1}}$) in the bulk sample. Similarly, in Ref.~\citen{zhao2022strain}, strain tuning was used to achieve AHC exceeding $100~\mathrm{(\Omega.cm)^{-1}}$, while in Ref.~\citen{liu2018electrical}, epitaxial strain was shown to alter the AHC in Mn$_3$Pt films by more than an order of magnitude.
\textcolor{black}{
In the chiral AFM, Mn$_3$Ge, anomalous Hall conductivity of $60~(\Omega.\mathrm{cm})^{-1}$ at room temperature and $380~(\Omega.\mathrm{cm})^{-1}$ at $\mathrm{5~K}$ has been measured, while the AHE was also shown to reverse sign with a rotation of $\mathrm{0.1~T}$ magnetic field.~\cite{kiyohara2016giant}}
\textcolor{black}{Very recently, a significantly large anomalous Hall conductivity of about $150~(\Omega.\mathrm{cm})^{-1}$ at room temperature and $530~(\Omega.\mathrm{cm})^{-1}$ at $\mathrm{10~K}$ has been reported for Mn$_{2.4}$Ga.~\cite{song2024large}
Similar to other chiral AFMs, the anomalous Hall conductivity in Mn$_{2.4}$Ga exhibits sign reversal with the rotation of the small external magnetic field.}

The ANE in which a temperature gradient results in a transverse voltage drop perpendicular to the heat flow and the magnetization is feasible in chiral AFMs. A Seebeck coefficient~\cite{ikhlas2017large} of $0.35~\mathrm{\mu V.K^{-1}}$ at room temperature and $0.6~\mathrm{\mu V.K^{-1}}$ at $200~\mathrm{K}$ has been measured in bulk single-crystal Mn$_3$Sn grown via the Bridgman method. 
Using \emph{ab initio} calculations, it was shown that the spin Nernst conductivity (SNC) in Mn$_3$Sn and Mn$_3$Ga is about five times larger than that of platinum~\cite{guo2017large}. Moreover, the anomalous Nernst conductivity (ANC) of Mn$_3$Ga at $300~\mathrm{K}$ was found to be $50\times$ that of bcc Fe. The large value of ANC and SNC in noncollinear AFMs is advantageous for implementing thermoelectric and spin caloritronic devices.~\cite{narita2017anomalous}

The momentum-dependent spin splitting in Mn$_3$Sn also leads to magnetic SHE~\cite{kimata2019magnetic} which has an anomalous sign change with the octupole polarity. 
Novel spin currents in chiral AFMs have also been studied via \emph{ab initio} methods. In particular, it was shown that spin currents in noncollinear AFMs have both longitudinal and transverse spin polarization components. The longitudinal spin current is similar to the spin-polarized current in FMs.
However, the transverse component 
is odd under time reversal and is thus distinct from the conventional SHE, which is even under time reversal. 
\textcolor{black}{The first experimental demonstration of the out-of-plane magnetic SHE and the associated out-of-plane field-like torque using spin-torque ferromagnetic resonance (ST-FMR) in single-crystal Mn$_3$Sn/Ni-Fe bilayers was reported in Ref.~\citen{kondou2021giant}. Using macroscopic symmetry analysis and microscopic quantum kinetic theory, Kondou \emph{et~al.} revealed that the origin of the magnetic SHE is likely the momentum-dependent spin splitting in Mn$_3$Sn.
In Ref.~\citen{hazra2023generation}, authors established the relationship between the in-plane polarized spin current in Mn$_3$Sn and its antiferromagnetic order. It was shown that for various thicknesses, in the range of $3-12~\mathrm{nm}$, of epitaxially grown Mn$_3$Sn on Ru buffered Al$_2$O$_3$, the in-plane polarized spin current was dependent on the AFM order, while the out-of-plane polarized spin current originated from the spin swapping scatterings at the interface between Mn$_3$Sn and Py.}
These novel spin currents~\cite{vzelezny2017spin} can help interpret SOT measurements in noncollinear AFMs, while also being exploited for new spintronics applications.

The electronic bands in Mn$_3$Sn show momentum-dependent spin splitting of the Fermi surface due to the broken time reversal symmetry,~\citep{dong2022tunneling, chen2023octupole} 
\textcolor{black}{which is postulated to result in the recently reported finite TMR ratio of about $2\%$ at $300~\mathrm{K}$ in
experimental Mn$_3$Sn/MgO/Mn$_3$Sn samples~\cite{chen2023octupole} and about 100\% in experimental Mn$_3$Pt/MgO/Mn$_3$Pt samples.~\cite{qin2023room} 
This is because the TMR values measured in the Mn$_3$Sn/MgO/Mn$_3$Sn tunnel junctions far exceed (by about two orders of magnitude) the TMR ratio theoretically estimated from the anisotropic spin polarization found in Mn$_3$Sn.~\cite{chen2023octupole}}
\textcolor{black}{A recent work has suggested that the rather low TMR value measured in Mn$_3$Sn/MgO/Mn$_3$Sn junctions, as compared to the Mn$_3$Pt/MgO/Mn$_3$Pt junctions, is likely due to the polycrystalline nature of the top Mn$_3$Sn electrode employed in these samples.~\cite{shao2024antiferromagnetic}
Another notable recent work has measured an even smaller magnetoresistance (MR) effect of 0.3\% at room temperature in granular Mn$_3$Sn-Ag films.~\cite{wang2023room} In the Mn$_3$Sn-Ag composite, the AFM/metal/AFM heterostructure functions as an all-AFM spin valve and thus exhibits a finite MR, which was found to be two orders of magnitude higher compared to the MR measured in Mn$_3$Sn films.
}

\textcolor{black}{Theoretical estimations presented in Refs.~\citen{dong2022tunneling, chen2023octupole, qin2023room}, on the other hand, suggested the possibility of higher TMR ratio in both Mn$_3$Pt and Mn$_3$Sn tunnel junctions. For example, a theoretical TMR value of 240\% was predicted for Mn$_3$Pt/SrTiO$_3$/Mn$_3$Pt in Ref.~\citen{qin2023room}. The authors argued that the lower experimental TMR value for all-Mn$_3$Pt junctions with MgO barrier compared to the theoretically projected TMR value of the junction with SrTiO$_3$ barrier is a result of grain boundaries and
dislocations caused by large lattice mismatch, interfacial roughness, and other defects. In the case of all-Mn$_3$Sn tunnel junctions with vacuum as a barrier, theoretical calculations showed a TMR of 300\% while a TMR of 124\% was predicted in Mn$_3$Sn junctions with HfO$_2$ as the tunneling barrier.
Very recent experimental investigations have provided credibility to such theoretical predictions as a TMR of 110\% and resistance-area (RA) product of $1.7~\mathrm{k \Omega.\mu m^2}$ at room temperature have been reported in Mn$_3$Pt/Al$_2$O$_3$/Mn$_3$Pt junctions sputter deposited on a thermally oxidized silicon substrate, thus making the process compatible with conventional silicon manufacturing.~\cite{shi2024electrically}} 
\textcolor{black}{Additional research is needed to understand the reliability of the insulating barrier at high temperatures and current values and enable high-performance AFM-based spintronics applications with fully electrical readout.}

Imaging the magnetic domains can provide crucial information regarding the reversal process of the chiral spin structure in a spatially resolved manner. Using magneto-optic Kerr effect (MOKE) method, Higo \emph{et~al.} measured a large zero-field Kerr rotation angle of $20~\mathrm{mdeg}$ at $300~\mathrm{K}$ in Mn$_3$Sn,~\cite{higo2018large} while a large polar MOKE signal, around $8.2~\mathrm{mdeg}$, and a large longitudinal MOKE signal, around $5.6~\mathrm{mdeg}$, have also been measured in Mn$_3$Ge single crystals.~\cite{wu2020magneto}
The Kerr angle value in chiral AFMs is comparable to that found in FMs and is a useful probe to study the dynamics of magnetic domains in chiral AFMs. In this regard, Uchimura \emph{et~al.} measured Kerr rotation angle in thin films of Mn$_{3+x}$Sn$_{1-x}$ ($-0.11\leq x \leq 0.14$) and identified that the symmetry requirements of MOKE are the same as those of AHE~\cite{uchimura2022observation}. Moreover, they showed that the domain reversal process begins by nucleation of reversed domains in the media, followed by domain expansion and preferential propagation along the [$11\bar{2}0$] direction. \textcolor{black}{
Uchimura \emph{et al.} also revealed the size of the nucleated domain to be less than 1$~\mathrm{\mu m}$, which is much smaller than that observed in the case of bulk samples.~\cite{higo2018large} Studies on current-driven nucleation and displacement of domain walls in chiral AFMs have recently appeared.~\cite{wu2024current, sugimoto2020electrical} Sugimoto \emph{et al.} identified a sub-micron size Bloch-like domain wall parallel to the Kagome plane in a microfabricated wire of Mn$_3$Sn.~\cite{sugimoto2020electrical} The nucleation of the domain wall was initiated by current-induced torques in a wedge-shaped device with thickness variation from 500$~\mathrm{nm}$ to 1$~\mathrm{\mu m}$ along the [0001] direction. It was found that the current density for the domain-wall nucleation was around $10^9~\mathrm{A/m^2}$, which is two-to-three orders of magnitude lower than that in the case of ferromagnets. Using MOKE, Wu \emph{et al.} observed N\'{e}el-like domain wall in Mn$_3$Ge, which could be accelerated up to 750$~\mathrm{m/s}$ with a current density of $7.56\times 10^{10}~\mathrm{A/m^2}$ without any magnetic fields.~\cite{wu2024current} The results on the imaging, nucleation, and propagation of magnetic-octupole domain walls in microfabricated samples of chiral AFMs are highly significant to drive their new domain-wall-based applications.}

% Like Mn$_3$Sn, Mn$_3$Ge is also a chiral AFM. However, unlike Mn$_3$Sn, Mn$_3$Ge does not have a helical phase transition, enabling unique studies of noncollinear magnetic textures at low temperatures. %Additionally, the AHC reported in Mn$_3$Ge can be three times as high as that in Mn$_3$Sn. 

Scanning thermal gradient microscopy (STGM) has also been used to image the local magnetic structure of Mn$_3$Sn films. STGM technique in the presence of magnetic fields can form the basis of writing domain patterns into Mn$_3$Sn as discussed in Ref.~\cite{reichlova2019imaging}. 
This work specifically measures the ANE voltage, $V_\mathrm{ANE}$, for a temperature gradient along the c-axis, which is established with the help of laser illumination. $V_\mathrm{ANE}$ is governed by the average octupole moment of the domains under the illuminated laser spot.

\section{Spintronic Devices Based on Chiral AFMs}\label{sec:devices}
To realize an energy-efficient spintronic device based on chiral AFMs, it is important to investigate
\textcolor{black}{the physics of SOT-driven dynamics of the}
%methods that can control the 
octupole moment in \textcolor{black}{both bulk and thin film samples of chiral AFMs.}
%a fully electrical manner. 
On the readout front, although
%Although 
AHE and anisotropic magnetoresistance can be utilized, the detected signals are generally weak and require amplification adding to the energy consumption and area of the spintronic device. Thus, electrical readout via the TMR is preferred as it provides a more robust signal and is typically compatible with circuit implementations utilizing silicon transistors as the peripherals. 

\begin{figure}[ht!]
  \centering
  \includegraphics[width = \columnwidth, clip = true, trim = 0mm 0mm 0mm 0mm]{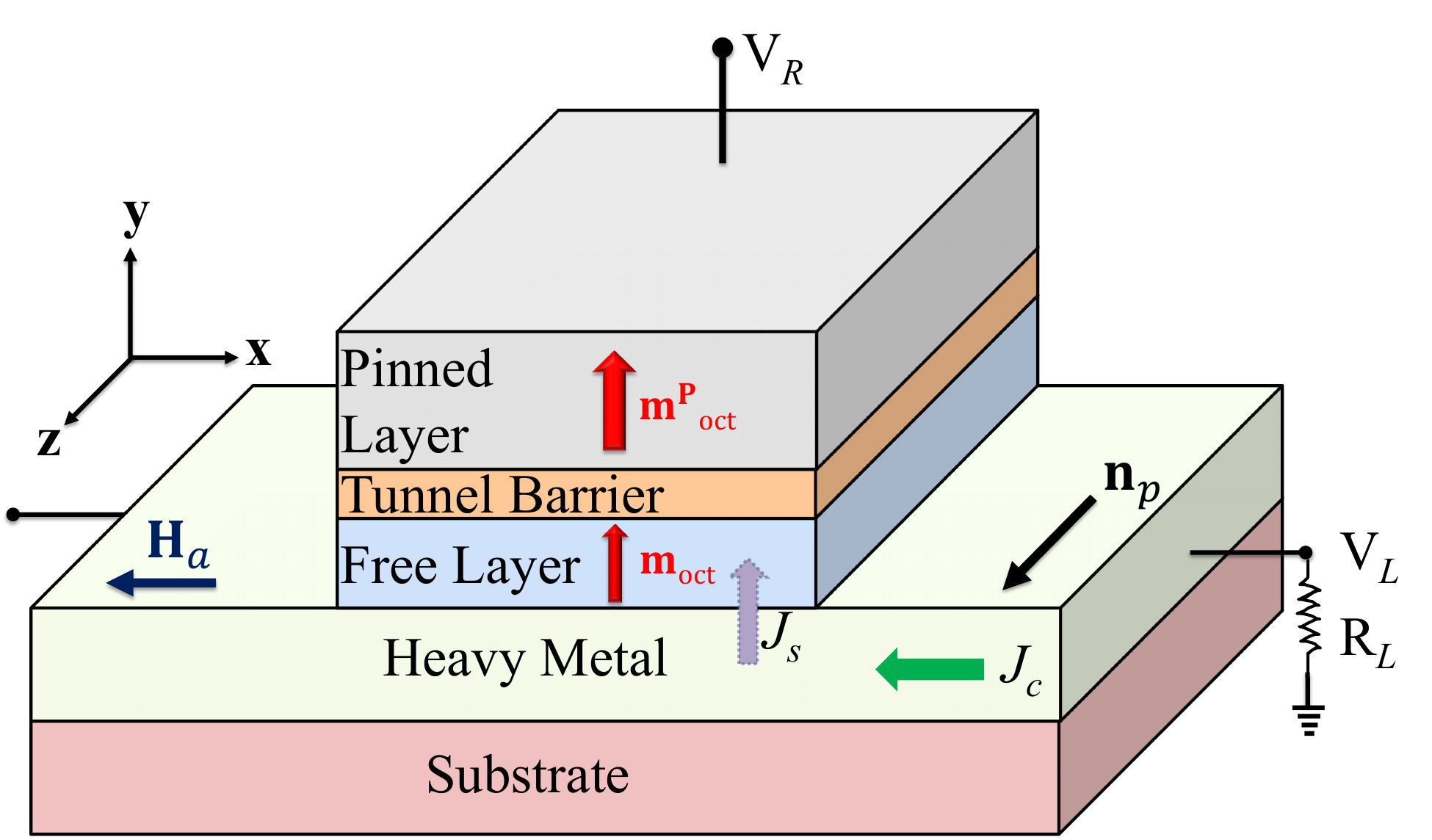}
  \caption{A conceptual three-terminal spintronic device using a negative chirality AFM (\emph{e.g.}, Mn$_3$Sn) as the active element. The AFM serves as both the `Free Layer' and the `Pinned Layer'. SOT generated in the heavy metal is used to manipulate the octupole moment of the free layer. 
  $J_c$ and $J_s$ are the input charge current density and the spin current density, respectively. $J_s = \theta_\mathrm{SH}J_c$, where $\theta_\mathrm{SH}$ is the spin Hall angle of the heavy metal. 
 If needed, $\vb{H}_a$ is the external magnetic field, applied to aid the deterministic switching of the magnetic octupole, $\vb{m}_\mathrm{oct}$--this is useful when the AFM layer is strained resulting in perpendicular magnetic anisotropy. The readout mechanism is via the TMR such that the parallel orientation of the pinned layer ($\vb{m}_\mathrm{oct}$) and the free layer ($\vb{m}_\mathrm{oct}$) results in a lower resistance state, while the anti-parallel configuration of the layers results in a higher resistance state.  
 For a fixed read voltage, $V_R$, and load resistance, $R_L$, the load voltage, $V_L$, has two different values capturing the low and high resistance states of the junction. 
 Previous works have utilized Si/SiO$_2$~\cite{tsai2020electrical, yan2022quantum, krishnaswamy2022time, yoo2024thermal}, sapphire,~\cite{yan2022quantum, pal2022setting, xu2023robust} and MgO~\cite{takeuchi2021chiral, higo2022perpendicular, yoon2023handedness} substrates.
  } 
  \label{fig:device}
\end{figure}

A conceptual spintronic device using a chiral AFM is illustrated in Fig.~\ref{fig:device}. Here, information can be written into the AFM domain via the SOT effect generated in the heavy metal (HM), while TMR can be utilized to transduce the magnetic state into a resistance value.  
In principle, this device can be used as a high-frequency signal generator/detector, emulator of a biological neuron, as well as a non-volatile memory element. The specific functionality depends on the device design, including material growth parameters, presence of strain, and the characteristics of the input stimuli (\emph{i.e.}, pulse width and amplitude of the SOT and the magnetic field). 
To explore the functionality of the spintronic device, it is important to consider both the order-parameter writing and reading processes and accordingly quantify the limits and challenges of the device for future computing applications.

In this section, we discuss the control of octupole moment in negative chirality AFMs via the SOT effect. Progress in SOT manipulation of bulk AFMs with six-fold symmetric energy landscape and thin-film strained AFMs with two-fold symmetric energy landscape is presented. 
In the case of Mn$_3$Sn, a prototypical negative chirality AFM,
the atomic spins are slightly canted toward the in-plane easy axes, resulting in a small net magnetization, which is six-fold degenerate in the Kagome plane.~\citep{markou2018noncollinear}
In the presence of an in-plane uniaxial strain, the two-fold degeneracy dominates, while the net magnetization, compared to the bulk system, is modified.~\citep{ikhlas2022piezomagnetic} 
\textcolor{black}{We also analyze the impact of Joule heating, thermal noise, and the limits of electrical readout of the information encoded in the magnetic octupole.}

\subsection{Current-induced dynamics in six-fold symmetric \texorpdfstring{Mn{$_3$}Sn}{Lg}}\label{sec:six-fold}
Recent experimental works have extensively investigated the SOT-driven dynamics in polycrystalline films of Mn$_3$Sn, with thicknesses ranging from $8.3~\mathrm{nm}$ to $100~\mathrm{nm}$, both with and without external magnetic fields.~\cite{tsai2020electrical, tsai2021large, tsai2021spin, takeuchi2021chiral, yan2022quantum, pal2022setting, krishnaswamy2022time, deng2023all, xu2023robust, zheng2024effective}
These films, owing to their polycrystalline nature, are composed of grains with different orientations of the Kagome plane, and therefore, exhibit different dynamics under the effect of SOT with a fixed spin polarization. 
In such polycrystalline films, each grain exhibits a six-fold symmetric energy landscape, similar to the case of bulk Mn$_3$Sn. 
The AHE signal measured in the Hall bar setup is, therefore, an average effect of the final states of all the different grains. 

\begin{figure}[h!]
    \centering
    \includegraphics[width = \columnwidth, clip = true, trim = 2mm 2mm 10mm 2mm]{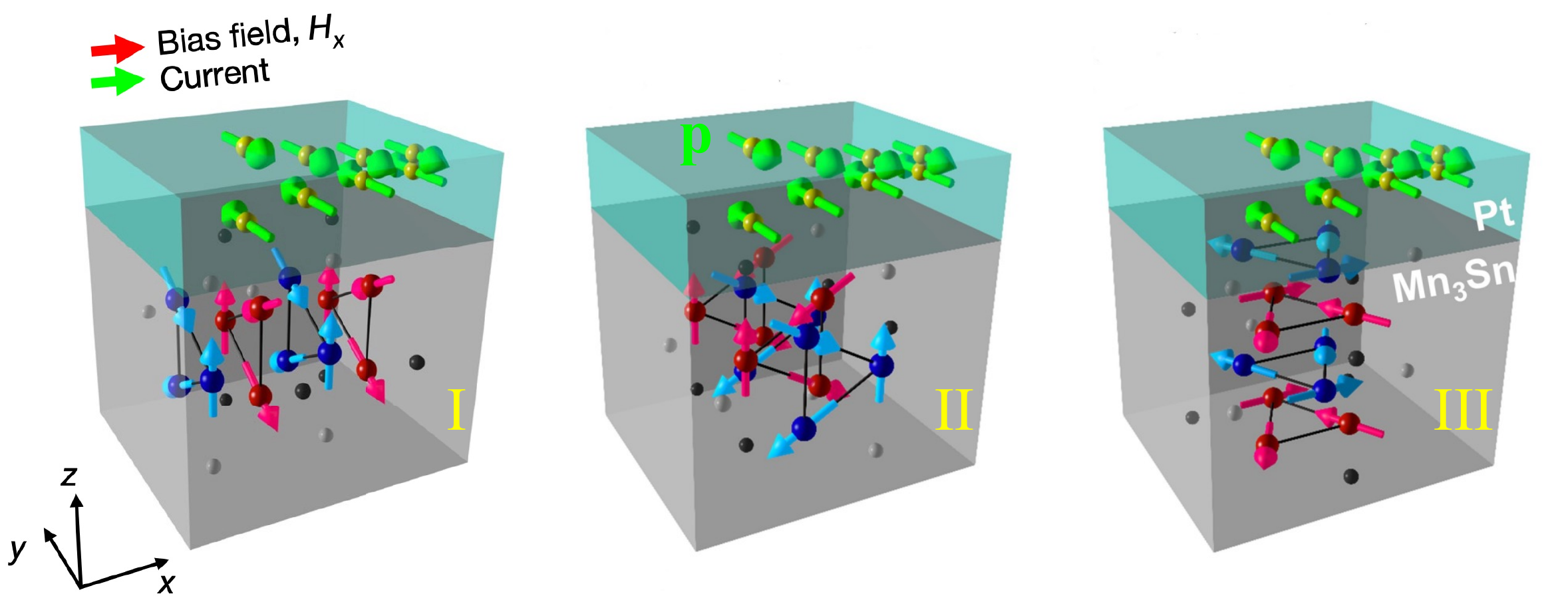}
    \caption{Three different configuration of the Kagome lattice with respect to the input charge current, $I_\mathrm{write}$, and spin polarization, $\vb{p}$. Magnetic field, $H_x$, is applied in all the three configurations.
    Reproduced with permission from Ref.~\cite{tsai2020electrical}.}
    \label{fig:config}
\end{figure}

To explain the observed Hall measurements, one or all of the three different device configurations shown in Fig.~\ref{fig:config} have been explored.
In configuration~I, the Kagome lattice is perpendicular to the electric current but parallel to spin polarization direction, $\vb{p}$. That is, $x \parallel [0001]$ ($\vb{p} \parallel [01\bar{1}0]$). In configuration~II, the Kagome lattice is parallel to the electric current but perpendicular to $\vb{p}$, which corresponds to $x\parallel [01\bar{1}0]$ ($\vb{p} \parallel [0001]$). Finally, in configuration~III, both the electric current and $\vb{p}$ are parallel to the Kagome lattice but perpendicular to each other, or $x \parallel [2\bar{1}\bar{1}0]$ ($\vb{p} \parallel [01\bar{1}0]$). 
In all the three cases, an in-plane symmetry-breaking magnetic field, $H_x$, is applied parallel to the current.
Among the three configurations, II maximizes the initial SOT efficiency, resulting in relatively low switching power consumption.~\cite{takeuchi2021chiral, higo2022perpendicular}

%%%%%%%%%%%%%%%%%%%%%%%%%%%%%%%%%%%%%%%%%%%%%%%%%%%%%%%%%%%%%%%%%%%%%%%%%%%%%%%%%%%%%%%%%%

%%%%%%%%%%%%%%%%%%%%%%%%%%%%%%%%%%%%%%%%%%%%%%%%%%%%%%%%%%%%%%%%%%%%%%%%%%%%%%%%%%%%%%%%%%
\begin{figure*}[ht!]
    \centering
    \includegraphics[width = 2\columnwidth, clip = true, trim = 5mm 5mm 5mm 5mm]{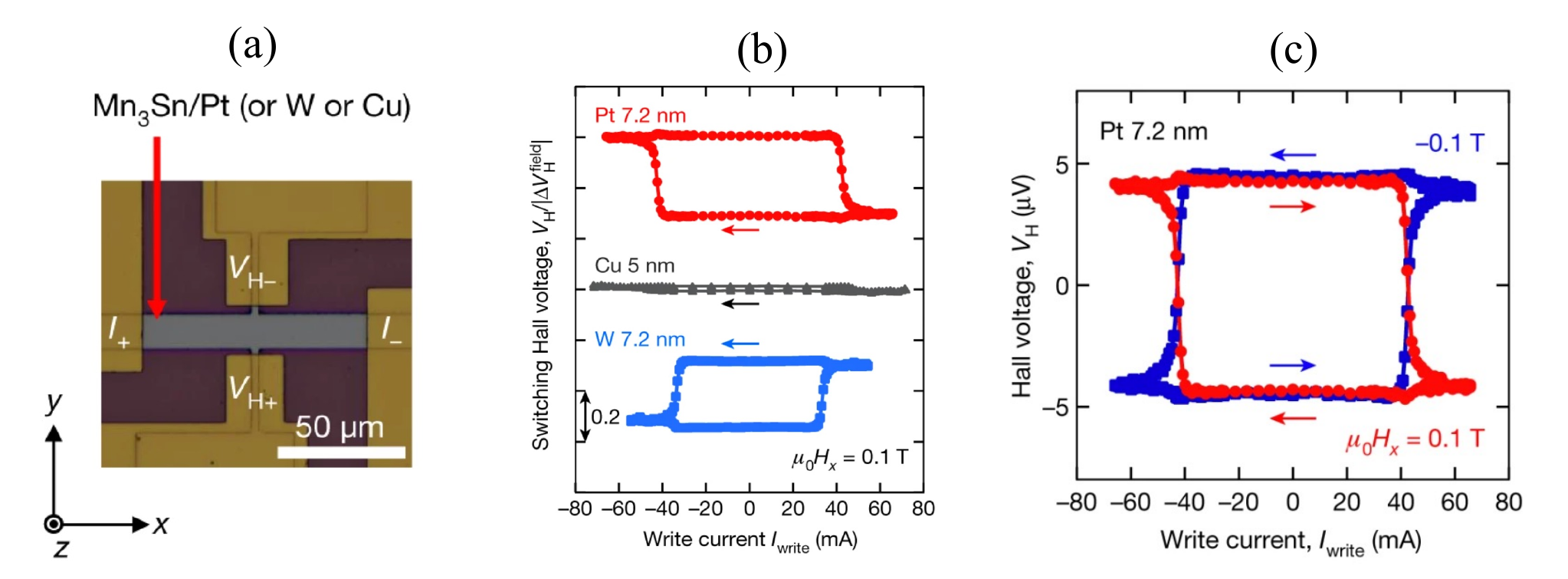}
    \caption{(a) Optical micrograph of the fabricated Mn$_3$Sn/HM bilayer Hall bar devices. Write and read currents are applied along the x direction under magnetic field $H_x$, $H_y$ or $H_z$ along the x, y or z direction, respectively. 
    (b) Hall voltage, $V_H$, versus write current, $I_\mathrm{write}$, for the Mn$_3$Sn/Pt ($7.2~\mathrm{nm}$), Mn$_3$Sn/Cu ($5~\mathrm{nm}$) and Mn$_3$Sn/W ($7.2~\mathrm{nm}$) devices at room temperature. The Hall voltage is normalized by the zero-field Hall voltage, $\Delta V^\mathrm{field}_H$, obtained from the magnetic field dependence of each sample.
    (c) Dependence of $V_H$ on $I_\mathrm{write}$ ($V_H$–$I_\mathrm{write}$ loops) for the Mn$_3$Sn/Pt ($7.2~\mathrm{nm}$) device under a magnetic field of $H_x = +0.1~\mathrm{T}$ (red) and $-0.1~\mathrm{T}$ (blue) along the x direction. 
    Reproduced with permission from Ref.~\cite{tsai2020electrical}.}
    \label{fig:tsai1}
\end{figure*}
In their seminal experimental work, Tsai \emph{et~al.} demonstrated magnetic field-assisted SOT-driven 60$^\circ$ deterministic switching between two adjacent stable states in the Kagome plane of Mn$_3$Sn.~\cite{tsai2020electrical}
To manipulate the order dynamics in Mn$_3$Sn., charge current pulses of duration $100~\mathrm{ms}$ and amplitude $(10^6-10^{7})~\mathrm{A/cm^2}$ were applied to a bilayer, which comprises a HM (Pt or W) layer and polycrystalline Mn$_3$Sn film of thickness $40~\mathrm{nm}$. 
An optical micrograph of the bilayer device setup to manipulate and detect the magnetic order in Mn$_3$Sn is shown in Fig.~\ref{fig:tsai1}(a). 
In this setup, an in-plane external magnetic field, $H_x$, and a charge current, $I_\mathrm{write}$, are both applied along the x-direction while the spin polarization, $\vb{p}$, corresponding to the spin current generated in the HM, is along the y-direction.
Experiments revealed that both the magnitude and the switching direction of the Hall voltage, $V_H$, at fixed $H_x$, were dependent on the HM layer. 
This is due to unique positive and negative spin Hall angles in Pt and W, respectively, leading to opposite switching directions in the two cases, as shown in Fig.~\ref{fig:tsai1}(b). 
Consequently, the final steady-state of the octupole moment in Mn$_3$Sn is different for Pt and W. 
The maximum change in $V_H$ was measured to be about 30\% (25\%) of $\Delta V_H^\mathrm{field}$ ($\Delta V_H^\mathrm{field} = V_H (+H_z  \to 0) - V_H (-H_z \to 0)$) for Pt (W). This was attributed to the polycrystalline nature of the film and the SOT-driven partial 60$^\circ$ switching. 
For a given HM, the polarity of $V_H$-$I_\mathrm{write}$ was found to reverse when $H_x$ was reversed, as shown in Fig.~\ref{fig:tsai1}(c).
%to be coupled to the direction of $I_\mathrm{write}$ when $H_x$ was reversed. 
Further, the change in Hall voltage for the different current directions, \emph{i.e.}, $\Delta V_H^\mathrm{current} = V_H (+I_\mathrm{write} \to 0) - V_H (-I_\mathrm{write} \to 0)$ was found to increase with $|H_x|$. 
Finally, no change in $V_H$ was observed when the magnetic field was applied along y- or z-direction.

To investigate the observed Hall voltage in the bilayer device setup, Tsai \emph{et~al.} numerically explored the SOT-driven dynamics in Mn$_3$Sn with each of the three grain configurations shown in Fig.~\ref{fig:config}. 
The numerical simulations revealed that deterministic switching by 60$^\circ$ was possible only in configuration~I when a symmetry-breaking magnetic field was applied perpendicular to the Kagome plane.
$H_x$ leads to a small out-of-(Kagome)-plane component of the octupole moment of Mn$_3$Sn (\emph{i.e.}, $m_{\perp}$ along the x-direction). The interplay of $m_{\perp}$ along with the appropriate magnitude and direction of the injected current leads to deterministic switching.
Here, the direction of switching is determined by the vector products ($m_{\perp} \cp \vb{p}$), or ($H \cp \vb{p}$), which determine the direction of torques. 
On the other hand, in configuration~II, numerical simulations revealed the possibility to generate up to THz oscillations of the order parameter.~\cite{fujita2017field, shukla2022spin}
Subsequently, probabilistic switching to one of the six-stable states was observed when the current pulse was turned off.~\cite{fujita2017field}
configuration~III was not studied in detail since it would not contribute to the AHE in the experiments.
This is because the octupole moment does not have any out-of-Kagome-plane component for the in-plane magnetic field. 
%%%%%%%%%%%%%%%%%%%%%%%%%%%%%%%%%%%%%%%%%%%%%%%%%%%%%%%%%%%%%%%%%%%%%%%%%%%%%%%%%%%%%%%%%%

\begin{figure*}[ht!]
    \centering
    \includegraphics[width = 2\columnwidth, clip = true, trim = 3mm 2mm 2mm 2mm]{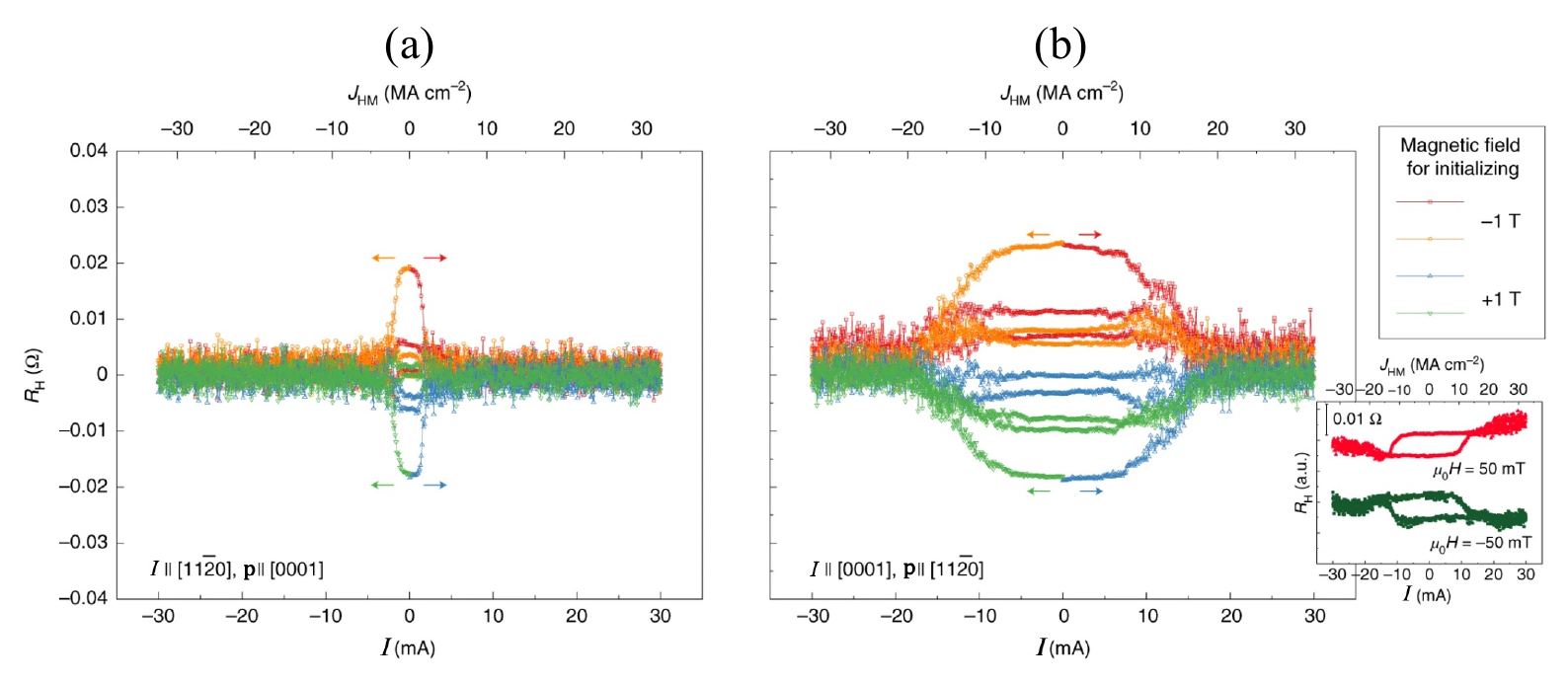}
    \caption{$R_H$–$I$ curves of a fabricated Hall device with $t_{\mathrm{Mn}_3\mathrm{Sn}} = 8.3~\mathrm{nm}$ for the configurations of (a) $I \parallel [11\bar{2}0]$ ($\vb{p} \parallel [0001]$) and (b) $I \parallel [0001]$ ($\vb{p} \parallel [11\bar{2}0]$), respectively. The top axis is the current density $J_\mathrm{HM}$ corresponding to the electric current $I$ in the bottom axis. The colours represent the initializing magnetic fields (the right panel) and the starting points of the measurements, with the latter indicated by arrows. The inset in (b) shows the bipolar switching under in-plane magnetic field $H$. Both the electric current $I$ and magnetic field $H$ are applied along the [0001] direction of Mn$_3$Sn.
    Reproduced with permission from Ref.~\cite{takeuchi2021chiral}.}
    \label{fig:takeuchi1}
\end{figure*}

%%%%%%%%%%%%%%%%%%%%%%%%%%%%%%%%%%%%%%%%%%%%%%%%%%%%%%%%%%%%%%%%%%%%%%%%%%%%%%%%%%%%%%%%%%
In another work, Takeuchi \emph{et~al.}~\cite{takeuchi2021chiral} fabricated different epitaxial $(1\bar{1}00)[0001]$ Mn$_3$Sn films, with thicknesses ranging from $8.3~\mathrm{nm}$ to $30~\mathrm{nm}$, on an $(110)[001]$ MgO substrate, and measured the change in the Hall resistance due to both external magnetic fields as well as charge currents within a Hall bar device structure.
To investigate charge current-driven SOT dynamics, they considered a device setup, where $\vb{p} \parallel [0001]$, similar to configuration~II. 
Using a large magnetic field of $|H| = 1~\mathrm{T}$, the Kagome lattice was initialized in either of the two uniform states, where the uncompensated magnetic moments in all magnetic domains were along the $[1\bar{1}00]$ direction.
When injected with current pulses of duration $100~\mathrm{ms}$ and amplitude $I$, in the absence of an external field, the Hall resistance, $R_H$, showed fluctuations for $|I|$ above a certain threshold current, irrespective of the initial state or the direction of current (Fig.~\ref{fig:takeuchi1}(a)).
The threshold current in this setup was found to be about an order lower than that required in the device setup similar to configuration~I, where $\vb{p} \parallel [11\bar{2}0]$ (Fig.~\ref{fig:takeuchi1}(b)).
In the presence of an in-plane external magnetic field, the latter setup demonstrated the expected deterministic switching~\cite{tsai2020electrical} shown in the inset of Fig.~\ref{fig:takeuchi1}(b). 

To study the microscopic origin of the observed fluctuations, Takeuchi \emph{et~al.} numerically explored the magnetic field-free SOT-driven dynamics in single-domain Mn$_3$Sn, with six minimum energy states, in both configurations~I and~II. 
Firstly, they found that the minimum current required to induce dynamics in configuration~II was indeed smaller than that required for configuration~I. 
This is because the antidamping-like (ADL) SOT in configuration~II has to overcome the weaker effective in-plane anisotropy. On the other hand, in configuration~I, the ADL SOT acts against the stronger exchange interaction.
Secondly, above the threshold current, the magnetic octupole demonstrated oscillatory behavior with a non-zero $m_{[0001]}$, which is the out-of-(Kagome)-plane component $m_{\perp}$. 
The frequency of these oscillation was shown to be current-tunable in the 100s of MHz to GHz range.~\cite{fujita2017field, shukla2022spin} 
When the current pulse was turned off, $m_{[0001]}$ decreased to zero and the magnetic octupole settled to the nearest minimum energy state.
This represents a probabilistic, not deterministic, switching behavior since the particular minimum energy state depends on the state of the oscillating magnetization, when the current pulse was turned off. For a fixed pulse duration, therefore, the final state would depend on the pulse magnitude, $I$. 
% The authors then attributed the observed fluctuations of the Hall resistance, above the threshold current, to non-uniform interactions between the magnetic domains in the single crystal Mn$_3$Sn. 
It is likely that the spin state of the single crystal Mn$_3$Sn fabricated by Takeuchi \emph{et~al.} had small non-uniformities between the magnetic domains, which persisted or changed when injected with current above the threshold. 
Consequently, different domains randomly settled into one of the six degenerate states, leading to the fluctuations in the Hall resistance as the observed ensemble average effect. 
This theory was supported by the observation of reduced fluctuations in the Hall resistance under the application of an out-of-plane magnetic field.
The applied field lifted the degeneracy of the system making one of the states more favorable to the SOT-driven switching. 
The fluctuations in the Hall resistance were also reduced in a wider device that accommodated more magnetic domains as the non-uniformities were averaged out. 
%%%%%%%%%%%%%%%%%%%%%%%%%%%%%%%%%%%%%%%%%%%%%%%%%%%%%%%%%%%%%%%%%%%%%%%%%%%%%%%%%%%%%%%%%%

\begin{figure*}[ht!]
    \centering
    \includegraphics[width = 2\columnwidth, clip = true, trim = 3mm 2mm 2mm 2mm]{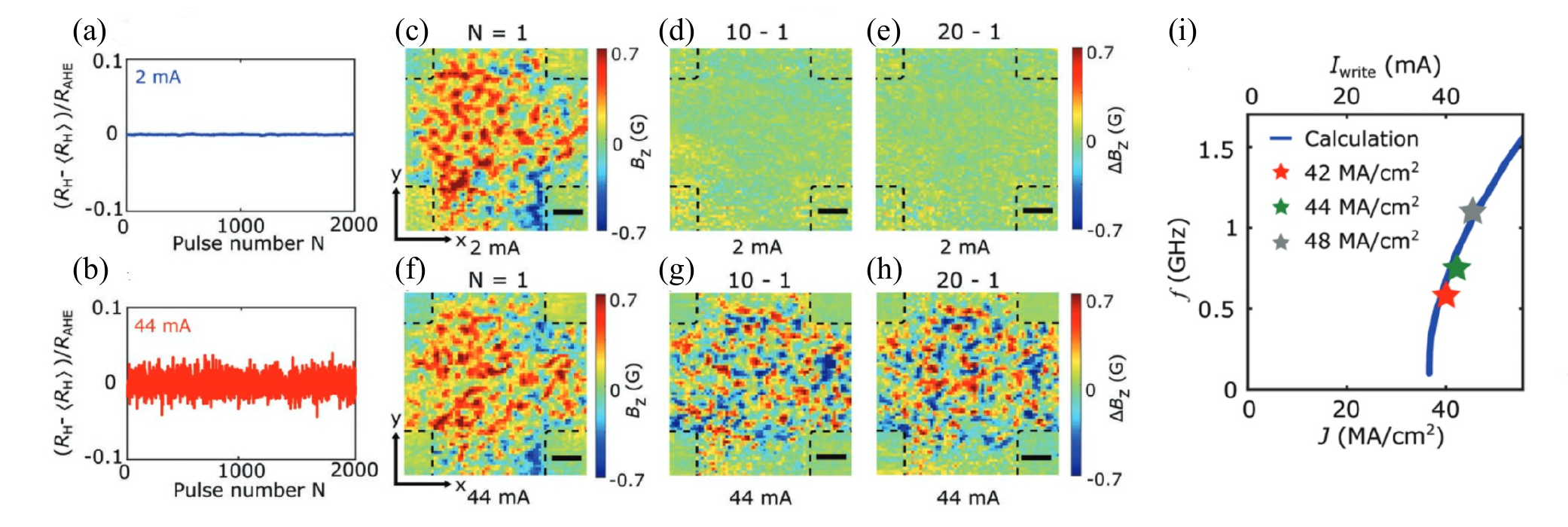}
    \caption{(a, b) Measured normalized Hall resistance, $(R_H - \langle R_H \rangle)/R_\mathrm{AHE}$, as a function of pulse number (N) for write current amplitudes of (c) $2~\mathrm{m A}$ and (d) $44~\mathrm{m A}$.
    (c, f) 2D images of $B_z$ measured after applying the first current pulse (N = 1) with amplitudes of (e) $2~\mathrm{m A}$ and (h) $44~\mathrm{m A}$, respectively.
    (d, g) SOT-induced variation of the stray field, $\Delta B_z$, recorded after applying a $2~\mathrm{m A}$ current pulse 10 (f) and 20 (g) times. 
    (e, h) SOT-induced variation of $\Delta B_z$ recorded after applying a $44~\mathrm{m A}$ current pulse 10 (i) and 20 (j) times.
    For (c–h) the black dashed lines mark the boundary of the patterned Mn$_3$Sn/Pt Hall cross, and the scale bar is $2.5~\mathrm{\mu m}$.
    (i) Theoretically calculated spin rotation frequency $f$ of Mn$_3$Sn as a function of the electric current density, $J$, flowing in the Pt layer. The experimentally measured $f$ for $J = 42~\mathrm{MA/cm^2}$ (the red marker), $44~\mathrm{MA/cm^2}$ (the green marker), and $48~\mathrm{MA/cm^2}$ (the gray marker) agree with the theoretical calculations. 
    Reproduced with permission from Ref.~\cite{yan2022quantum}.}
    \label{fig:Yan1}
\end{figure*}

%%%%%%%%%%%%%%%%%%%%%%%%%%%%%%%%%%%%%%%%%%%%%%%%%%%%%%%%%%%%%%%%%%%%%%%%%%%%%%%%%%%%%%%%%%
Although the work by Takeuchi \emph{et~al.} was significant in understanding oscillatory and switching dynamics in $(1\bar{1}00)[0001]$ Mn$_3$Sn thin films, the oscillating signal or its frequency could not be detected experimentally.
To address these challenges, Yan \emph{et~al.}~\cite{yan2022quantum} used a diamond microchip with nitrogen vacancy (NV) centers implanted on the chip's bottom surface. 
The NV centers are optically active defects in diamond and act as spin sensors, and therefore, enable imaging the effect of charge current on the local spin structure at microscopic length scales.
The diamond chip was placed atop a bilayer Hall device comprising HM (Pt or W) and $50~\mathrm{nm}$ thick polycrystalline Mn$_3$Sn. 
When millisecond charge current pulses of different amplitudes were applied to this device in the absence of an external magnetic field, the Hall resistance demonstrated characteristic fluctuations, similar to those reported by Takeuchi \emph{et~al.}, if $I_\mathrm{write}$ exceeded a certain threshold current.
This can be observed from Fig.~\ref{fig:Yan1}(a, b), where the anomalous Hall resistance normalized to the zero-field Hall resistance, $\frac{(R_H - \langle R_H \rangle)}{R_\mathrm{AHE}}$, is presented as a function of the write current pulse number N.
Enhanced fluctuations can be observed for $44~\mathrm{mA}$ while no such fluctuations are noticed in the case of $2~\mathrm{mA}$, indicating the existence of a threshold current, which was found to be around $36~\mathrm{MA/cm^2}$.
While the polycrystalline Mn$_3$Sn could consist of grains with different configurations, the fluctuations in Fig.~\ref{fig:Yan1}(a, b) are expected to be contributed by grains whose Kagome plane is perpendicular to $\vb{p}$ (configuration~II). 
This is because the threshold current to induce such fluctuations in grains with configuration~I is about an order of magnitude higher~\cite{takeuchi2021chiral} and no AHE would be contributed by grains with configuration~III.
Thus, the fluctuations reported in Fig.~\ref{fig:Yan1}(b) are expected due to the SOT-driven chiral oscillations of the magnetic octupole moment.~\cite{takeuchi2021chiral}

To establish a real-space correlation between the SOT and the chiral oscillations, Yan \emph{et~al.} performed NV wide-field magnetometry after every write current pulse.
Two-dimensional (2D) images of the spatially varying stray field, generated due to the net magnetization exhibited by Mn$_3$Sn, are presented in Fig.~\ref{fig:Yan1}(c-h).
The stray field measured after the first write current pulse (N = 1) in the case of both $2~\mathrm{mA}$ and $44~\mathrm{mA}$ demonstrate multi-domain characteristic of the polycrystalline Mn$_3$Sn film, as shown in Fig.~\ref{fig:Yan1}(c, f).
Figure~\ref{fig:Yan1}(d, e) show that the spatial distribution of the stray field after  10 and 20 current pulses remains largely unchanged for current pulses with amplitude $2~\mathrm{mA}$. 
On the other hand, for current pulses with amplitude $44~\mathrm{mA}$, the stray field distribution show significant change after 10 and 20 current pulses, as shown in Figure~\ref{fig:Yan1}(g, h).
In this case the stray field averaged spatially over the entire Hall cross-region remains constant during the entire measurement sequence.
These results corroborates the role of SOT-driven chiral oscillations, followed by probabilistic switching, in multi-domain Mn$_3$Sn used in configuration~II.
Special care was taken to remove the effects of Joule heating from these results.
Further, utilizing NV relaxometry measurements, frequency of the current-driven chiral oscillations for write current pulses above the threshold were also obtained, as shown by symbols in Fig.~\ref{fig:Yan1}(i).
These extracted data point agree well with the numerical calculations based on single-domain coupled Landau-Lifshitz Gilbert (LLG) equations, especially for higher currents. 
Finally, NV wide-field magnetometry performed in the presence of an in-plane external magnetic field along the direction of current yielded stray field, which averaged spatially over the entire Hall cross demonstrated periodic variation signifying deterministic switching, previously discussed in Ref.~\citen{tsai2020electrical}.
%%%%%%%%%%%%%%%%%%%%%%%%%%%%%%%%%%%%%%%%%%%%%%%%%%%%%%%%%%%%%%%%%%%%%%%%%%%%%%%%%%%%%%%%%%

\begin{figure}[h!]
    \centering
    \includegraphics[width = \columnwidth, clip = true, trim = 0mm 0mm 0mm 0mm]{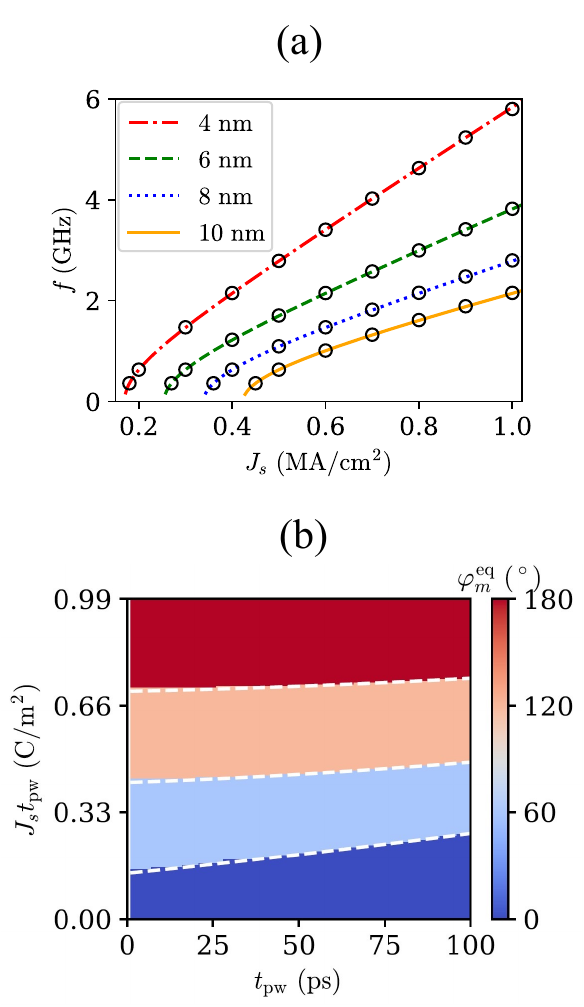}
    \caption{(a) Oscillation frequency as a function of the applied spin current, $J_s$, for different thickness of the Mn$_3$Sn layer. Numerical result from the solution of coupled LLG equations (symbols) agree very well with the analytical results obtained from Eq.~(\ref{eq:time_period}) (lines).
    (b) The final steady-state of the magnetic octupole as a function of the pulse duration, $t_\mathrm{pw}$, and the total injected spin charge density, $J_s t_\mathrm{pw}$. 
    The overlaid dashed white lines represent the analytic bounds of Eq.~(\ref{eq:delay}).
    Here, the thickness of the AFM film is $d_a = 4~\mathrm{nm}$.
    Other parameters are $J_E = 2.4 \times 10^8~\mathrm{J/m^3}$, $D_M = 2 \times 10^7~\mathrm{J/m^3}$, $K_e = 3 \times 10^6~\mathrm{J/m^3}$, and $M_s = 1.3 \times 10^6~\mathrm{A/m}$.
    Adapted from Ref.~\cite{shukla2023order}.}
    \label{fig:shukla1}
\end{figure}

%%%%%%%%%%%%%%%%%%%%%%%%%%%%%%%%%%%%%%%%%%%%%%%%%%%%%%%%%%%%%%%%%%%%%%%%%%%%%%%%%%%%%%%%%%
To get further insight into the dependence of the field-free SOT-driven dynamics pertinent to configuration~II on the intrinsic energy scale of Mn$_3$Sn and its material parameters, our group has theoretically investigated single-domain Mn$_3$Sn with six stable states.~\cite{shukla2023order} 
On the numerical front, we explored chiral oscillation and probabilistic switching dynamics under the effect of d.c. and pulsed currents, respectively, and found the results to be similar to those presented in Ref.~\citen{takeuchi2021chiral}.
Then using a minimal energy landscape model, we obtained an analytic model of the threshold spin current to excite the dynamics as $J_s^\mathrm{th} = d_a \frac{2e}{\hslash}\frac{\qty(3J_E + 7\sqrt{3}D_M)K_U^3}{9\qty(J_E + \sqrt{3}D_M)^3}$, where $e$ and $\hslash$ are the charge of an electron and reduced Plank constant while $d_a$, $J_E$, $D_M$, and $K_U$ are the  thickness, exchange energy constant, DMI constant, and uniaxial energy constant of Mn$_3$Sn, respectively. 
For d.c. input spin current density $J_s \geq J_s^\mathrm{th}$, the frequency of oscillation showed a non-linear (linear) behavior with current for small (large) $J_s$. This behavior was accurately captured by the equation 
\begin{flalign}\label{eq:time_period}
    \begin{split}
        f = \frac{\hslash}{2e} \frac{\gamma J_s}{2\pi \alpha_G M_s d_a} \sqrt{1 - \qty(\frac{J_s^\mathrm{th}}{J_s})^2},
    \end{split}
\end{flalign}
where $\gamma$, $\alpha_G$, and $M_s$ are the gyromagnetic ratio, Gilbert damping constant, and saturation magnetization, respectively.
On the other hand, the final steady-state, owing to the probabilistic switching behavior, for current pulses of duration $t_\mathrm{pw}$ and amplitude $J_s$ was shown to be dependent on the net input spin charge density, $J_s t_\mathrm{pw}$, as
\begin{flalign}\label{eq:delay}
    \begin{split}
        J_s t_\mathrm{pw} &= \alpha_G \frac{2e}{\hslash}\frac{M_s d_a}{\gamma}\frac{\frac{(2n-1) \pi}{2} + \tan^{-1}{\qty(\frac{J_s^\mathrm{th}/J_s}{\sqrt{1-\qty(J_s^\mathrm{th}/J_s)^2}})} }{3\sqrt{1 - \qty(J_s^\mathrm{th}/J_s)^2}}. 
    \end{split}
\end{flalign}
For $J_s \gg J_s^\mathrm{th}$ and small pulse duration, the final state depends strongly on $J_s$ whereas for $J_s \gtrsim J_s^\mathrm{th}$ and large pulse duration, the final state is controlled by $t_\mathrm{pw}$. 
Figure~\ref{fig:shukla1} shows an excellent agreement between the analytic models of Eqs.~\ref{eq:time_period} and~\ref{eq:delay} and the numerical data obtained from the coupled solution of LLG equation.
The effects of Joule heating were not considered in this modeling approach.
%%%%%%%%%%%%%%%%%%%%%%%%%%%%%%%%%%%%%%%%%%%%%%%%%%%%%%%%%%%%%%%%%%%%%%%%%%%%%%%%%%%%%%%%%%

%%%%%%%%%%%%%%%%%%%%%%%%%%%%%%%%%%%%%%%%%%%%%%%%%%%%%%%%%%%%%%%%%%%%%%%%%%%%%%%%%%%%%%%%%%
Although the aforementioned investigations attributed the current-induced changes in single crystal as well as polycrystalline Mn$_3$Sn to ADL SOT while neglecting any significant role of Joule heating, Takeuchi \emph{et~al.} evaluated the temperature rise in the case of devices with configuration~I to be quite large.
In addition, their numerical simulations suggested the threshold current in the case of configuration~I to be at least one order of magnitude higher than that observed experimentally. 
Furthermore, the spin diffusion length in Mn$_3$Sn is only about $1~\mathrm{nm}$~\cite{muduli2019evaluation} but Mn$_3$Sn films used in the experiments were thicker than $8~\mathrm{nm}$.
These peculiarities suggested the contribution of a different switching mechanism, in addition to SOT, in these experiments.
To this end, Pal \emph{et~al.}~\cite{pal2022setting} and Krishnaswamy \emph{et~al.}~\cite{krishnaswamy2022time} independently investigated the current-driven dynamics in bilayer of HM and polycrystalline Mn$_3$Sn films of thicknesses $30~\mathrm{nm}$ to $100~\mathrm{nm}$, in the presence of an in-plane external magnetic field collinear with current pulse.
Both these works showed the demagnetization of Mn$_3$Sn film, owing to its temperature rise above the N\'eel temperature, as an important step in the switching process.

\begin{figure*}[ht!]
    \centering
    \includegraphics[width = 2\columnwidth, clip = true, trim = 5mm 5mm 5mm 0mm]{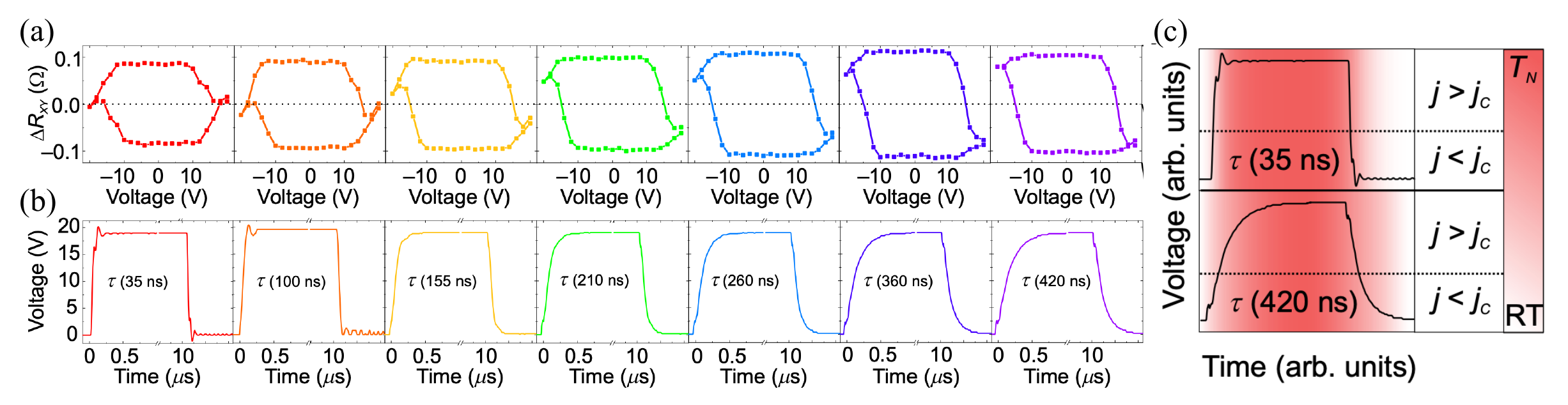}
    \caption{ (a) Switching loops of Mn$_3$Sn/Pt as a function of pulse voltage with rise and fall time, $\tau$, increasing from left to right. 
    Short $\tau$ such as $35~\mathrm{ns}$ leads to tristable switching whereas long $\tau$ of $420~\mathrm{ns}$ leads to bistable switching behavior.
    (b) Corresponding pulse shape.
    (c) Schematic voltage pulse with temperature profile indicated by the red shading.
    Reproduced with permission from Ref.~\cite{krishnaswamy2022time}.}
    \label{fig:krishnaswamy1}
\end{figure*}

In particular, Krishnaswamy \emph{et~al.} showed that the current-induced switching was a function of the pulse fall time, $\tau$. 
This can be observed from Fig.~\ref{fig:krishnaswamy1}(a, b), where bistable switching, typical of configuration~I, is possible only for longer fall times, such as $\tau = 420~\mathrm{ns}$.
On the other hand, shorter fall times ($\tau < 400~\mathrm{ns}$) lead to tristable switching, \emph{i.e.}, a third state with small or zero AHE signal is detected in the steady state above the critical voltage, whereas below the critical voltage, two stable states, typical of bistable switching, are obtained.  
When a current pulse of duration $10~\mathrm{\mu s}$ is applied, the temperature rises above the N\'eel temperature, which was found to be $390~\mathrm{K}$ in $50~\mathrm{nm}$ thick polycrystalline film, in 10s of $\mathrm{ns}$.
When the current is reduced, the Mn$_3$Sn film cools down below its N\'eel temperature due to the diffusion of heat into the substrate.
As depicted in Fig.~\ref{fig:krishnaswamy1}(c), for longer fall times, the current is above a critical value, $j_c$, when Mn$_3$Sn film cools down below the N\'eel temperature, resulting in bistable deterministic switching 
On the other hand, for $\tau = 35~\mathrm{ns}$, the current decreases to values below $j_c$ while the temperature of Mn$_3$Sn film is still above the the N\'eel temperature, leading to random freeze-out of Mn$_3$Sn grains, resulting in zero AHE signal. 
The magnitude and direction of the detected AHE signal was shown to be dependent on the external magnetic field as well as the HM, similar to Fig.~\ref{fig:tsai1}(b, c), thereby making the switching process a function of heat and SOT. 
The polycrystalline films of Mn$_3$Sn would likely consist of grains representing configuration~II; however, no coherent chiral oscillations were observed in experiments, perhaps due to strong damping or heat-induced demagnetization.

\begin{figure*}[ht!]
    \centering
    \includegraphics[width = 2\columnwidth, clip = true, trim = 5mm 0mm 5mm 0mm]{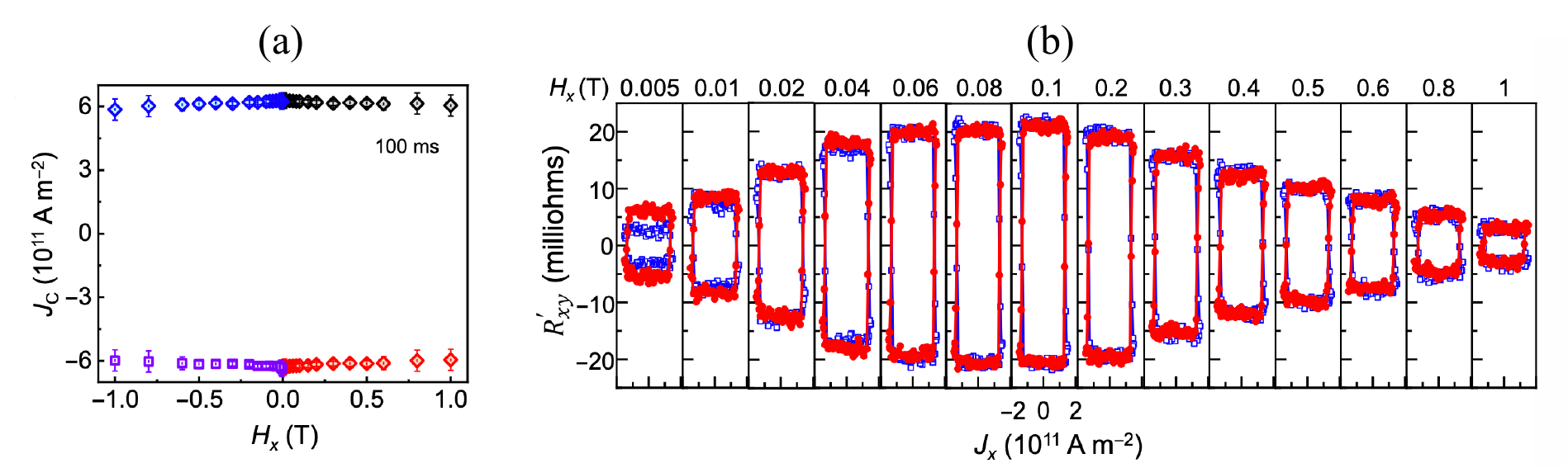}
    \caption{(a) Dependence of $J_c$ on $H_x$ for a pulse duration of $100~\mathrm{ms}$. Black ($+H_x$, $-J_x$), red ($+H_x$, $+J_x$), purple ($-H_x$, $-J_x$), and blue ($-H_x$, $+J_x$) show negative and positive current sweep for positive and negative $H_x$, respectively.
    (b) Variation of $R^{'}_{xy}$ on $H_x$ during current-induced switching.
    Reproduced with permission from Ref.~\cite{pal2022setting}.}
    \label{fig:pal1}
\end{figure*}

On the other hand, Pal \emph{et~al.} proposed seeded SOT to explain the observed field-assisted bistable switching in their $30~\mathrm{nm}$ to $100~\mathrm{nm}$ thick Mn$_3$Sn films. 
In this mechanism, the current-induced Joule heating randomizes the magnetic domains in Mn$_3$Sn as its temperature rises above the N\'eel temperature, in the first step. 
Then the SOT deterministically switches a thin region of Mn$_3$Sn (of the order of the spin diffusion length) near the HM/AFM interface, if the current is above a critical value while the thin region cools down. 
The interface region then acts as the seed for the deterministic switching of the rest of the AFM as the film cools down. 
Switching efficiency showed strong dependence on the fall time---longer fall times facilitated deterministic switching---similar to that described in Ref.~\citen{krishnaswamy2022time}.
At a fixed pulse duration, the critical current density, $J_c$, was found to be almost independent of the in-plane magnetic field, $H_x$, as shown in Fig.~\ref{fig:pal1}(a), presenting a strong evidence towards the role of heating Mn$_3$Sn above its N\'eel temperature for deterministic switching. 
On the contrary, the Hall resistance, $R^{'}_{xy}$, exhibited a non-monotonic dependence on $H_x$, as shown in Fig.~\ref{fig:pal1}(b).
Numerical simulations revealed that the out-of-(Kagome)-plane component of the net magnetization, $m_{\perp}$, increases with $H_x$. 
Pal \emph{et~al.} attributed the initial increase in $R^{'}_{xy}$, for $H_x \leq 0.1~\mathrm{T}$, to the increase in $m_{\perp}$, whereas they explained the subsequent decrease in $R^{'}_{xy}$ for higher $H_x$ as a consequence of decreasing energy barrier between the six different minimum energy states.
Other evidences for the role of Joule heating in the observed deterministic bistable switching included a decrease in $J_c$ with an increase in pulse duration (device temperature) at fixed device temperature (pulse duration).
Finally, unlike pure SOT, where $J_c$ would increase with thickness of the magnetic layer, here $J_c$ was found to decrease with the thickness of Mn$_3$Sn layer while the power density stayed constant.
%%%%%%%%%%%%%%%%%%%%%%%%%%%%%%%%%%%%%%%%%%%%%%%%%%%%%%%%%%%%%%%%%%%%%%%%%%%%%%%%%%%%%%%%%%

\begin{figure*}[ht!]
    \centering
    \includegraphics[width = 2\columnwidth, clip = true, trim = 0mm 0mm 0mm 0mm]{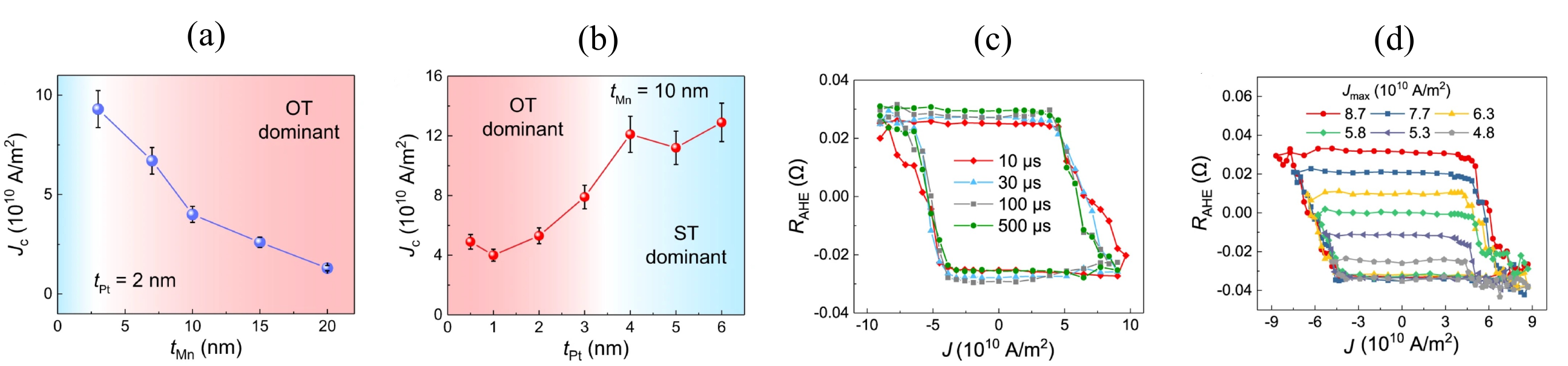}
    \caption{\textcolor{black}{(a) Critical switching current density, $J_c$, as a function of Mn layer thickness, $t_\mathrm{Mn}$, with a fixed Pt layer thickness of $t_\mathrm{Pt} = 2~\mathrm{nm}$. $J_c$ decreases with $t_\mathrm{Mn}$ as the contribution from OT (red region) increases.
    (b) $J_c$ as a function of $t_\mathrm{Pt}$ with a fixed Mn thickness ($t_\mathrm{Mn} = 10~\mathrm{nm}$). When $t_\mathrm{Pt} \leq 3~\mathrm{nm}$, the switching dynamics are dominated by OT (red region) whereas for $t_\mathrm{Pt} > 3~\mathrm{nm}$, spin-torque (blue region) dominates the switching dynamics.
    The error bars in (a) and (b) are obtained from multiple switching loops for each sample.
    (c) Anomalous Hall resistance, $R_\mathrm{AHE}$, as a function of the amplitude of the applied current pulse, $J$, for different pulse width.
    (d) Minor switching loops obtained by limiting the maximum current value, $J_\mathrm{max}$, in negative range.
    Reproduced with permission from Ref.~\cite{zheng2024effective}.
    }}
    \label{fig:Zheng1}
\end{figure*}

%%%%%%%%%%%%%%%%%%%%%%%%%%%%%%%%%%%%%%%%%%%%%%%%%%%%%%%%%%%%%%%%%%%%%%%%%%%%%%%%%%%%%%%%%%
\textcolor{black}{
A very recent work by Zheng~\emph{et~al.} demonstrates the field-assisted current-induced 60$^\circ$ deterministic switching of the AHE signal with current density as low as $1~\mathrm{MA/cm^2}$, due to an efficient charge-to-spin current conversion \textcolor{black}{with assistance of orbital Hall effect~(OHE)}.~\cite{zheng2024effective}
\textcolor{black}{OHE refers to the generation of a transverse orbital current (OC) in response to an external electric field. Unlike spin current, OC is an emergent concept and relies on the orbital character of the Bloch states rather than the electronic states of the constituent atoms.~\cite{go2021orbitronics} OC without an accompanying spin current is feasible in many materials,~\cite{bernevig2005orbitronics, phong2019optically} while in others OC can mediate the spin-Hall~\cite{tanaka2008intrinsic, kontani2009giant} and the valley-Hall effect.~\cite{bhowal2021orbital} OC can be used to control the dynamics of proximal magnetic layers. In~\cite{zheng2024effective}, OC generated in a source layer is first converted into a spin current by means of a heavy metal with a large SOC,~\cite{ding2020harnessing} which enables highly efficient manipulation of the magnetization of Mn$_3$Sn.}
Towards this, a trilayer device setup comprising polycrystalline Mn$_3$Sn~($40~\mathrm{nm}$)/Pt~($0-6~\mathrm{nm}$)/Mn~($0-20~\mathrm{nm}$)/AlN~($5~\mathrm{nm}$) is deposited on thermally oxidized silicon substrates (SiO$_x$/Si).
% Orbital angular momentum, which is generated in the Mn layer and subsequently converted to spin angular momentum in the Pt layer, is proposed as the primary source of the current-driven AHE switching of Mn$_3$Sn grains in configuration~I, where the input current (along $[0001]$ direction) is perpendicular to the Kagome plane.
\textcolor{black}{In this structure, the Mn layer serves as the generator of OC as it is known to possess a 
gigantic orbital Hall conductivity,~\cite{jo2018gigantic} while Pt with a large SOC is used to convert the OC into a spin current, which induces 
% and further converted into spin current in the Pt layer, which has a high SOC. The resulting OT is 
% proposed as the primary source of the 
current-driven switching of Mn$_3$Sn grains in configuration~I (\emph{i.e.}, the input current (along $[0001]$ direction is perpendicular to the Kagome plane).
}
As shown in Fig.~\ref{fig:Zheng1}(a), the critical switching current density, $J_c$, for a thin Pt layer of thickness $t_\mathrm{Pt} = 2~\mathrm{nm}$, is found to decrease monotonically with an increase in Mn layer thickness, likely due to an increased contribution from orbital-torque (OT).
On the other hand, for an Mn layer of thickness $t_\mathrm{Mn} = 10~\mathrm{nm}$, $J_c$ is found to non-monotonically change with $t_\mathrm{Pt}$ (Fig.~\ref{fig:Zheng1}(b)), likely due to a stronger orbital-torque~(spin-torque) contribution for thinner~(thicker) Pt layer. 
Unlike the work by Pal \emph{et~al.},~\cite{pal2022setting} $J_c$ is found to be almost independent of the pulse duration (Fig.~\ref{fig:Zheng1}(c)), probably due to a smaller or negligible contribution from thermally-activated switching mechanisms.  
However, similar to Figs.~\ref{fig:krishnaswamy1}(a) and~\ref{fig:pal1}(b), tristable switching behavior and non-monotonic dependence of the Hall resistance on magnetic field point to a contribution from Joule heating.~\cite{krishnaswamy2022time} 
Although the high OT efficiency is promising in the proposed trilayer structure, the observed results warrant further examination of the setup for a better insight into the switching mechanisms at play.
Finally, Fig.~\ref{fig:Zheng1}(d) presents stable minor switching loops that are obtained by limiting the maximum current amplitude, $J_\mathrm{max}$, in the negative direction. 
These non-volatile intermediate states are current tunable as well as recoverable to the initial state, and therefore raise the possibility of using Mn$_3$Sn-based devices as a memristor.}

\begin{figure}[ht!]
    \centering
    \includegraphics[width = \columnwidth, clip = true, trim = 0mm 0mm 0mm 0mm]{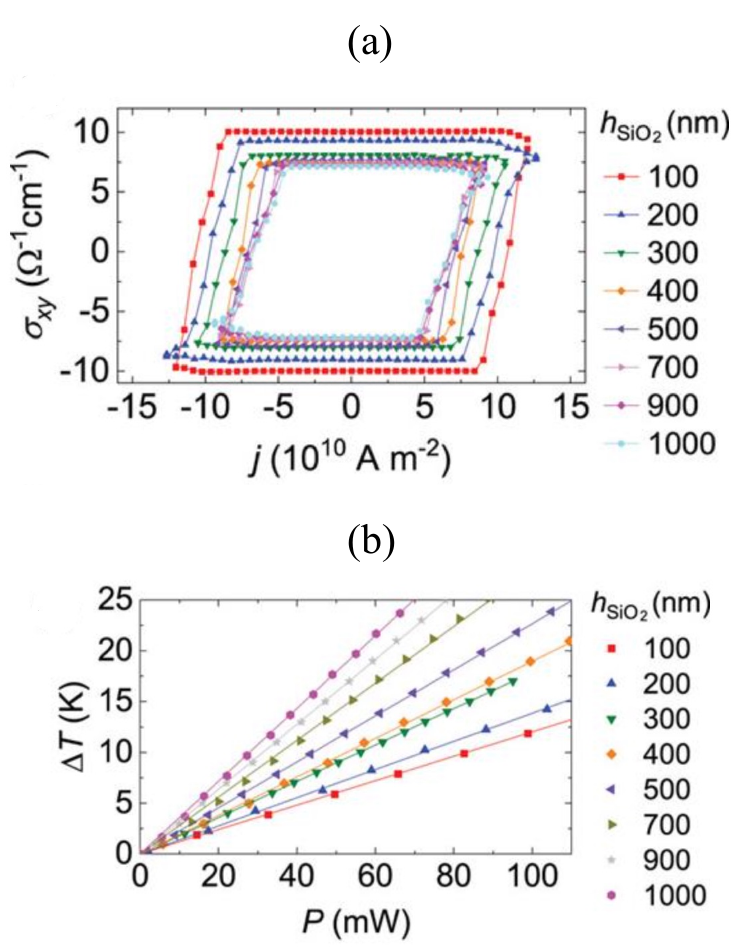}
    \caption{\textcolor{black}{(a) Hysteresis curves obtained for current pulses of different magnitudes, $|j|$, for different thicknesses of the SiO$_2$ layer, $h_\mathrm{SiO_{2}}$. 
    (b) Temperature rise, $\Delta T$, as a function of the input power, $P$, for different $h_\mathrm{SiO_{2}}$.
    Reproduced with permission from Ref.~\cite{yoo2024thermal}.
    }}
    \label{fig:Yoo1}
\end{figure}

%%%%%%%%%%%%%%%%%%%%%%%%%%%%%%%%%%%%%%%%%%%%%%%%%%%%%%%%%%%%%%%%%%%%%%%%%%%%%%%%%%%%%%%%%%
\textcolor{black}{To further investigate the main mechanism behind the current-driven switching dynamics in a bilayer of polycrystalline Mn$_3$Sn and W, very recently, Yoo \emph{et~al.}~\cite{yoo2024thermal} systematically studied a structure similar to that shown in Fig.~\ref{fig:tsai1}(a). 
In particular, they deposited W($7.1~\mathrm{nm}$)/Mn$_3$Sn($34.4~\mathrm{nm}$)/MgO($2~\mathrm{nm}$) films on Si/SiO$_2$ substrates with varying SiO$_2$ thicknesses, $h_\mathrm{SiO_{2}}$, ranging from $100~\mathrm{nm}$ to $1~\mathrm{\mu m}$, and base temperatures, $T_0$, ranging from $200~\mathrm{K}$ to $400~\mathrm{K}$. 
The AHC, $\sigma_{xy}$, was found to be tunable between about $\pm 30~\mathrm{(\Omega.cm)^{-1}}$, when an externally applied out-of-plane magnetic field was tuned between $\pm 2~\mathrm{T}$,
and showed negligible variation with increase in $h_\mathrm{SiO_{2}}$. However, for $T_0 > 260~\mathrm{K}$, $\sigma_{xy}$ and the coercive field reduced with an increase in temperature, likely due to a decrease in the effective magnetic anisotropy. 
On the other hand, when current pulses of varying magnitudes, $j$, but fixed duration of about $105~\mathrm{ms}$ and fall time of about $150~\mathrm{\mu s}$ were applied parallel to an in-plane magnetic field of magnitude $0.1~\mathrm{T}$, $\sigma_{xy}$ exhibited bidirectional switching between about $\pm 10~\mathrm{(\Omega.cm)^{-1}}$ for $h_\mathrm{SiO_{2}} = 100~\mathrm{nm}$ (Fig.~\ref{fig:Yoo1}(a)). 
When $h_\mathrm{SiO_{2}}$ was increased to $1~\mathrm{\mu m}$, the range of $\sigma_{xy}$ reduced to about $\pm 7~\mathrm{(\Omega.cm)^{-1}}$, with the transition occurring at a lower $|j|$.
This reduction in the switching threshold current with increasing $h_\mathrm{SiO_{2}}$ alludes to an important role of the substrate in the switching dynamics, likely via Joule heating.
Indeed, further investigations revealed that for a fixed input power, $P$, which is proportional to the magnitude of the current pulse, the temperature rise, $\Delta T$, was higher for thicker SiO$_2$ layer. Alternatively, same temperature rise was obtained at lower $P$ for thicker SiO$_2$ layer (Fig.~\ref{fig:Yoo1}(b)). 
Here, the choice of pulse duration was to ensure temperature rise to steady state values for Si/SiO$_2$ substrates while the fall time was chosen to prevent formation of multi-stable states (Fig.~\ref{fig:krishnaswamy1}). 
These results provide further evidence and support the claim that current-driven switching in bilayer of polycrystalline Mn$_3$Sn and HM proceeds via demagnetization above the N\'eel temperature, followed by a slow cooling down below it.~\cite{krishnaswamy2022time}
}
%%%%%%%%%%%%%%%%%%%%%%%%%%%%%%%%%%%%%%%%%%%%%%%%%%%%%%%%%%%%%%%%%%%%%%%%%%%%%%%%%%%%%%%%%%
Recently, Xu \emph{et~al.}~\cite{xu2023robust} utilized current pulses of duration $2~\mathrm{ms}$ but different amplitudes and investigated the dynamics in single crystals of Mn$_3$Sn, in the presence of an in-plane collinear magnetic field.
For Mn$_3$Sn in configuration~I, they demonstrated a weak dependence of the critical current on $H_x$ while it was found to decrease with an increase in the film thickness, similar to Ref.~\citen{pal2022setting}.
Thus alluding to the role of Joule heating in the bistable deterministic switching dynamics in their devices.
They also showed the non-monotonic behavior of the Hall resistance with $H_x$, suggesting an increase in $m_{\perp}$ with $H_x$. 
Other expected outcomes such as the dependence of the Hall resistance on the magnitude and direction of $H_x$ were also presented.
Unexpectedly, their Mn$_3$Sn device in configuration~II showcased deterministic switching, considered to be forbidden by previous works.~\cite{tsai2020electrical, takeuchi2021chiral, yan2022quantum, krishnaswamy2022time} 
Although they attribute this behavior to imperfect growth of their Mn$_3$Sn crystals, \textcolor{black}{recent numerical works have suggested the possibility of deterministic switching in configuration~II, ignoring the effects of Joule heating.~\cite{shukla2024impact, xu2024deterministic}
Nevertheless,} further experimental investigations are required to evaluate the possibility of chiral oscillation and deterministic switching in configuration~II.

% Another possibility could be that such grains deterministically switched for the applied field and input current, as demonstrated in recent experimental work by T. Xu \emph{et~al.},~\cite{xu2023robust} and explored via numerical simulations by Z. Xu \emph{et~al.}.~\cite{xu2024deterministic}
% Finally, recent works have also reported switching dynamics in a single layer of polycrystalline Mn$_3$Sn, suggesting spin current generation in the bulk of the AFM to be the cause of the switching dynamics.~\cite{xie2022magnetization}
%%%%%%%%%%%%%%%%%%%%%%%%%%%%%%%%%%%%%%%%%%%%%%%%%%%%%%%%%%%%%%%%%%%%%%%%%%%%%%%%%%%%%%%%%%

%%%%%%%%%%%%%%%%%%%%%%%%%%%%%%%%%%%%%%%%%%%%%%%%%%%%%%%%%%%%%%%%%%%%%%%%%%%%%%%%%%%%%%%%%%
\subsection{SOT-induced dynamics in two-fold symmetric \texorpdfstring{Mn{$_3$}Sn}{Lg}}\label{sec:two-fold}
\begin{figure*}[ht!]
    \centering
    \includegraphics[width = 2\columnwidth, clip = true, trim = 0mm 0mm 0mm 0mm]{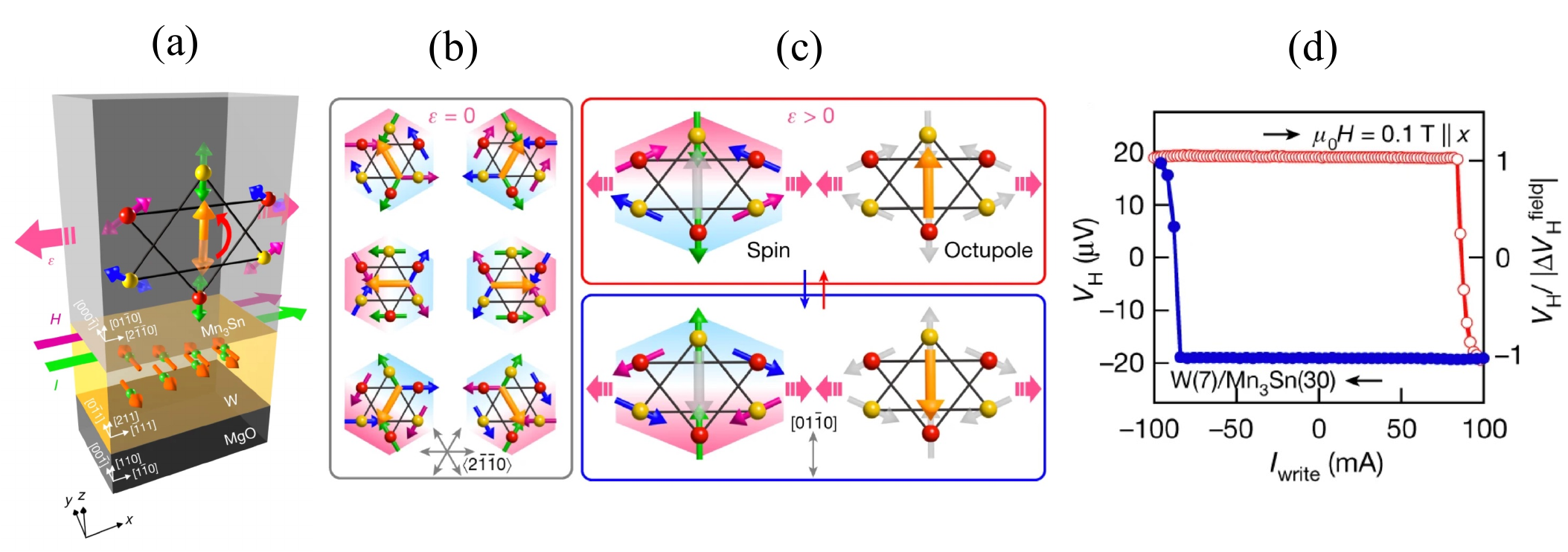}
    \caption{(a) Schematic illustrating the bidirectional switching of the perpendicular magnetic octupole (orange arrows) of Mn$_3$Sn by the SOT. The three Mn moments on the Kagome easy plane are shown by arrows with different colours. Red and yellow spheres represent Mn atoms at z = 0 and 1/2, respectively. An electrical current $I$ (green arrow) flowing in the W layer generates a spin current (green spheres) whose polarization vector (red arrow) is perpendicular to the Kagome plane and induces the SOT on the Mn$_3$Sn layer. 
    $\varepsilon$ (pink arrows) and H (magenta arrows) schematically show the epitaxial, uniaxial tensile strain and the magnetic field, respectively. 
    (b) Chiral spin structure on the Kagome bilayers with no strain ($\varepsilon = 0$), which has six degenerate states with the octupole pointing along the six-equivalent $\langle2\bar{1}\bar{1}0 \rangle$ direction. 
    (c) Chiral PMA binary states on the Kagome bilayers stabilized by the application of the uniaxial tensile strain ($\varepsilon > 0$). 
    (d) The Hall voltage, $V_H$ (left) and volume fraction of the current controllable $V_H/|\Delta V_H^\mathrm{field}|$ (right) versus write current $I_\mathrm{write}$ of the W/Mn$_3$Sn heterostructure at room temperature. The bias magnetic field of $0.1~\mathrm{T}$ is applied along the x direction. 
    Reproduced with permission from Ref.~\cite{higo2022perpendicular}.
    }
    \label{fig:Higo1}
\end{figure*}

% In a recent groundbreaking work, Ikhlas \emph{et~al.}~\cite{ikhlas2022piezomagnetic} demonstrated the possibility to tune the magnetization in Mn$_3$Sn via strain.
Recently, Higo \emph{et~al.} demonstrated that a $30~\mathrm{nm}$ thick film of Mn$_3$Sn, used in a bilayer heterostructure comprising HM and Mn$_3$Sn, grown epitaxially on $(110)[001]$ MgO substrate, exhibited perpendicular magnetic anisotropy (PMA).~\cite{higo2022perpendicular}
As shown in Fig.~\ref{fig:Higo1}(a), the Mn$_3$Sn layer was found to be oriented as MgO~$(110)[001] \parallel$ Mn$_3$Sn~$(01\bar{1}0)[0001]$ with perpendicular magnetic octupole moment along the $[01\bar{1}0]$ direction. 
The lattice mismatch between the Mn$_3$Sn layer and the MgO substrate, for a thin HM layer, led to an in-plane tensile strain of about $0.2\%$ in the Kagome plane along the $[2\bar{1}\bar{1}0]$ direction.
Consequently, the six degenerate minimum energy states, with the octupole pointing along the six-equivalent $\langle2\bar{1}\bar{1}0 \rangle$ direction, as shown in Fig.~\ref{fig:Higo1}(b), reduced to two degenerate minimum energy states shown in Fig.~\ref{fig:Higo1}(c).

When write current pulses of duration $100~\mathrm{ms}$ and amplitude $I_\mathrm{write}$ were applied to this bilayer structure, in the presence of a symmetry-breaking magnetic field, $H_x$, parallel to the current, the Hall voltage, $V_H$, displayed bidirectional deterministic switching behavior above a critical current, as shown in Fig.~\ref{fig:Higo1}(d)~(left axis).
The corresponding critical current density was found to be $J_c^w = 14~\mathrm{MA/cm^2}$.
Strikingly, the entire volume of the Mn$_3$Sn film was found to switch due to the applied current pulse, as shown by the ratio on the right axis of Fig.~\ref{fig:Higo1}(d), where $\Delta V_H^\mathrm{field}$ is the Hall voltage change due to a magnetic field applied in the out-of-plane direction, $H_z$.
This ratio was limited to only $(25-30) \%$ in all previous works.~\cite{tsai2020electrical, takeuchi2021chiral, yan2022quantum}
No switching was observed for magnetic fields perpendicular to the Kagome plane, $H_y$, as it did not break the two-fold symmetry.
This, however, changed for devices, where the Kagome plane was rotated from the current direction ($x$) by an angle. This is because $H_y$ was not perpendicular to the rotated Kagome plane anymore, instead it broke the symmetry, and enabled deterministic switching.
Finally, in the absence of any external field, $V_H$ showed fluctuations, similar to those described in Ref.~\citen{takeuchi2021chiral}, likely representing probabilistic switching dynamics.

\begin{figure*}[ht!]
    \centering
    \includegraphics[width = 2\columnwidth, clip = true, trim = 0mm 0mm 0mm 0mm]{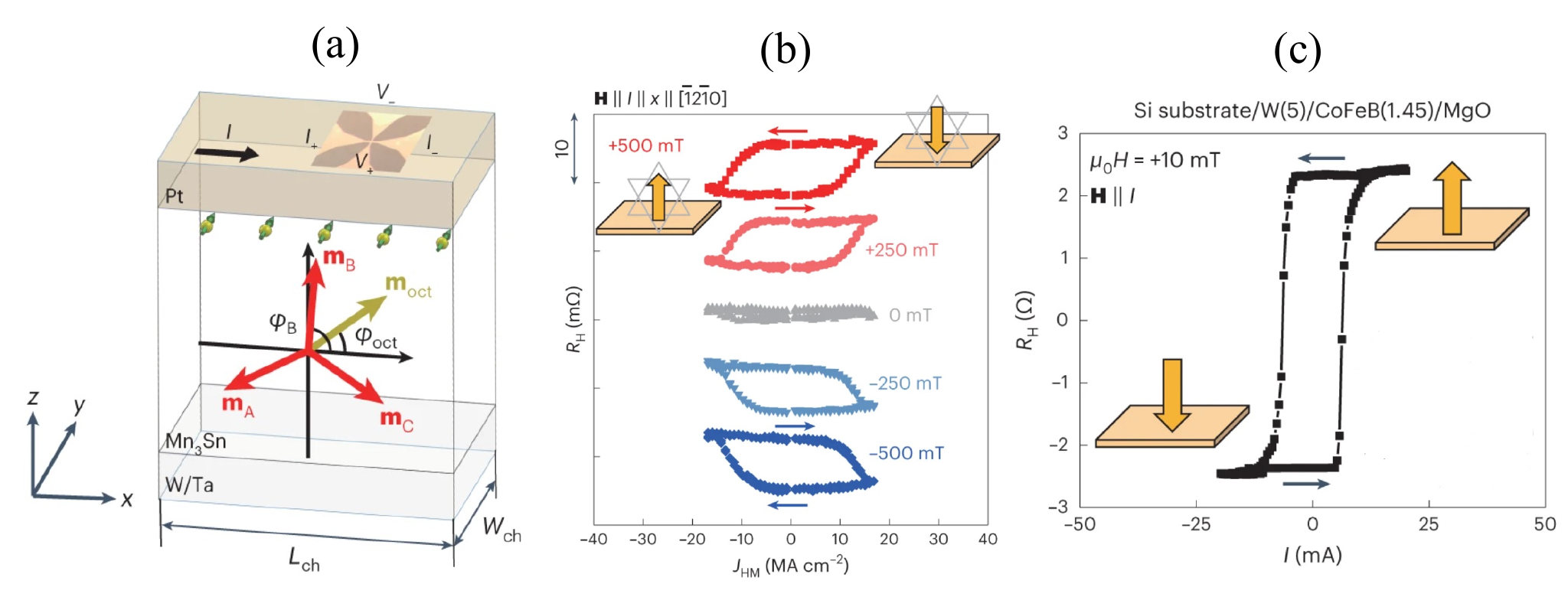}
    \caption{(a) Schematic of the transport measurement reported in Ref.~\cite{yoon2023handedness}. The in-plane spins (yellow spheres and green arrows) are injected from Pt to Mn$_3$Sn. The sublattice moments and the octupole moment are defined as $\vb{m}_{(A,B,C)}$ and $\vb{m}_\mathrm{oct}$ with angles of $\varphi_{(A,B,C)}$ and $\varphi_\mathrm{oct}$, respectively. An optical micrograph of the fabricated Hall bar with a channel width W$_\mathrm{ch}$ and a length L$_\mathrm{ch}$ of $15 - 30~\mathrm{\mu m}$ and $90~\mathrm{\mu m}$, respectively, is shown on the top of the stacks with the definition of the current and voltage directions.
    (b) $R_H - I$ curves for different magnetic field, applied parallel to the injected current, $I$, in the x-direction.
    (c) $R_H - I$ curve measured in W($5~\mathrm{nm}$)/CoFeB($1.45~\mathrm{nm}$)/MgO stacks deposited on a Si substrate, where the bottom W layer injects spins with the same polarization onto the CoFeB layer as those from the top Pt layer onto the Mn$_3$Sn layer. 
    The schematics in (b) and (c) illustrate the switched magnetic states, represented by the octupole moment and the magnetization vector, respectively, in Mn$_3$Sn and CoFeB, respectively.
    Reproduced with permission from Ref.~\cite{yoon2023handedness}.
    }
    \label{fig:Yoon1}
\end{figure*}

It is well known that in the case of FMs and collinear AFMs the experimentally measurable magnetic order, \emph{i.e.}, magnetization vector and N\'eel vector, respectively, rotate in the same direction as the constituting atomic spins.~\cite{liu2011spin, cheng2014spin}
Although Higo \emph{et~al.} demonstrated a field-assisted SOT-induced deterministic switching dynamics in PMA Mn$_3$Sn,~\cite{higo2022perpendicular} similar to that in FMs, there was no discussion regarding the switching direction of the experimentally measured magnetic octupole.
To address this important issue, Yoon and Zhang \emph{et~al.} considered a heterostructure, comprising HM layer (Pt) and $8.8~\mathrm{nm}$ thick PMA Mn$_3$Sn film, and applied a symmetry-breaking in-plane magnetic field along a current pulse of duration $100~\mathrm{ms}$ and amplitude $I$.~\cite{yoon2023handedness} 
The measured Hall resistance, $R_H$, in configuration~II (Fig.~\ref{fig:Yoon1}(a)), showed deterministic switching (Fig.~\ref{fig:Yoon1}(b)), for currents above a certain threshold current ($J_\mathrm{co} \approx 15~\mathrm{MA/cm^2}$), similar to Ref.~\citen{higo2022perpendicular}. 
However, only $30\%$ of the magnetic domains were found to switch for the SOT-driven dynamics, which is different from the $100\%$ switching shown on the right axis of Fig.~\ref{fig:Higo1}(d). 
They attribute the low ratio of switched magnetic domains to possible non-uniform current distribution in the Hall cross device or tilting of the octupole moment from the perpendicular orientation.
They observed almost no change in $R_H$ for configuration~I, even for currents twice as large as that used in configuration~II.
This observation is similar to that reported in Refs.~\citen{takeuchi2021chiral, yan2022quantum}, where the threshold currents for devices in configuration~I were found to be about an order higher than those for configuration~II (see also Sec.~\ref{sec:six-fold}).
Finally, the temperature rise for $15~\mathrm{MA/cm^2}$ was found to be only about $4~\mathrm{K}$, thus eliminating any contribution of Joule heating from the switching dynamics, unlike Refs.~\citen{pal2022setting, krishnaswamy2022time}, where the temperature increased by $140~\mathrm{K}$ for a $1~\mathrm{ms}$ long current pulse of magnitude $17~\mathrm{MA/cm^2}$, resulting in deterministic switching (Fig.~\ref{fig:pal1}(b)).
The small temperature rise could be due to a thinner Mn$_3$Sn layer ($8.8~\mathrm{nm}$) as compared to the $30~\mathrm{nm}$ thick Mn$_3$Sn layer used by Pal \emph{et~al.}, where the critical current was found to decrease with the thickness.
Another reason for the wide variation in the temperature rise could be the different substrates in the two device setups---sapphire in Ref.~\citen{pal2022setting} and (110)MgO in Refs.~\citen{higo2022perpendicular, yoon2023handedness}.

To compare the aforementioned switching dynamics in configuration~II to that in FMs, Yoon and Zhang \emph{et~al.} measured the field-assisted SOT-driven changes in $R_H$ of a W/CoFeB bilayer. 
As shown in Fig.~\ref{fig:Yoon1}(c), the polarity of the switched magnetization in the CoFeB FM was found to be opposite to that of the octupole moment in Mn$_3$Sn (Fig.~\ref{fig:Yoon1}(b)), for the same polarization of the injected spin current.
This surprising observation, pertinent to the opposite rotation of the octupole moment as compared to the direction of rotation of a ferromagnetic moment, was termed as the ``handedness anomaly''.
Through careful examination of the measurements, they concluded that the effect of the field-like torque on the SOT-induced dynamics in the Mn$_3$Sn film was negligible.
Also, owing to the epitaxial nature of the Mn$_3$Sn film, the likelihood of self-generated spin-polarized current~\cite{xie2022magnetization} in Mn$_3$Sn was negligible.
Consequently, the observed SOT-driven switching in PMA Mn$_3$Sn could be completely attributed to the ADL torque. 
This indicates that the direction of the ADL SOT on the magnetic octupole is opposite to its direction on the magnetization of a FM, and therefore, opposite to the direction of the ADL SOT on the constituting sublattice vectors of Mn$_3$Sn.

To get a deeper insight into the ADL SOT-induced dynamics in PMA Mn$_3$Sn in configuration~II, Higo \emph{et~al.} numerically investigated solutions of single-domain coupled LLG equations for a range of input currents and magnetic fields.
They showed that, in the steady state, the octupole moment either stayed near the initial minimum energy state or ground state, switched to the other ground state, or exhibited chiral oscillations. The final steady-states were found to be a function of the initial ground state, the direction and magnitude of the input stimuli.
Yoon and Zhang \emph{et~al.}, on the other hand, developed an effective energy landscape model of the magnetic octupole moment, wherein the in-plane epitaxial tensile strain manifested as perturbations in exchange and uniaxial anisotropy energy interactions. This results in clear two-fold and four-fold effective anisotropies, in addition to the intrinsic six-fold anisotropy.  
They utilized this model to accurately fit and describe the quasi-static second-harmonic Hall measurements of PMA Mn$_3$Sn, used in configuration~II, as a function of the direction of the symmetry-breaking magnetic field and the input current, keeping their respective magnitudes fixed.
The dependence of the different steady-states on the intrinsic energy landscape and the material parameters of PMA Mn$_3$Sn, however, was not discussed in either of these works. 
The time scales associated with switching and oscillation dynamics, and their relation to the material properties as well as the external stimuli, were also not addressed.

A thorough analysis of the different regimes of operation in strained PMA Mn$_3$Sn films, and their dependence on the material parameters and external stimuli, have been presented in our recent work.~\cite{shukla2024impact}. 
Our numerical simulations revealed that in the absence of a magnetic field the magnetic octupole either stays in the initial well or oscillates around the spin polarization, depending on the magnitude of applied current.
On the other hand, for a non-zero magnetic field, the magnetic octupole can be deterministically switched to the other state, if the injected current satisfies $J_c^\mathrm{th1}< J_c < J_c^\mathrm{th2}$, where $J_c^\mathrm{th1}$ and $J_c^\mathrm{th2}$ represent the lower and upper current bounds, respectively.
For currents above $J_c^\mathrm{th2}$, the octupole exhibits chiral oscillations, whose frequency is current tunable.
No switching was observed for currents below $J_c^\mathrm{th1}$, instead stationary states in the initial ground states are obtained. 
Our results showed some differences from those reported by Higo \emph{et~al.}.
To investigate these differences, we utilized an effective energy landscape model of the magnetic octupole moment, which clearly depicts the dependence of the energy barrier separating the two minimum energy states on the in-plane tensile strain.~\cite{liu2017anomalous, li2022free, yoon2023handedness} 
We found that $J_c^\mathrm{th1}$ and $J_c^\mathrm{th2}$ were the minimum currents required to overcome the maximum effective in-plane anisotropy in the two energy wells. Since the external magnetic field breaks the symmetry, $J_c^\mathrm{th1}$ ($J_c^\mathrm{th2}$) decreases (increases) with the magnitude of the field. 
We then derived analytic expressions of the threshold currents, which were found to be independent of damping constant but linearly proportional to the magnitude of the field, the final stationary steady states as well as the switching time as functions of the input stimuli and the material parameters. 
Our models show excellent agreement against numerical data, and therefore, provide important insight into the dynamics of the material, which would be useful to experimentalists and designers. 
Finally, we varied the Gilbert damping constant and found that for damping constants below a certain value ($\alpha \lesssim 2 \times 10^{-3}$ for the set of material parameters considered in our work), the dynamics is similar to that reported by Higo \emph{et~al.}.  
For these low values of damping constants, analytic expressions deviate from the numerical results.  
To resolve these differences, careful analysis involving the exchange interaction due to the out-of-(Kagome)-plane component of magnetic octupole is required.~\cite{yoon2023handedness}

%%%%%%%%%%%%%%%%%%%%%%%%%%%%%%%%%%%%%%%%%%%%%%%%%%%%%%%%%%%%%%%%%%%%%%%%%%%%%%%%%%%%%%%%%%

%%%%%%%%%%%%%%%%%%%%%%%%%%%%%%%%%%%%%%%%%%%%%%%%%%%%%%%%%%%%%%%%%%%%%%%%%%%%%%%%%%%%%%%%%%
\subsection{Joule heating in the presence of SOT}\label{sec:joule_heating}
As discussed in Refs.~\citen{pal2022setting, krishnaswamy2022time, xu2023robust}, Joule heating likely plays an important role in the current-induced switching dynamics in the bilayer devices, comprising HM and Mn$_3$Sn, especially in configuration~I. 
Temperature rise above the N\'eel temperature could also be the reason behind the failure in detection of chiral oscillations, in both six-fold and two-fold symmetric Mn$_3$Sn, in configuration~II.~\cite{krishnaswamy2022time}
In general, Joule heating induced temperature rise as well as the rate of change of the temperature depend on the properties of the conducting and substrate layers in addition to the input current pulse.
% These include the amplitude and duration of the input current pulse, the electrical properties and the size of the layer that the current flows through, the ambient temperature, and the thermal properties of the conducting (magnetic and non-magnetic) media and the substrate.  
In our recent work,~\cite{shukla2023order} we considered a one-dimensional (1D) heat flow model~\cite{coffie2003characterizing} to investigate the current-induced temperature change in the device setup shown in Fig.~\ref{fig:device}.

Although the current pulse is applied to the HM, a part of the current leaks into the metallic Mn$_3$Sn layer, leading to heat generation in both the layers.
To evaluate the current distribution between the two layers a parallel resistor model was employed.
Per this model, the total generated heat depended on the individual resistivity and dimension of both layers.
Both the HM and Mn$_3$Sn were assumed to be at the same temperature and any spatial variation in the temperature was neglected.
The generated heat was assumed to flow through the substrate towards the bottom electrode.
Consequently, the temperature at the junction between the bilayer and the substrate, $T_j$, was found to be strongly dependent on the properties of the substrate such as its thermal conductivity ($\kappa_\mathrm{sub}$), thickness ($d_\mathrm{sub}$), mass density ($\rho_\mathrm{sub}$), and specific heat capacity ($C_\mathrm{sub}$).
Therefore, the rise in the junction temperature, $\Delta T_j (t)$, for current density pulse of magnitude $J_c$ and duration $t_\mathrm{pw}$, was calculated as~\citep{coffie2003characterizing} 
\begin{flalign}\label{eq:DeltaTj}
    \begin{split}
        &\Delta T_j (t) = \Delta T_M\left(\qty(1 - \frac{8}{\pi^2} \sum_{k = \mathrm{odd}}\frac{\exp(-k^2 t/\tau_\mathrm{th})}{k^2}) \right. \\
        &\left. - \theta(t - t_\mathrm{pw}) \qty(1 - \frac{8}{\pi^2} \sum_{k = \mathrm{odd}}\frac{\exp(-k^2 (t - t_\mathrm{pw})/\tau_\mathrm{th})}{k^2})\right),
    \end{split}
\end{flalign}
where $\tau_\mathrm{th} = \frac{4 d_\mathrm{sub}^2 \rho_\mathrm{sub} C_\mathrm{sub}}{\pi^2 \kappa_\mathrm{sub}}$ is the thermal rate constant, $\theta(\cdot)$ is the Heaviside step function, and
\begin{equation}
    \Delta T_M = \frac{d_\mathrm{sub} J_c^2}{\kappa_\mathrm{sub}}\frac{\qty(\rho_\mathrm{Mn_3Sn} d_\mathrm{Mn_3Sn} + \qty(\frac{\rho_\mathrm{Mn_3Sn}}{\rho_\mathrm{HM}}\frac{d_\mathrm{HM}}{d_\mathrm{Mn_3Sn}})^2 \rho_\mathrm{HM} d_\mathrm{HM})}{\qty(1 + \frac{\rho_\mathrm{Mn_3Sn}}{\rho_\mathrm{HM}} \frac{d_\mathrm{HM}}{d_\mathrm{Mn_3Sn}})^2}
\end{equation}
is the maximum possible rise in $T_j$ due to $J_c$. 
Here, $\rho_\mathrm{Mn_3Sn}$ ($\rho_\mathrm{HM}$) and $d_\mathrm{Mn_3Sn}$ ($d_\mathrm{HM}$) are the electrical resistivity and thickness of Mn$_3$Sn (HM) layer.

\begin{figure}[ht!]
    \centering
    \includegraphics[width = \columnwidth, clip = true, trim = 0mm 0mm 0mm 0mm]{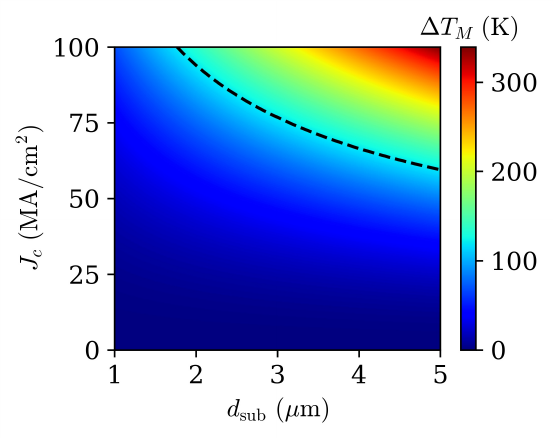}
    \caption{Maximum possible rise in the junction temperature, $\Delta T_M$, for different thickness of substrate MgO, $d_\mathrm{sub}$, and input charge current density, $J_c.$ 
    $\Delta T_\mathrm{M}$ increases with both $d_\mathrm{sub}$ and $J_c$, as expected.
    The dashed black line corresponds to $\Delta T_M = 120~\mathrm{K}$, which is the maximum temperature rise before the antiferromagnetic order in Mn$_3$Sn changes.
    Here, the thickness of the AFM film is $d_a = 4~\mathrm{nm}$.
    Adapted from Ref.~\cite{shukla2023order}.}
    \label{fig:shukla2}
\end{figure}

Figure~\ref{fig:shukla2} shows $\Delta T_M$ as a function of $d_\mathrm{sub}$ and $J_c$.
Here, the thickness of Mn$_3$Sn (W) film is $d_\mathrm{Mn_3Sn} = 4~\mathrm{nm}$ ($d_\mathrm{W} = 7~\mathrm{nm}$)~\cite{higo2022perpendicular} and $\rho_\mathrm{Mn_3Sn} = 330~\mathrm{\mu \Omega~cm}$ ($\rho_\mathrm{W} = 43.8~\mathrm{\mu \Omega~cm}$).~\cite{higo2022perpendicular} 
The substrate is assumed to be MgO and its relevant material parameters~\cite{meinert2018electrical}  are $\kappa_\mathrm{sub} = 40~\mathrm{W/(m~K)}$, $\rho_\mathrm{sub} = 3580~\mathrm{kg/m^3}$, and $C_\mathrm{sub} = 930~\mathrm{J/(kg~K)}$.
In Fig.~\ref{fig:shukla2}, an increase in $J_c$, increases the generated heat in the metallic layers, which in turn, increases $\Delta T_M$. 
On the other hand, an increase in $d_\mathrm{sub}$, increases the thermal resistance in the direction of heat flow, thereby slowing down the heat removal from the system and increasing $\Delta T_M$.
The dashed black curve, corresponding to $\Delta T_M = 120~\mathrm{K}$, represents the maximum temperature rise before Mn$_3$Sn gets demagnetized.
Consequently, no chiral oscillations would be observed in the region above the dashed curve. 
Here, the room temperature and the N\'eel temperature were assumed to be $300~\mathrm{K}$ and $420~\mathrm{K}$, respectively.
Although not shown here, for a fixed $J_c$ and $d_\mathrm{sub}$, $\Delta T_M$ would increase for thicker $d_\mathrm{HM}$ or $d_\mathrm{Mn_3Sn}$ since the generated heat would increase with the respective thickness. 
This explains the observed reduction in critical current density with an increase in AFM thickness.~\cite{pal2022setting} 
The generated heat is expected to be lower if Pt was used as the HM since it has lower resistivity.~\cite{tsai2020electrical} 
Finally, for the same values of $J_c$ and $d_\mathrm{sub}$, $\Delta T_M$ would be higher in the case of both sapphire and SiO$_2$ substrates. This is due to their lower $\kappa_\mathrm{sub}$,~\cite{cahill1990thermal} and therefore, higher thermal resistance ($\propto 1/\kappa_\mathrm{sub}$).

\begin{figure}[ht!]
    \centering
    \includegraphics[width = \columnwidth, clip = true, trim = 0mm 0mm 0mm 0mm]{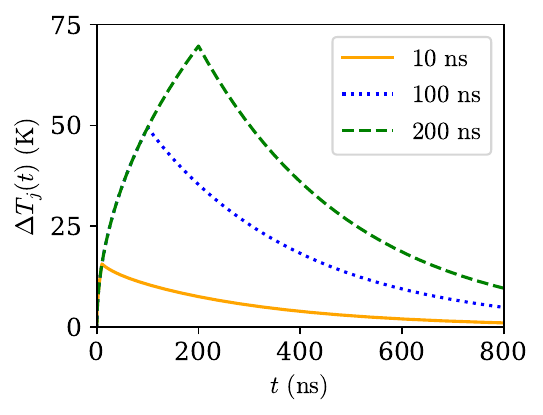}
    \caption{Temperature rise of HM-Mn$_3$Sn bilayer as a function of time for the charge current pulse of amplitude $J_c = 76.8~\mathrm{MA/cm^2}$ but different durations, as indicated in the legend. 
    Here, $d_\mathrm{sub} = 3~\mathrm{\mu m}$, $\tau_\mathrm{th} \approx 303~\mathrm{ns}$, and $\Delta T_M = 120~\mathrm{K}$ for MgO substrate.
    $\Delta T_j (t)$ is small if $t_\mathrm{pw} \ll \tau_\mathrm{th}$ as in the cases of $10~\mathrm{ns}$, and increases with $t_\mathrm{pw}$. \textcolor{black}{Here, $\Delta T_j (t)$ is evaluated using a heat flow model discussed in Ref.~\cite{coffie2003characterizing}.}}
    \label{fig:shukla3}
\end{figure}

Other factors that affect $\Delta T_j (t)$ are $\tau_\mathrm{th}$ and $t_\mathrm{pw}$, which decide the rate of change of the temperature and the total generated heat, respectively.
For fixed $J_c$ and $d_\mathrm{sub}$, $\Delta T_j (t)$ reaches to about 95\% of the respective $\Delta T_M$ in $t \approx 3 \tau_\mathrm{th}$; therefore, applying $t_\mathrm{pw} \ll \tau_\mathrm{th}$ could ensure lower $\Delta T_j (t)$. 
In the case of $d_\mathrm{sub} = 3~\mathrm{\mu m}$, $\tau_\mathrm{th} \approx 304~\mathrm{ns}$ while $J_c = 76.8~\mathrm{MA/cm^2}$ leads to $\Delta T_M = 120~\mathrm{K}$ for the MgO substrate.
As shown in Fig.~\ref{fig:shukla3}, $\Delta T_j (t) \approx 16~\mathrm{K}$ for $t_\mathrm{pw} = 10~\mathrm{ns}$ ($\ll \tau_\mathrm{th}$) but increases to about $70~\mathrm{K}$ for $t_\mathrm{pw} = 200~\mathrm{ns}$.
Short input pulses ($t_\mathrm{pw} \ll \tau_\mathrm{th}$) could also enable the application of $J_c$ above the limit imposed by the dashed black curve in Fig.~\ref{fig:shukla2}, and hence facilitate faster switching of the magnetic octupole, if applicable.
The heat generated in the bilayer, due to a fixed $J_c$, would reach the bottom electrode faster for a thinner substrate; therefore, $\tau_\mathrm{th}$ decreases with $d_\mathrm{sub}$;
consequently, $\Delta T_j (t)$ reaches $\Delta T_M$ faster for lower $d_\mathrm{sub}$.
This is expected since the maximum possible $\Delta T_M$, at a fixed $J_c$, is smaller for lower $d_\mathrm{sub}$ (Fig.~\ref{fig:shukla2}).
Finally, for the same value of $d_\mathrm{sub}$, $\tau_\mathrm{th}$ would be higher in the case of both sapphire and Si/SiO$_2$ substrates, due to their lower thermal diffusivity $\alpha = \frac{\kappa_\mathrm{sub}}{\rho_\mathrm{sub} C_\mathrm{sub}}$.
This slow heat dissipation could be behind the dependence of deterministic switching on fall time, as reported in Ref.~\citen{pal2022setting, krishnaswamy2022time}, where sapphire and SiO$_2$ were used as substrates, respectively.
%%%%%%%%%%%%%%%%%%%%%%%%%%%%%%%%%%%%%%%%%%%%%%%%%%%%%%%%%%%%%%%%%%%%%%%%%%%%%%%%%%%%%%%%%%

%%%%%%%%%%%%%%%%%%%%%%%%%%%%%%%%%%%%%%%%%%%%%%%%%%%%%%%%%%%%%%%%%%%%%%%%%%%%%%%%%%%%%%%%%%
\subsection{Role of Thermal Fluctuations in \texorpdfstring{Mn{$_3$}Sn}{Lg}}
\label{sec:thermal}
Recent experimental~\cite{tsai2020electrical, takeuchi2021chiral, yan2022quantum, pal2022setting, krishnaswamy2022time, higo2022perpendicular, xu2023robust, yoon2023handedness} and theoretical works,~\cite{dasgupta2022tuning, shukla2023order, xu2024deterministic, shukla2024impact} as discussed in Secs.~\ref{sec:six-fold} and~\ref{sec:two-fold}, have made significant progress in elucidating the current-driven dynamics in Mn$_3$Sn with six equivalent minimum energy states as well as in Mn$_3$Sn with PMA, both in the presence and absence of a symmetry breaking magnetic field. 
The promising results make Mn$_3$Sn a strong candidate for CMOS compatible, compact and low power memory devices as well as GHz-THz signal sources. 
{A practical realization of such devices, however, requires an investigation into the effect of thermal fluctuations, at different temperatures, on the long-term stability of a Mn$_3$Sn memory element and on the linewidth of a coherent signal source.
Therefore, a detailed investigation of the thermal stability factor, $\Delta$, which depends on the energy barrier between the stable states, corresponding to thin Mn$_3$Sn films is needed.}

Towards this, Sato \emph{et~al.}, in a very recent work,~\cite{sato2023thermal} evaluated the thermal stability of $20~\mathrm{nm}$ thick epitaxial films of Mn$_3$Sn of varying diameters $D$, ranging from $175~\mathrm{nm}$ to $1000~\mathrm{nm}$, each of which exhibited PMA due to tensile strain from the MgO substrate.
They applied magnetic field pulses of duration $1~\mathrm{s}$ but different amplitudes in-the-(Kagome)-plane perpendicular to the interface, and explored the switching probability of the bistable magnetic order. 
Then using the N\'eel-Arrhenius thermal activation model and assuming an attempt frequency of $1~\mathrm{GHz}$, typically used for FMs, they evaluated, from the switching probability curves, the thermal stability factor of the different samples.
They found that epitaxial Mn$_3$Sn films, characterized by dominant two-fold and weak six-fold anisotropies, of diameters below $\sim 300~\mathrm{nm}$ exhibited single-domain behavior, \emph{i.e.}, $\Delta \propto D^2$. 
On the other hand, for $D \gtrsim 300~\mathrm{nm}$, the epitaxial films are expected to be multi-domain in nature since $\Delta$ showed negligible change with $D$. 
At $D = 300~\mathrm{nm}$, $\Delta$ is evaluated to be in the range of about $100-150$.
This result, however, is sensitive to the value of the attempt frequency, and needs further analysis for higher attempt frequencies. 

Kobayashi \emph{et~al.} investigated the dependence of the critical switching voltage on the duration of the applied pulse, $t_p$, for a bilayer comprising Pt and $40~\mathrm{nm}$ thick Mn$_3$Sn, grown on Si/SiO$_2$ substrate.~\cite{kobayashi2023pulse} 
A symmetry-breaking magnetic field, $H_x$, was also applied parallel to the current/voltage pulse.
The Hall resistance measured as a function of the applied voltage, for $t_p$ in the range of $10~\mathrm{\mu s}$ to $1~\mathrm{ms}$ and different $H_x$, showed tristable switching behavior, similar to Fig.~\ref{fig:krishnaswamy1}(a), indicating clearly that Joule heating played a role in switching dynamics.~\cite{pal2022setting, krishnaswamy2022time, xu2023robust}
The voltage at which the Hall resistance changed to zero, $V_\mathrm{critical}$, was found to decrease monotonically with an increase in $t_p$, as expected.~\cite{pal2022setting}
Nevertheless, ignoring the effects of Joule heating, and using a switching-time model, pertinent to FMs in the thermally-activated regime, given as~\cite{koch2004time} 
\begin{equation}\label{eq:Vcrit}
    V_\mathrm{critical} = V_\mathrm{c0}\qty[1 - \qty(\frac{1}{\Delta}\log{\qty(\frac{t_p}{t_0})})^{1/n}],
\end{equation}
where $V_\mathrm{c0}$ is the zero temperature critical voltage, $1/t_0$ is the attempt frequency while $n$ is the exponent of the thermally-assisted switching dynamics, 
Kobayashi \emph{et~al.} estimated $\Delta = 131.1 \pm 0.5$ at $300~\mathrm{K}$.
Here, it was assumed that $t_0 = 1~\mathrm{ns}$ and $n = 1$, which are typically used for FMs.~\cite{koch2004time}
The resonant frequency in noncollinear chiral Mn$_3$Sn was recently measured to be approximately $0.9~\mathrm{THz}$~\cite{miwa2021giant} while a previous work
measured the attempt frequency in MnIr, which is a collinear AFM, in the THz regime.~\cite{vallejo2010measurement}
Another recent work theoretically calculated the attempt frequency in Mn$_3$Ir to be greater than $10~\mathrm{THz}$.~\cite{jenkins2019magnetic}
The attempt frequency in Mn$_3$Sn, therefore, could also be in the THz regime.~\cite{matsuo2022anomalous} 
Consequently, assuming $t_0 = 1~\mathrm{ps}$, the estimated value of $\Delta$ increases by about 6.9, resulting in $\Delta \approx 138$.
Including the effects of Joule heating in the calculation is expected to lead to further changes in $\Delta$.

Previous theoretical works involving STT-driven dynamics in single-domain PMA FMs~\cite{taniguchi2011thermally, pinna2012thermally} showed that a careful Fokker-Plank- or ensemble LLG-based approach was required to understand the switching behavior in the long pulse regime. 
In particular, these works found $n$ to be equal to 2, and discussed the importance of its accuracy for an accurate estimation of $\Delta$.
Similarly, solving an ensemble of coupled LLG equations, in the presence of Gaussian random thermal noise,~\cite{go2022noncollinear} could help decipher $n$, accurately, in the case of SOT-driven switching dynamics in Mn$_3$Sn nanomagnets with six-fold as well as two-fold symmetry. 
This would then enable an accurate estimation of $\Delta$, and therefore the long-term retention capability of single-domain Mn$_3$Sn nanodots \textcolor{black}{required for memory elements.
Deterministic switching in the case of memory element requires the write error rate ($1 - P_\mathrm{switch}$, where $P_\mathrm{switch}$ is the switching probability) to be negligible. On the other hand, $P_\mathrm{switch} = 0.5$ is required to build a true random number generator.~\cite{shukla2023true} Increased thermal fluctuations due to an increase in operating temperature due to Joule heating, for example, could change $P_\mathrm{switch}$, and consequently, the reliability of the device operation.~\cite{shukla2023true} An accurate estimation of $n$ in Eq. (\ref{eq:Vcrit}) would enable an accurate estimation of the change in $P_\mathrm{switch}$ with temperature, and thus help circuit designers consider the area and energy overhead of \textcolor{black}{spintronics-based}
probabilistic \textcolor{black}{and neuromorphic} computing paradigms.}
%%%%%%%%%%%%%%%%%%%%%%%%%%%%%%%%%%%%%%%%%%%%%%%%%%%%%%%%%%%%%%%%%%%%%%%%%%%%%%%%%%%%%%%%%%

\textcolor{black}{
With the exception of a few studies that focused on thick epitaxial layers, almost all other experimental investigations on field- and current-driven dynamics in noncollinear AFMs have focused on bulk polycrystalline AFM layers with large diameters. Such thick films with large diameters are not compatible with next-generation CMOS technology. Instead, thin epitaxial AFMs layer that can be scaled down to 10s of nanometers are preferable for spintronics application. 
However, in Ref.~\citen{sato2023thermal}, Sato \emph{et~al.} predicted that scaling down the diameter below $100~\mathrm{nm}$ could reduce $\Delta$ below $20$. 
Furthermore, elevated temperatures likely reduce the effective magnetic anisotropy of the system.~\cite{yoo2024thermal} 
Consequently, the energy barrier between the bistable states of Mn$_3$Sn nanodots could become comparable to thermal energy, $k_\mathrm{B} T$, where $k_\mathrm{B}$ is the Boltzmann constant while $T$ is the temperature, making the magnetic octupole moment fluctuate between the two states without any external perturbation, similar to the case of FMs~\cite{bean1959superparamagnetism, lopez2002transition} and collinear AFMs.~\cite{richardson1991origin, flipse1999magnetic} 
Such Mn$_3$Sn nanoparticles could find applications in biomedical applications~\cite{wu2019magnetic} as well as spintronics-based probabilistic and neuromorphic computing paradigms.~\cite{vodenicarevic2017low, camsari2017stochastic, debashis2018design, camsari2019p, borders2019integer, camsari2020charge} 
Therefore, it is important to theoretically analyze and understand the various thermal and switching properties of such nanodots, similar to the case of a ferromagnetic or collinear antiferromagnetic nanoparticles.~\cite{coffey2012thermal, rozsa2019reduced, hayakawa2021nanosecond}
}

%%%%%%%%%%%%%%%%%%%%%%%%%%%%%%%%%%%%%%%%%%%%%%%%%%%%%%%%%%%%%%%%%%%%%%%%%%%%%%%%%%%%%%%%%%
\subsection{Readout of order parameter}
As discussed in Sec.~\ref{sec_physphen}, several methods may be used to detect the state of the octupole moment in Mn$_3$Sn and related chiral AFMs. Of these methods, TMR offers the potential of providing a large readout signal that would be compatible with CMOS circuitry. In chiral AFMs, the ferroic order of the cluster magnetic octupole breaks the macroscopic time-reversal symmetry leading to a finite TMR.~\citep{dong2022tunneling, chen2023octupole}   
For the setup shown in Fig.~\ref{fig:device}, the low and high resistance states correspond to angles $0^\circ$ and $180^\circ$ between the magnetic octupoles of the pinned and free Mn$_3$Sn layers, respectively.~\cite{chen2023octupole}
The octupole moment of the pinned layer, $\vb{m}^\mathrm{P}_\mathrm{oct}$, is fixed and not expected to change for the currents applied to the device.
On the other hand, the dynamics of the order parameter of the free layer, $\vb{m}_\mathrm{oct}$, depends on the magnitude of $H_a$ and $J_c$, as described in detail in Secs.~\ref{sec:six-fold} and~\ref{sec:two-fold}.
The different steady-state response of $\vb{m}_\mathrm{oct}$, with respect to $\vb{m}^\mathrm{P}_\mathrm{oct}$, is then measured as $V_L$ across the resistive load, $R_L$, for a given read voltage, $V_R$. 
Here, $V_R$ is sufficiently small, such that the current due to it does not disturb the steady states of either $\vb{m}^\mathrm{P}_\mathrm{oct}$ or $\vb{m}_\mathrm{oct}$.

Chen \emph{et~al.} reported a TMR of 2\% at room temperature~\cite{chen2023octupole} in a trilayer device comprising epitaxial Mn$_3$Sn($12~\mathrm{nm}$)/MgO($3.3~\mathrm{nm}$)/\textcolor{black}{Mn$_3$Sn}($42~\mathrm{nm}$). 
The RA product of this device, however, was extremely large on the order of $10^{8}~\mathrm{\Omega~\mu m^2}$ while the corresponding resistance states were also quite large (about $2~\mathrm{M \Omega}$), making the device CMOS incompatible.
Although lowering the RA product reduced the device resistance, it also reduced the already small TMR of the trilayer device. Similarly, reducing the thickness of the tunnel barrier lowers both the resistance and the TMR.
\textcolor{black}{Chen \emph{et~al.} found} the $c$-axis of the epitaxial Mn$_3$Sn film to be at an angle of about $30^\circ$ with the normal direction of the film. Therefore, the current due to $V_R$ passed through the trilayer at an angle of about $60^\circ$ with respect to the Kagome plane.
Theoretical calculations revealed that with vacuum as the tunnel barrier and current perpendicular (parallel) to the Kagome plane, TMR increased to about $130\%$ ($150\%$) for a barrier thickness of about $0.45~\mathrm{nm}$ ($0.65~\mathrm{nm}$). 
In both the cases, the RA product was found to be $\leq 1~\mathrm{\Omega~\mu m^2}$ in the parallel as well as the antiparallel orientations of the two octupole moments.
Calculations suggested that both the TMR and RA product would increase to higher values with the thickness of the barrier, at least for thicknesses smaller than $1.1~\mathrm{nm}$.

As shown by Dong \emph{et~al.} in their theoretical work,~\citep{dong2022tunneling} large TMR of about $300\% $ at room temperature, for vacuum as a tunnel barrier ($0.6~\mathrm{nm}$ thick) and current perpendicular to the Kagome plane of the AFM tunnel junction, is possible. 
In this calculation, the RA product increased from approximately $0.05~\mathrm{\Omega \vdot \mu m^2}$ in the parallel configuration to $0.2~\mathrm{\Omega \vdot \mu m^2}$ in the antiparallel configuration.
On the other hand, when $\vb{m}_\mathrm{oct}$ was at $60^\circ$ and $120^\circ$ with respect to $\vb{m}^\mathrm{P}_\mathrm{oct}$, the RA product was found to be approximately $0.07~\mathrm{\Omega \vdot \mu m^2}$ and $0.11~\mathrm{\Omega \vdot \mu m^2}$, respectively.
Assuming same values of RA product for the device setup of Fig.~\ref{fig:device}, where current is parallel to the Kagome plane, we estimated $V_L = 91~\mathrm{mV}$ ($71~\mathrm{mV}$) in the parallel (antiparallel) configuration, for $V_R = 100~\mathrm{mV}$, $R_L = 50~\mathrm{\Omega}$ and the area of the tunnel barrier $\mathcal{A} = 100 \times 100~\mathrm{nm^2}$. 
Finally, the relative orientations of $60^\circ$ ($120^\circ$) leads to intermediate value of $V_L$ as $88~\mathrm{mV}$ ($82~\mathrm{mV}$).
Further improvement in $V_L$ could be achieved by increasing $V_R$ to a higher value as long as the current flowing through the tunnel junction does not disturb the octupole moments.
Improvements in the RA product of the tunnel junction without compromising the TMR could also help enhance the values of $V_L$.
The TMR of $300\% $ with vacuum as the tunnel barrier, for a fixed RA product, \textcolor{black}{does not necessarily} represents the upper limit.~\cite{dong2022tunneling} Calculations showed that a more realistic tunnel junction with monolayer hexagonal hafnia reduced the TMR to $124\%$. 
In the case of AHE, the Hall voltage was found to be on the order of a few $\mathrm{\mu V}$ for 10s of $\mathrm{\mu A}$ of read current,~\cite{shukla2023order} which is quite small,  making TMR as the better choice for readout.

% \begin{figure*}[h!]
%     \centering
%     \includegraphics[width = \textwidth]{Applications.pdf}
%     \caption{\textcolor{black}{Schematic of chiral AFM's prospective applications, on artificial synapses, spiking neurons, probabilistic computing, memory devices, thermoelectrics and quantum computing.}}
%     \label{fig:applications}
% \end{figure*}

\section{Applications} \label{sec:application}
Their rich bandstructure physics and the range of dynamical responses in the presence of charge current and magnetic field make Mn$_3$Sn and other negative chirality materials an excellent choice for new spintronic computing applications. 
Of particular interest in this regard are biologically inspired computing and probabilistic computing paradigms. As we discuss below, the SOT-induced incoherent oscillations of the order parameter in Mn$_3$Sn have striking resemblance to the time-domain response of spiking biological neurons. On the other hand, low-energy barrier Mn$_3$Sn thin films or nanodots could exhibit thermally induced switching which could be used for implementing p-bits, which are an important building block for stochastic computing hardware and Boltzmann machines. 
%The model used for the simulations \textcolor{black}{for the probabilistic switching} reported here can be consulted in the appendix. 
\textcolor{black}{Other applications include non-volatile memory devices, high-frequency signal generators/detectors, 
ANE-based thermoelectric generators and sensors and quantum computing applications.}

% \begin{figure*}
%     \centering
%     \includegraphics[width = 0.95\linewidth]{Zheng2.pdf}
%     \caption{\textcolor{black}{(a) Long-term depression (LTD) and potentiation (LTP) achieved with good linearity via 
%     120 negative and positive current pulses, respectively, in an Mn$_3$Sn/Pt/Mn sample driven by OT.  
%     (b) Schematic of the constructed ANN with $100 \times 100$ memory cells for image recognition task.
%     (c) Image accuracy rates as a function of learning epochs using three kinds of devices. An ideal device corresponds to one where the non-linearity (NL) index for weight update is set to zero, whereas the NL index for spin-torque (ST)-driven device is 0.686 and that of the OT-driven device is only 0.166. Reproduced with permission from Ref.~\cite{zheng2024effective}.}}
%     \label{fig:AFM_synapse}
% \end{figure*}

\begin{figure*}
    \centering
    \includegraphics[width = 0.95\linewidth]{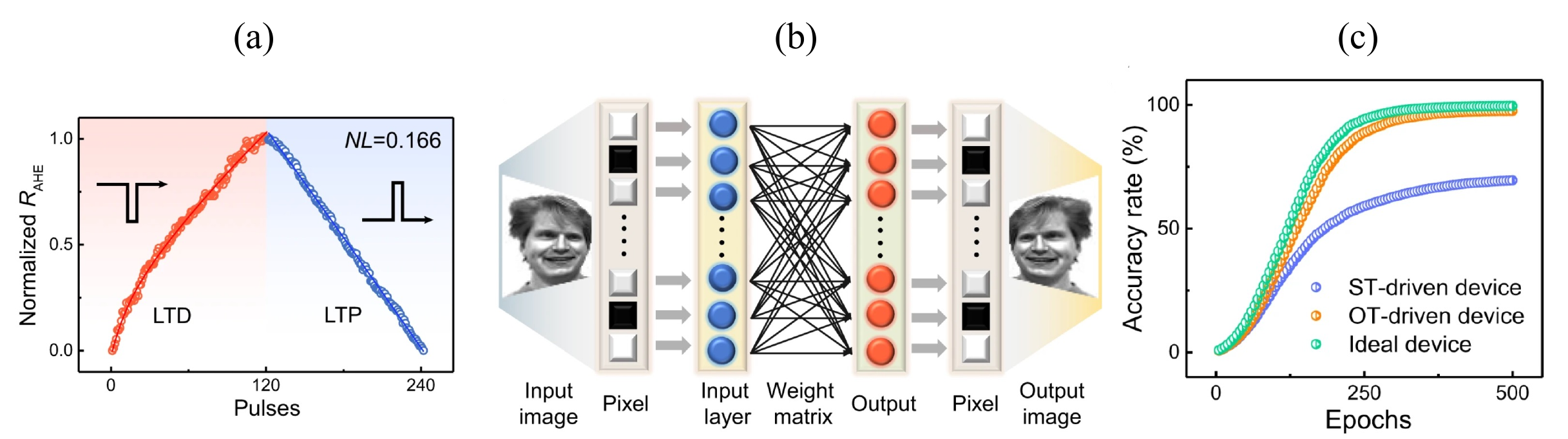}
    \caption{\textcolor{black}{(a) Long-term depression (LTD) and potentiation (LTP), depicted by the increase and decrease of the AHE resistance, $R_{\mathrm{AHE}}$, for 120 negative and 
 positive current pulses, respectively, is achieved in an OT-driven Mn$_3$Sn/Pt/Mn sample. Here, charge current is converted into orbital current in the Mn layer while Pt layer assists in orbital-current-to-spin-current conversion.  $R_{\mathrm{AHE}}$ displays a good linear dependence on the number of current pulses.
    (b) Schematic of the constructed ANN with $100 \times 100$ memory cells for image recognition task.
    (c) Image accuracy rates as a function of learning epochs using three kinds of devices. An ideal device corresponds to one where the non-linearity (NL) index for weight update is set to zero, whereas the NL index for spin-torque (ST)-driven device, comprising bilayer of Mn$_3$Sn/Pt, is 0.686 and that of the OT-driven device is only 0.166, which is much closer to an ideal device~(NL = 0) with linear LTD and LTP.
     Reproduced with permission from Ref.~\cite{zheng2024effective}.}}
    \label{fig:AFM_synapse}
\end{figure*}

\textcolor{black}{
\subsection{Neuromorphic computing}
Neuromorphic computing leverages the dynamical principles of the brain to implement computing tasks in hardware for AI applications. The basic building blocks of a neuromorphic computer include neurons and synapses. A neuron integrates the membrane potential frequently and fires (\emph{i.e.}, generates a spike) once the integrated potential exceeds a threshold. Synapses capture the strength of the connection between neurons and are associated with learning and memorizing information in the brain.
The dynamics of spintronic devices is inherently non-linear, while these devices can also be fabricated at the nanoscale.~\cite{zhou2021prospect} As such, spintronics-based neurons and synapses have been proven to be much more area- and energy-efficient compared to CMOS-only implementations.~\cite{kumar2023ultralow} Domain wall motion, skyrmions, spin torque oscillators, and magnetic tunnel junctions have been utilized toward the hardware emulation of neurons and synapses.~\cite{fukami2018perspective} Of these implementations, antiferromagnetic neurons, which can generate voltage spikes when excited by a spin torque above a threshold, can mimic the spiking response of biological neurons.~\cite{parthasarathy2021precessional} Likewise, spintronic synapses utilizing domain-wall motion in magnetic materials can result in nonvolatile multi-state resistance states.~\cite{roy2024spintronic} In the following, we discuss theoretical and experimental advances in utilizing spin-torque driven dynamics in Mn$_3$Sn to emulate neurosynpatic functionality in hardware.}

\subsubsection{\textcolor{black}{Artificial synapses}}
\textcolor{black}{
Several emerging devices, including resistive RAM, phase-change RAM, and spintronics devices have been used as synapses to store non-volatile weights in 
artificial neural networks (ANNs).
A spintronic synapse based on domain-wall motion must exhibit multiple resistance states that are 
%domain-wall synapse requires multiple resistance states that can be 
efficiently manipulated by spin torque. To pin the domain walls in desired positions and mitigate cycle-to-cycle drifts, artificial pinning of domain walls via magnetic and geometric modification of the device has been previously explored.~\cite{kumar2023ultralow} Current-controlled domain-wall motion in ferro-~\cite{sengupta2016short} and ferri-magnetic~\cite{liu2022compensated} materials
%Domain wall motion 
has also been used to implement long-term depression (LTD) and long-term potentiation (LTP) in ANNs. 
}
\textcolor{black}{
%In artificial neural networks (ANN), 
LTD is the process in which the efficiency of synaptic connections between neurons decreases with time, while
LTP refers to the increase in the synaptic strength. 
%LTD and LTP have been achieved through current-induced magnetization switching processes in ferro-~\cite{sengupta2016short} and ferri-magnetic~\cite{liu2022compensated} materials, especially adopting domain wall motion. 
The linearity of the measured output is significant for the accuracy of ANNs~\cite{park2020emerging} and requires precise control of domain-wall motion, which in turn needs a large footprint along the direction of propagation for a quasi-1D confinement of domain-wall dynamics. Moreover, material defects~\cite{wang2021multi}  may restrict the improvement of linearity because of random pinning and depinning of domain walls.
}

\textcolor{black}{In the work by Zheng \emph{et~al.},~\cite{zheng2024effective} OT generated in a trilayer setup comprising Mn$_3$Sn/Pt/Mn was utilized to switch the AHE signal (Fig.~\ref{fig:Zheng1}). In this work, 120 positive (negative) current pulses of amplitude $4.5~\mathrm{MA/cm^2}$ each, along the $[0001]$ ($[000\bar{1}]$) direction, which is perpendicular to the Kagome plane, were employed to mimic the LTD (LTP) process.~\cite{zheng2024effective}  
Figure~\ref{fig:AFM_synapse}(a) shows the quasi-linear dependence of normalized anomalous Hall resistance, $R_{\mathrm{AHE}}$, on the number of applied pulses. $R_{\mathrm{AHE}}$ can be treated as the weight, $G$, of the artificial synapse in the ANN. 
The performance of an ANN crossbar array of size $100 \times 100$ with Mn$_3$Sn-based synapses was also evaluated. 
An image recognition task, as depicted in Fig.~\ref{fig:AFM_synapse}(b), was implemented on the ANN crossbar array. 
As shown in Fig.~\ref{fig:AFM_synapse}(c), the learning accuracy of the synaptic network comprising OT-driven Mn$_3$Sn devices was found to be approximately 97.5\%, which was only slightly lower than that of the ideal device with perfect linearity and significantly higher than that of spin-torque-driven Mn$_3$Sn devices. The lower accuracy of spin-torque Mn$_3$Sn synapses could be due to the inhomogeneous switching of the sample as a result of the higher Joule heating effect.}

\textcolor{black}{In a recent work by Wu \emph{et al.}, N\'{e}el-like domain wall motion was experimentally probed via MOKE in Mn$_3$Ge and Mn$_3$Sn.~\cite{wu2024current} It was found that current densities as low as $7.5\times 10^6~\mathrm{A/cm^2}$ could result in domain-wall velocity of 750$~\mathrm{m/s}$ in Mn$_3$Ge. Similar to ferromagnets, the motion of the domain wall was found to occur through thermal creep for currents below $\sim 2.5\times 10^6~\mathrm{A/cm^2}$ in Mn$_3$Ge. The velocity of N\'{e}el-like domain walls was found to be higher than that of Bloch-like domain walls in both Mn$_3$Ge and Mn$_3$Sn. The mobility of domain walls in Mn$_3$Ge and Mn$_3$Sn was extracted to be more than an order of magnitude higher than that in ferromagnetic materials. 
For a sample size of 6$~\mathrm{\mu m}$ width, 45$~\mathrm{\mu m}$ length, and 800$~\mathrm{nm}$ thickness, the energy consumption during the application of nanosecond pulses was estimated to be $\sim 25~\mathrm{nJ}$. Thus, for a scaled domain-wall synapse, the energy consumption could be on the order of 100's of $\mathrm{fJ}$.
The high domain-wall velocity and mobility at low critical current densities demonstrates the superior potential of chiral AFMs to realize ultra-fast and energy-efficient synaptic devices in a spintronic ANN. 
Yet, future research must systematically investigate the impact of materials properties and device design on the energy consumption and latency of chiral AFM-based synapses.  
 }

\subsubsection{Spiking neurons}
In pursuit of enhanced energy efficiency compared to ANNs, spiking neural networks (SNNs)~\cite{yamazaki2022spiking} mimic more closely the spiking behaviors of biological brains. The fundamental computing units of SNNs, known as spiking neurons, accumulate potential in response to input and fire when this potential surpasses a predefined threshold. This ``Integration-Firing'' dynamics is essential to the functionality of a spiking neuron. Implementing these neurons with beyond-CMOS technology could further improve the energy efficiency and operational frequency of SNNs.~\cite{nikonov2019benchmarking}

\textcolor{black}{
Several prior works have explored the use of antiferromagnetic materials to implement artificial spiking neurons.~\cite{khymyn2017antiferromagnetic, khymyn2018ultra, shukla2022spin, bradley2023artificial, bradley2024pattern, bradley2023antiferromagnetic, bradley2024antiferromagnetic} For example, Khymyn \emph{et al.}~\cite{khymyn2018ultra} have shown that the thresholding behavior of spin torque induced spiking dynamics can be used in ``future neuromorphic signal processing circuits working at clock frequencies of tens of gigahertz.'' Likewise, Bradley \emph{et al.} showed that antiferromagnetic neurons could generate spikes with a duration on the order of picoseconds, while consuming $\sim 1~\mathrm{fJ}$ per spiking operation.~\cite{bradley2023artificial} Moreover, antiferromagnetic neurons are also shown to possess built-in features, such as response latency, refraction, and inhibition, that resemble the features of biological neurons. More recently, a spiking neural network with antiferromagnetic neurons was designed for pattern recognition using spike pattern association neuron (SPAN) algorithm. The network was trained under 1$~\mathrm{\mu s}$ for MNIST pattern recognition task, while the energy consumption of the network was estimated to be around 1$~\mathrm{pJ}$ during inference.~\cite{bradley2024pattern}
}

\begin{figure*}[ht!]
    \centering
    \includegraphics[width = 0.95\linewidth]{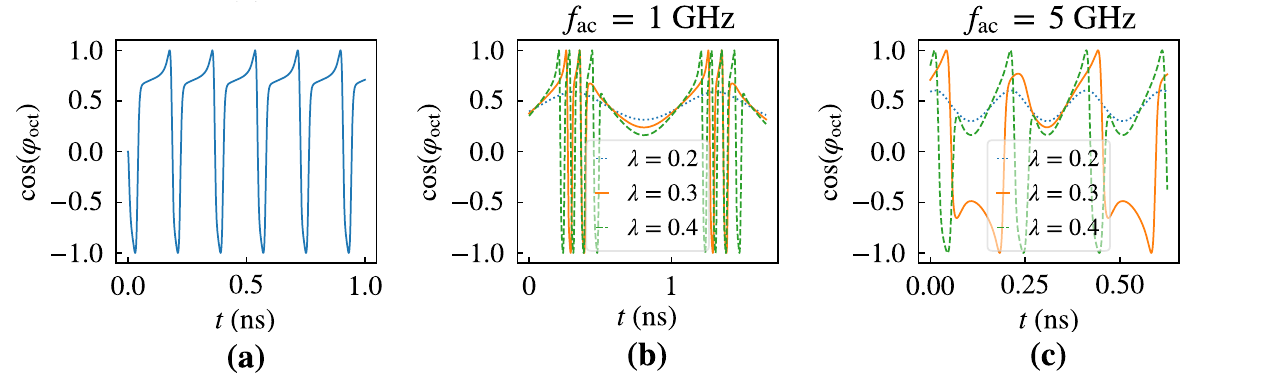}
    \caption{The dynamics of octupole's azimuthal angle versus time (a) applied d.c. current $J = 2.7~\mathrm{MA/cm^2}$ where the threshold current density $J_s^{\mathrm{cr1}} = 2.6~\mathrm{MA/cm^2}$. (b) Applied a.c. current $J = 0.8J_s^{\mathrm{cr1}} + \lambda J_s^{\mathrm{cr1}} \cos(2\pi f_{\mathrm{ac}}t)$ and $f_{\mathrm{ac}} = 1~\mathrm{GHz}$ and (c) $f_{\mathrm{ac}} = 5~\mathrm{GHz}$.} 
    \label{fig:computing}
\end{figure*}

Mn$_3$Sn is a promising material for building spiking neurons, owing to a Dirac-comb-like nonlinear dynamics for high damping and currents near the  threshold current.~\cite{shukla2022spin}
As illustrated in Fig.~\ref{fig:computing}(a), for a two-fold anisotropy Mn$_3$Sn, when the input spin current density slightly exceeds a critical value, denoted by $J_s^\mathrm{cr1}$, the in-plane component of the magnetic octupole measured via the TMR, $\propto \cos{\qty(\varphi_\mathrm{oct})}$, ($\varphi_\mathrm{oct}$ is the azimuthal angle of the magnetic octupole)
demonstrates spiking behavior in the time domain.
\textcolor{black}{Moreover, the device automatically resets after the pulsed input current is turned off.}
\textcolor{black}{In the classic Leaky-Integrate-and-Fire model of spiking neurons, the superposition of current flowing from connected synapses mimics the integration process, while the spiking behavior, when the input exceeds a set threshold value, corresponds to the firing operation of a biological neuron. The activity of the artificial AFM neurons, made using Mn$_3$Sn as the active element, is suppressed for a period of time, known as the refractory period, immediately after a spiking event as depicted in Fig.~\ref{fig:computing}(a), which agrees well with the functionality of biological spiking neurons.}

Previous works~\cite{khymyn2018ultra, sulymenko2018ultra, shukla2022spin} have shown that the rate of firing of these neurons can be controlled if the d.c. current is superimposed with a small a.c. current such the net current just exceeds $J_s^\mathrm{cr1}$. 
Furthermore, the co-application of a subthreshold direct current (d.c.) stimulus (e.g., $0.8J_s^{\mathrm{cr1}}$) with an a.c. component (e.g., $\lambda J_s^{\mathrm{cr1}} \cos(2\pi f_{\mathrm{ac}}t)$) can engender a richer spectrum of spiking behaviors. Both the amplitude and the frequency of the a.c. signal can control the nature of the spikes, \emph{i.e.}, it might be possible to generate single spikes as well as bursts of spikes as shown in Fig.~\ref{fig:computing}(b)\&(c).
Such neuron emulators could then be coupled to each other to get interesting dynamics, where the rate of firing of the second neuron could be controlled by that of the first. If the coupling is bidirectional in nature, unique dynamics, which can then be used for computation, e.g. collective state computing and oscillatory Hopfield network~\cite{csaba2020coupled} is possible.

\textcolor{black}{
Previous work has benchmarked the performance of Mn$_3$Sn neuron at the device and system level, by embedding it within an SNN to perform recognition tasks.~\cite{cruz2022performance, rakheja2024iccad} It has been shown that an Mn$_3$Sn neuron consumes roughly 12$~\mathrm{aJ}$ energy per spike for a time-domain input signal of 90$~\mathrm{ps}$ duration. The low energy consumption results from the low critical current required to excite a spiking response in chiral AFMs. 
In terms of network-level performance, the EDP of spintronics-based SNNs including Mn$_3$Sn-based neurons was found to as low as 5.30$\times 10^{-20}~\mathrm{J \cdot s}$, while that of analog-CMOS-based SNNs was 8.64$\times 10^{-16}~\mathrm{J \cdot s}$ 
for a small LeNet workload. }
\textcolor{black}{
The performance of Mn$_3$Sn neurons is significantly better than its spintronics counterparts. For example, it was shown in~\cite{srinivasan2017magnetic} that a stochastic SOT-controlled MTJ consumed 
1.16$~\mathrm{fJ}$ energy for 1$~\mathrm{ns}$ input current pulse, while a DW-based neuron consumed 0.14$~\mathrm{fJ}$ energy for 2-ns input current pulse.~\cite{chen2018magnetic}
}
% A phase-change memory synapse utilizing a nanopillar-based Ge$_2$Sb$_2$Te$_5$ active element consumed $\sim$1.8 pJ per paired-spike with $\sim$50~ns spikes and $\sim$1~$\mu$s inter-spike interval.~\cite{go2021fast} 
A comparison of various neuromorphic device technologies, including resistive, phase-change, ferroelectric, magnetic, and electrochemical RAM, for deep neural networks is presented in a recent article by Lelmini \& Ambrogio.~\cite{ielmini2019emerging} The authors showed that magnetic, ferroelectric, and Li-ion electrochemical RAM consumed 100$~\mathrm{fJ/bit}$, while magnetic and resistive RAM could operate under 10$~\mathrm{ns}$ for training. 
An SNN based in-memory accelerator with a spintronic computational RAM (\emph{i.e.}, STT-MRAM or SOT-MRAM) was shown to reduce the energy consumption by $\sim 150\times$ compared to 
 %when compared to 
a representative
application-specific integrated-circuit solution.~\cite{cilasun2021spiking} At the element level, it was shown that SOT-based computational RAM used $\sim$150$~\mathrm{fJ}$ energy for spiking at $\sim500~\mathrm{ns}$ latency resulting in a spike operation energy-delay product (EDP) of 7.85$\times 10^{-20}~\mathrm{J \cdot s}$.
\textcolor{black}{Despite the superior performance of Mn$_3$Sn neurons,} 
further research is needed to understand the impact of CMOS peripherals, scalability limits, and mapping of algorithms on to the hardware to harness the full potential of Mn$_3$Sn for \textcolor{black}{neuromorphic} computing. 

\subsection{Probabilistic computing}
While traditional computing is characterized by its pursuit of deterministic outputs, probabilistic computing has emerged as a novel paradigm integrating randomness and uncertainty into the computing process. Insights from neuroscience suggest that the human brain optimizes for energy efficiency at the expense of reliability by incorporating stochasticity in computation. This approach proves particularly effective in addressing specific problems such as factorization, satisfiability, and combinatorial optimization.~\cite{borders2019integer} 
In this context, the fundamental computing unit, known as the p-bit, functions akin to a controllable random number generator, with its output being a function of external input (\emph{i.e.}, SOT).

Low barrier MTJs are proposed as viable candidates for p-bits in diverse applications, including invertible logic gates and Boltzmann machines,~\cite{camsari2017stochastic} as well as Gaussian random number generators.~\cite{debashis2022gaussian} 
Their suitability arises from the ease of inducing stochastic behaviors with the aid of thermal noise. 
\textcolor{black}{Although there are no direct experimental reports on p-bits using Mn$_3$Sn, previous studies have evaluated its thermal stability and behavior in the presence of thermal noise. Specifically, \textcolor{black}{the effect of thermal noise during}
field-driven~\cite{sato2023thermal} and voltage-driven~\cite{kobayashi2023pulse} %probabilistic 
switching has been 
experimentally investigated in Mn$_3$Sn. To enable p-bit applications with Mn$_3$Sn, such as machine-learning-inspired Bayesian networks, Boltzmann machines, and invertible Boolean logic,~\cite{camsari2020charge} future work must systematically explore 
%but these works are not systematic in its 
the thermal stability and thermally driven dynamics of the octupole moment, \textcolor{black}{particularly for scaled devices in which the energy barrier separating the stable states is comparable to the thermal energy}. This field is rapidly advancing, both experimentally and theoretically, and \textcolor{black}{
additional research that} provides early insights into the underlying physics of stochastic switching and its application toward probabilistic computing \textcolor{black}{is urgently needed.}}

% Similar to ferromagnetic p-bits,~\cite{camsari2020charge}
% correlated probabilistic Mn$_3$Sn devices can be used for machine-learning-inspired Bayesian networks and Boltzmann machines, while they can also be used for implementing invertible Boolean logic and solving combinatorial optimization problems. 

\subsection{\textcolor{black}{High frequency signal generation and detection}}
\textcolor{black}{Magnetic materials that can operate at high frequencies are highly desirable for pushing the range and bandwidth of spintronics-based signal generators and detectors. Toward this,
antiferromagnetic materials are promising candidates as they can display ultra-fast dynamics resulting from their
strong exchange interactions.~\cite{baltz2018antiferromagnetic, sulymenko2017terahertz} 
While typically ferromagnetic materials have been adopted as GHZ-spin-torque oscillators (STO),~\cite{chen2016spin} microwave and radio frequency signal generators,~\cite{he2021integrated} and magneto-optical devices,~\cite{liu2021applications} it is expected that AFMs can be used as the active element of similar applications. 
For these applications, the oscillations in the order parameter can be extracted via the TMR in the device setup of Fig.~\ref{fig:device}. The TMR is sensitive to the orientation of the order parameter, \emph{i.e.}, the octupole moment in chiral AFMs and the N\'{e}el vector in collinear AFMs and can be utilized for the electrical readout. 
Several proposals for terahertz (THz) signal generation, detection, and signal analysis have appeared in the literature.~\cite{sulymenko2018terahertz, gomonay2018narrow, parthasarathy2021precessional, sulymenko2018electrodynamic, khymyn2017antiferromagnetic} For example, in Ref.~\citen{artemchuk2020terahertz}, a nanoscale spectrum analyzer, operational in the THz regime, is conceptualized using a four-layer tunnel junction comprising a lower Pt layer, a conducting AFM, an MgO barrier, and an upper Pt layer. The oscillations in the N\'{e}el vector are converted into an electrical signal by means of the tunneling anisotropic magnetoresistance. Output a.c. voltage levels on the order a milli-volt were shown to be feasible in this structure.  
Although these prior works are based on collinear AFMs, we also expect high-frequency dynamics in chiral AFMs. 
Theoretical proposals on electrical control of the  
THz rotation of the octupole moment in chiral AFMs have recently appeared in the literature.~\cite{zhao2021terahertz, lund2023voltage, hu2024current}}

\textcolor{black}{Lund~\emph{et al.}~\cite{lund2023voltage} theoretically predicted voltage-controlled high-bandwidth oscillations in noncollinear and Kagome AFMs, where the oscillation frequency is continuous from zero up to the THz range, modulated via the SOT. We also theoretically and numerically investigated the spin-current-driven auto oscillations of Mn$_3$Sn and Mn$_3$Ir in the THz regime, revealing the threshold current density and the dependence of oscillation frequency on magnetic parameters.~\cite{shukla2022spin} Our work also presented several device setups to electrically extract the generated a.c. signal due to the oscillation of the octupole moment. 
However, we must acknowledge that in the case of current-driven dynamics of the octupole moment, large currents are required to increase the oscillation frequency of the octupole moment, where Joule heating may become prohibitive for practical realization of such high-frequency AFM-based spintronic devices.
Nonetheless, the unique spin arrangement and the physics of chiral AFMs permits highly tunable dynamics of the octupole moment, which may eventually be utilized to develop tightly integrated oscillators in a nanoscale platform.}

\textcolor{black}{On the experimental front, 
Miwa~\emph{et al.}~\cite{miwa2021giant} observed field-induced magnetic octupole order oscillation in Mn$_3$Sn through time-resolved MOKE, with a resonant frequency as high as $0.86~\mathrm{THz}$ under a $2~\mathrm{T}$ external field applied along the $[2\bar{1}\bar{1}0]$ direction. 
Ultrafast photocurrents~\cite{hamara2023ultrafast} have also been generated in an Mn$_3$Sn thin film excited by optical pump-THz emission spectroscopy. The magnitude and direction of these photocurrents could be controlled by the polarization and incidence angle of the optical pump, but the currents remained robust to external field perturbations. THz time-domain spectroscopy has been utilized to determine the dynamical properties of Mn$_3$Sn at ultra-high frequencies in Ref.~\citen{matsuda2020room}, where the authors established that Mn$_3$Sn's 
room-temperature anomalous Hall conductivity was approximately $20~\mathrm{(\Omega.cm)^{-1}}$ for frequencies up to $2~\mathrm{THz}$. 
These phenomena collectively highlight the broad potential of chiral AFMs for the generation, manipulation, and sensing of high-frequency signals.}

\subsection{\textcolor{black}{Non-volatile memory}}
\textcolor{black}{Non-volatile memory~(NVM) has attracted attention not only for its persistent data storage, but also for in-memory computing~\cite{sebastian2020memory, bavikadi2020review} and neuromorphic computing.~\cite{byun2023recent, burr2017neuromorphic} Memory devices with faster speed, higher density, lower power consumption, and higher bandwidth are in high demand, and materials, such as ferromagnets, ferroelectrics, and phase change materials, are all candidates for novel memory designs. 
The large anomalous Hall conductivity and small coercivity of Mn$_3$Sn make it ideally suited for implementing low-power spintronic memory devices. Potentially, the speed of Mn$_3$Sn memory can be much higher than that of ferromagnet-based NVMs, while the lack of stray fields in AFMs can also lead to higher memory density. A qualitative sketch in Fig.~\ref{fig:mrma_mtj} highlights the key differences between NVMs implemented using ferromagnet-MTJs versus AFM-MTJs. 
\begin{figure}[h!]
    \centering
    \includegraphics[width=3.5in]{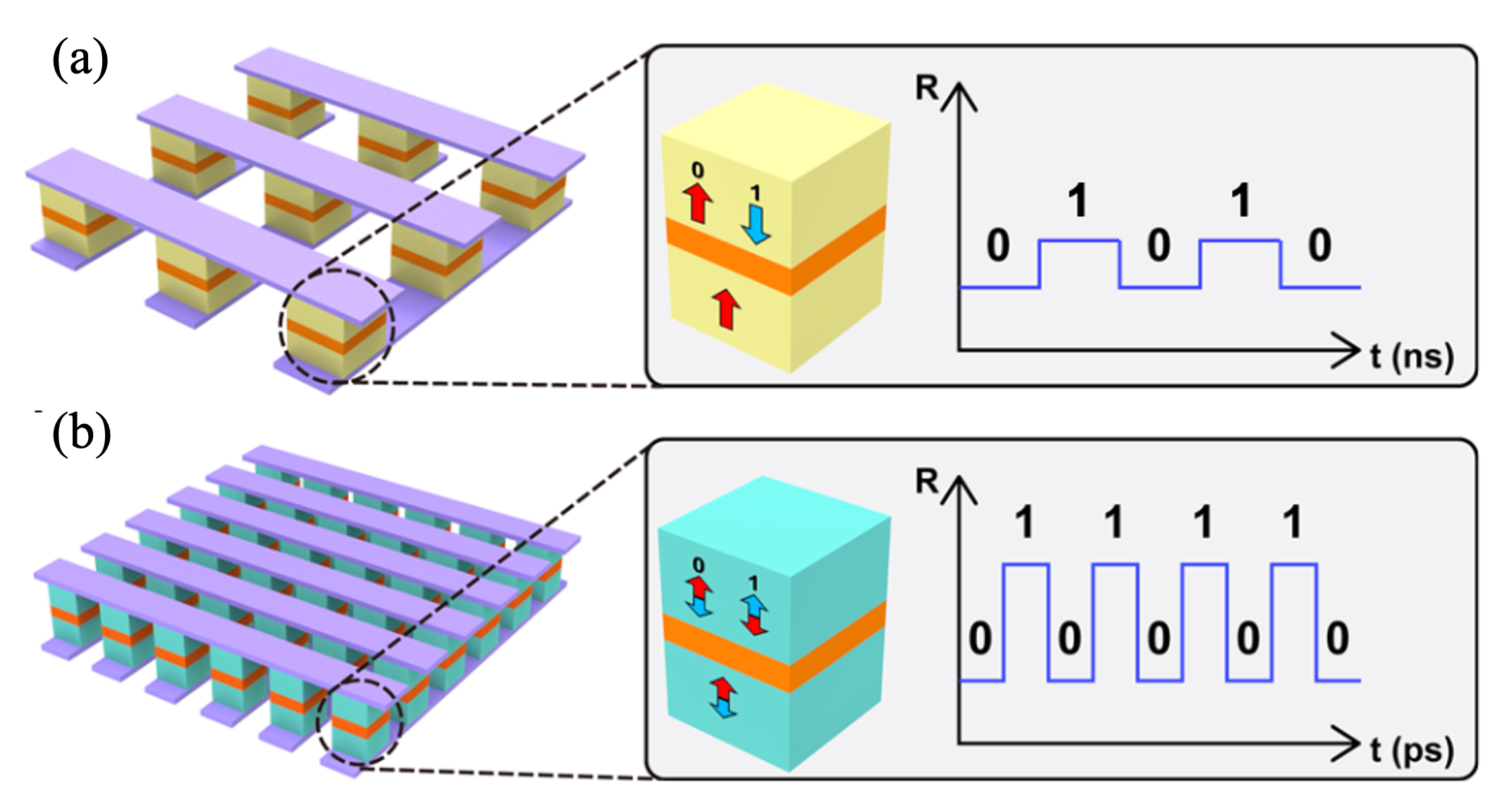}
    \caption{\textcolor{black}{An array of MTJs using (a) ferromagnets and (b) antiferromagnets as the active NVM element. Unlike ferromagnet-MTJs, AFM-based MTJs can be more tightly integrated leading to denser memory arrays, while they are also expected to operate at much faster speeds. Taken from~\cite{shao2024antiferromagnetic}.}}
    \label{fig:mrma_mtj}
\end{figure}
Yet, there are a few key challenges that must be resolved. For example, although the low coercive field of Mn$_3$Sn promotes energy-efficient write operations, it adversely affects the memory stability to external magnetic fields. To increase the memory density, device miniaturization, while ensuring reliable read and write operations is also essential. 
An important result in the context of NVMs was reported in 2021 by Tsai \emph{et al.}~\cite{tsai2021large} 
The authors showed that by optimizing the annealing conditions after the deposition of the heavy metal in Mn$_3$Sn/W, a large Hall resistance of $0.35~\Omega$ and a large read signal of $1~\mathrm{mV}$ could be achieved in the bilayer samples. 
This is an important step forward for realizing future memory technologies based on Mn$_3$Sn. 
Recently, Matsuo~\emph{et~al.}~\cite{matsuo2022anomalous} measured anomalous Hall signals in 100-$\mathrm{nm}$-scale Mn$_3$Sn grains in films thinner than $20~\mathrm{nm}$ at room temperature and at zero fields. This result signifies that Mn$_3$Sn samples can retain their topological features at the nanoscale limit, thus paving the path for miniaturized, high density memory applications that utilize AFMs as their key element. On the other hand, the large magnetic spin Hall effect of Mn$_3$Sn can generate a large out-of-plane spin accumulation,~\cite{kondou2021giant} which may enable the switching of a proximal ferromagnet with perpendicular magnetic anisotropy (PMA) for energy-efficient SOT NVM applications. For example, Hu \emph{et al.} demonstrated highly efficient, deterministic switching of Ni/Cr multilayers with PMA using the unconventional SOT generated in Mn$_3$Sn.~\cite{hu2022efficient} A switching current density of $4.6\times 10^6~\mathrm{A/cm^2}$ was found for Mn$_3$Sn/Cu/Ni-Cr samples, which was less than half of the switching current density of $\beta-$Ta/Cu/Ni-Cr samples, highlighting the potential of Mn$_3$Sn as an SOT source wherein an out-of-plane spin current with collinear spin polarization induced by in-plane charge current is feasible.
Thus, Mn$_3$Sn and related non-collinear AFMs can also be used as the SOT-generating layer in low-power non-volatile SOT memory devices.} 

\subsection{\textcolor{black}{ANE-based devices}}
\textcolor{black}{
The ANE in Mn$_3$Sn can be used for thermoelectric generation and for on-chip cooling. As a thermoelectric generator, Mn$_3$Sn can convert waste heat into usable electrical power, while it can also harvest heat from integrated circuits to provide localized cooling and thus improve the circuit reliability. 
Compared to a Seebeck thermopile, which consists of a pillar structure of p- and n-type semiconductors, the ANE-based thermopile has a planar structure and is thus simpler to fabricate. 
Owing to the lack of stray fields from AFMs, Mn$_3$Sn-based thermopiles can lead to densely integrated thermoelectric modules for providing efficient coverage of the heat source.~\cite{ikhlas2017large} 
Hideki \emph{et~al.}~\cite{narita2017anomalous} demonstrated the microfabrication of a thermoelectric element comprising Ta/Al$_2$O$_3$/Mn$_3$Sn layered structure, where Ta was used as the heating element, while Al$_2$O$_3$ enabled the diffusion of heat toward Mn$_3$Sn. 
The anomalous Nernst coefficient of the microfabricted Mn$_3$Sn was extracted to be $0.27~\mathrm{\mu V/K}$, which is similar to the Nernst coefficient of bulk Mn$_3$Sn. 
This result indicates the potential of developing efficient and dense thermopiles utilizing the topological features of microfabricated Mn$_3$Sn. A planar tilted Nernst thermopile using Mn$_3$Sn as the active element, illustrated in Fig.~\ref{fig:thermopile}, was demonstrated in Ref.~\citen{li2021monomaterial}.
In this design, the tiled structure simultaneously acts as the n- and p-legs of a conventional thermopile and thus the need for extrinsic contacts is obviated. Other applications of the ANE in Mn$_3$Sn include logic devices that utilize temperature gradients as the information token, sensor devices based on temperature gradients,~\cite{tanaka2023roll} efficient thermal management and spin current generation via the closely related spin-Seebeck effect,~\cite{yu2021large} and enabling efficient writing of heat-assisted magnetic memory devices.
\begin{figure}[h!]
    \centering
    \includegraphics[width=3in]{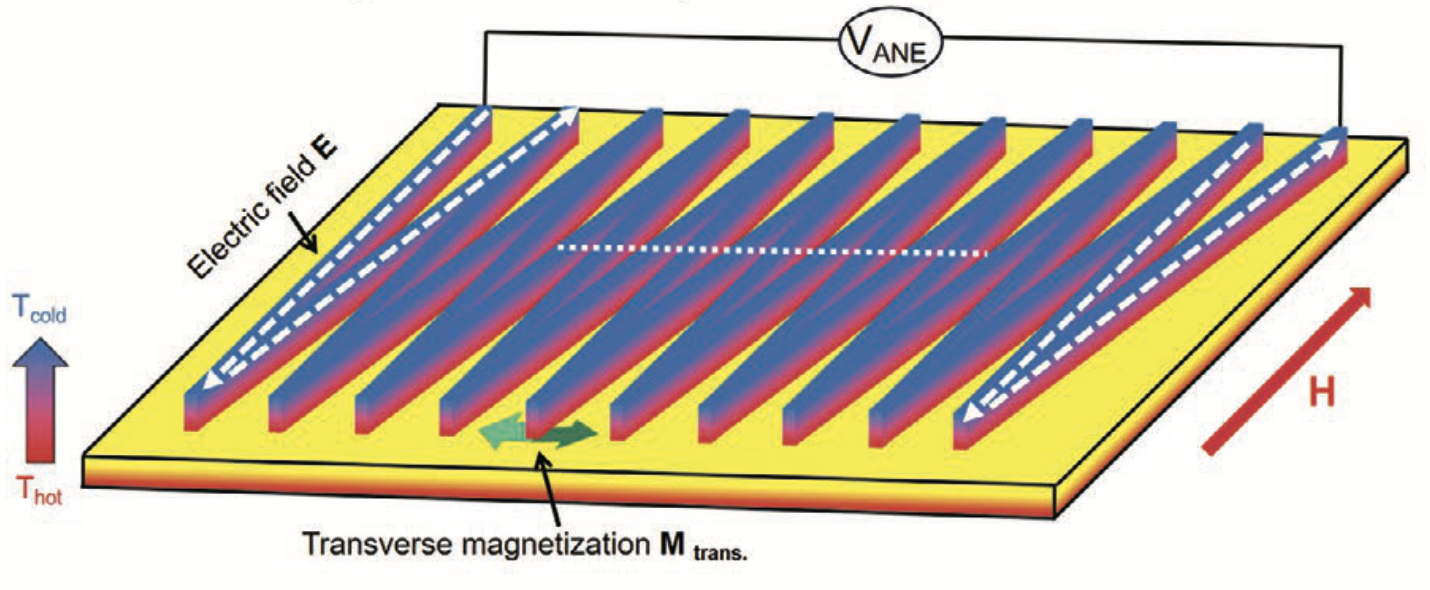}
    \caption{\textcolor{black}{A conceptual thermopile presented in Ref.~\cite{li2021monomaterial}, where the transverse magnetization direction is automatically alternated between adjacent elements. A sizeable transverse magnetization results during the domain reversal process in Mn$_3$Sn and thus Mn$_3$Sn can be used as the active element of the thermopile.}}
    \label{fig:thermopile}
\end{figure}
}

\subsection{\textcolor{black}{Quantum computing}}
\textcolor{black}{Quantum computing,~\cite{horowitz2019quantum} owing to the superposition of quantum states, is a promising solution for executing certain tasks that are intractable on classical computers. Superconducting qubits, typically implemented using Josephson junctions (JJs),~\cite{nielsen2010quantum} are among the most promising candidates for realizing scalable quantum computers. Superconductor/ferromagnet (SC/FM) heterostructures~\cite{cai2023superconductor} are promising platforms for future superconducting spintronics and quantum computation applications because of their unique physical properties, such as spin-triplet superconductivity,~\cite{bergeret2005odd} superconducting order parameter oscillation, and topological superconductivity.~\cite{lyuksyutov2005ferromagnet} However, such SC/FM structures require delicate interface engineering, including careful electronic energy band matching and circumventing out-of-plane stray fields.}

\begin{figure}[h!]
    \centering
    \includegraphics[width=3.5in]{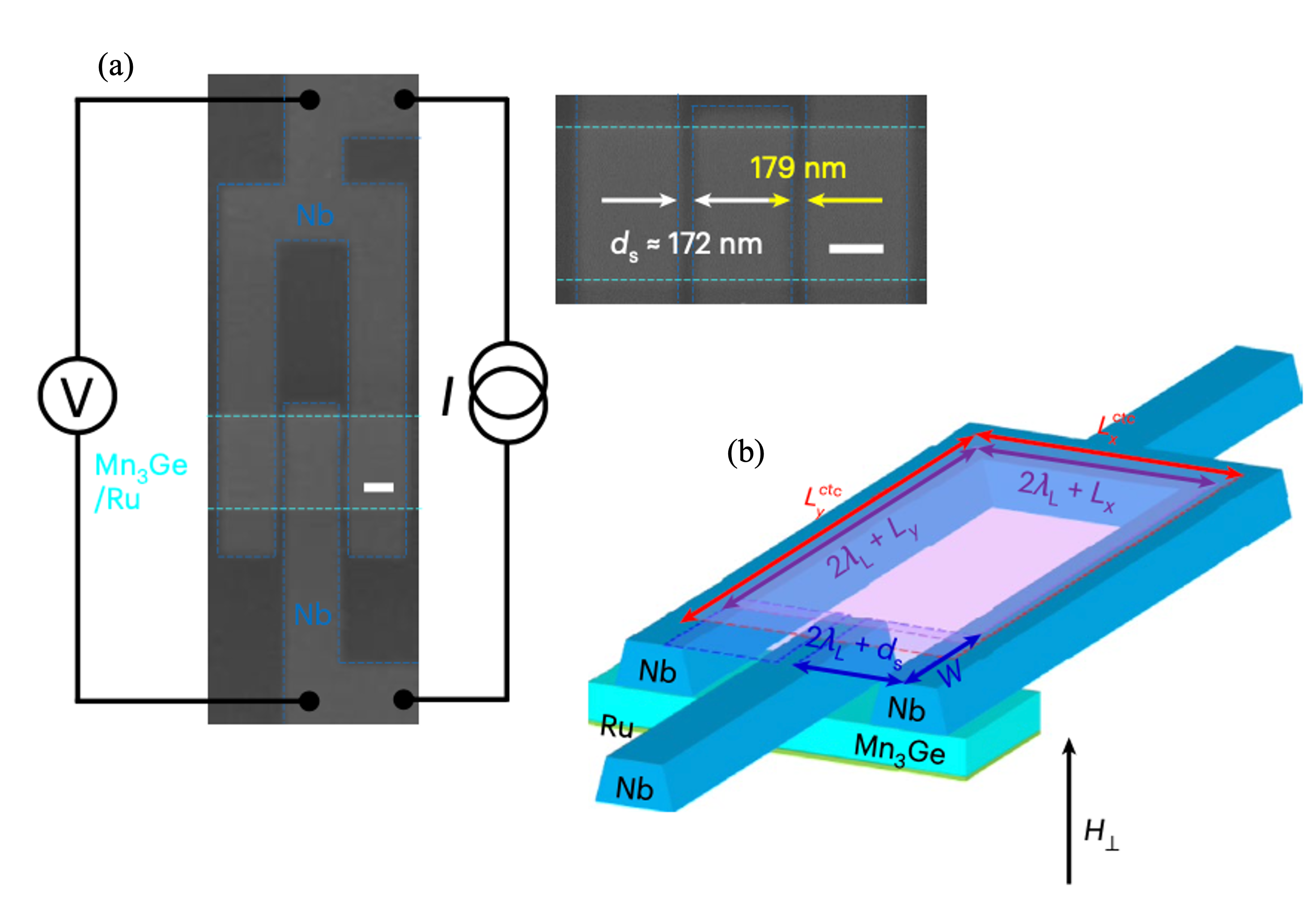}
    \caption{\textcolor{black}{(a) Scanning electron micrograph of the Mn$_3$Ge JJ and (b) measurement setup of the fabricated d.c. SQUID with two Nb/Mn$_3$Ge/Nb JJs separated by a distance, $d_s$, and connected through a single layer of Mn$_3$Ge. $\lambda_L$ is the London length and $L_{x,y}^\mathrm{ctc}$ denotes the center-to-center spacing between the tracks defining the opposite sides of the SQUID.}}
    \label{fig:JJ}
\end{figure}

\textcolor{black}{The topological chiral antiferromagnet Mn$_3$Ge~\cite{jeon2023chiral} has been shown to be effective in forming JJs, as well as in further applications in spin-triplet supercurrent spin valves and direct-current superconducting quantum interference devices (d.c. SQUIDs). In Ref.~\citen{jeon2023chiral}, a JJ was implemented using a 
Nb/Mn$_3$Ge/Nb structure and a d.c. SQUID is also fabricated containing two chiral Mn$_3$Ge antiferromagnetic spin-triplet JJs. An external field perpendicular to the kagome plane changes the associated Berry curvature~\cite{xiao2010berry} around the Fermi energy and subsequently enables triplet amplitude manipulation through an extremely small magnetic field ($<2~\mathrm{mT}$). 
These results are the first step toward chiral AFM applications for manipulating spin-triplets and logic circuits, while the lack of stray fields in AFMs can enable highly integrated superconducting spintronic logic circuits.~\cite{linder2015superconducting}}

\section{Conclusions and outlook}
Noncollinear AFMs with Kagome lattices exhibit properties such as spin momentum locking, topologically protected surface states, large spin Hall conductivity, and magnetic spin Hall effect arising from the topology. In particular, research pertaining to the physics and applications of the negative chirality noncollinear AFM, Mn$_3$Sn, have intensified in the last few years. 
Mn$_3$Sn can only be stabilized in excess of Mn, whereas with a lower Mn concentration the system gets contaminated with Mn$_2$Sn. Below its Ne\'{e}l temperature of $420-430~\mathrm{K}$, the negative chirality 120 degrees spin structure is stabilized due to the 
combination of exchange coupling and DM interaction between the Mn moments. 
With Mn deficiency, a helix phase emerges at intermediate temperatures where Mn moments form a spiral spin structure propagating along its c-axis.
At much lower temperatures ($\leq 50~\mathrm{K}$), spin glass like texture is also feasible. 
Due to their many intriguing transport properties including large spin Hall effect, anomalous Hall effect, and FM-like spin polarized charge current, negative chirality AFM metals can enable the development of AFM spintronic devices and applications along a similar route as their FM counterparts. 
\textcolor{black}{Furthermore, a comprehensive understanding of these chiral materials can also illuminate the pathways for experimental, theoretical, and applied research and development in the field of altermagnetism.}

In this paper, we presented recent developments, both on theoretical and experimental fronts, related to understanding Mn$_3$Sn's non-trivial bandstructure properties, which lead to strong magnetotransport signatures in the material. 
Anomalous Hall effect in Mn$_3$Sn, comparable to that in elemental FMs like Co, Ni, Fe has been experimentally reported. 
Magnetic switching of Mn$_3$Sn has been demonstrated in bilayers 
comprising the AFM and heavy metal and current densities ($\sim (10^{6}-10^{7}~\mathrm{A/cm^2}$) comparable to that used for switching FMs in FM/heavy metal bilayers have been obtained. In this structure, readout voltages $> 1~\mathrm{mV}$ could be detected by optimizing the annealing conditions and the seed layer during the fabrication process. 
Large magneto-optic Kerr signal, around $20~\mathrm{mdeg}$ at $300~\mathrm{K}$, has been reported in Mn$_3$Sn. Spin-torque ferromagnetic resonance measurements have been used to quantify the field-like torque due to the magnetic spin Hall effect and values greater than that of conventional heavy metals like Pt, Ta, and W have been found in Mn$_3$Sn. 
NV relaxometry measurements have been performed to determine the frequency of chiral oscillations of the order parameter in Mn$_3$Sn as a function of the input current. 
More recently, a finite TMR, around $2\%$ at $300~\mathrm{K}$, has been experimentally reported in all-AFM MTJs with Mn$_3$Sn electrodes and Mgo tunnel barrier. This discovery sets the stage for developing advanced CMOS-compatible spintronic devices using Mn$_3$Sn as the active magnetic element. Studies of strained Mn$_3$Sn thin films with two fold symmetric energy landscape suggest that they can be used as memory elements or high-frequency signal generators and detectors depending on the magnitude of input spin current and the magnetic field. 

Theoretical models of critical input current to induce different dynamics with and without magnetic fields in six-fold and two-fold symmetric Mn$_3$Sn can be valuable to gain physical insight into the effect of material parameters on the order parameter dynamics. Likewise, models of oscillation frequency and switching time versus input current can shed light on the performance metrics and scaling limits of Mn$_3$Sn-based spintronic devices. 
The effect of Joule heating in Mn$_3$Sn cannot be ignored, especially when the input current significantly exceeds the threshold current for a given dynamical response. However, by employing pulsed SOT in the short pulse regime, the issue of Joule heating can be reduced. Understanding the impact of thermal noise on the order parameter in Mn$_3$Sn is crucial for realizing non-volatile memory as well as p-bits, which serve as the fundamental unit of probabilistic computing hardware. Toward this, thermal stability of Mn$_3$Sn thin films with perpendicular anisotropy has been studied experimentally as well as theoretically by employing the N\'{e}el-Arrhenius model. 
However, further research in this area is needed to quantify Mn$_3$Sn's attempt frequency and the effective temperature in the presence of SOT. 

To move the research field further and to develop practical spintronic applications using chiral AFMs, experimental techniques to probe time-domain dynamics without Joule heating are needed. Advances in improving TMR in all-AFM tunnel junctions are also critical for ensuring robust readout signals that are compatible with CMOS circuitry. Additional work is needed to elucidate the role of thermal noise in Mn$_3$Sn thin films of various dimensions. Theoretical calculations will be indispensable to shed light on the exotic bandstructure physics of these materials and provide a theoretical rigor to experimental investigations. Meanwhile, the state of analytical models must be further advanced to enable technology-device co-design and benchmarking. 
Development of prototypes of antiferromagnetic spintronic devices that are CMOS-compatible can allow this field to rapidly advance and lead to novel computing paradigms with superior performance compared to their ferromagnetic-only counterparts. 

\section*{Author Contributions}
Ankit Shukla: methodology, investigation, formal
analysis, investigation, data curation, and writing – original
draft. Siyuan Qian: investigation, data creation, and
writing – review. Shaloo Rakheja: supervision, conceptualization, methodology, resources, funding acquisition, and writing – review
and editing. All authors discussed the results and the implications of this manuscript. All authors have given approval to the
final version of the manuscript.

\section*{Conflicts of interest}
There are no conflicts to declare.

\section*{Acknowledgements}
This research was partially supported by the NSF through the University of Illinois at Urbana-Champaign Materials Research Science and Engineering Center DMR-1720633. The authors also acknowledge the support of AFRL/AFOSR, under AFRL Contract No. FA8750-21-1-0002 and ZJU-UIUC Joint Research Center on Dynamic Research Enterprise for Multidisciplinary Engineering Science (DREMES).

%%%END OF MAIN TEXT%%%

%\section{Appendix}
%\subimport{./}{appendix.tex}

\balance


%%%REFERENCES%%%
\bibliography{rsc} %You need to replace "rsc" on this line with the name of your .bib file
\bibliographystyle{rsc} %the RSC's .bst file

\end{document}